\documentclass[a4paper]{article}
\usepackage{anysize}
\marginsize{2cm}{2cm}{2cm}{2cm}

\usepackage[utf8]{inputenc}

\usepackage[title]{appendix}

\usepackage{textcomp}
\usepackage{graphicx}
\usepackage{physics}
\usepackage{mathtools}
\usepackage{adjustbox}

\usepackage{yhmath}

\usepackage{tikz}
\usetikzlibrary{patterns,calc,matrix,backgrounds,fadings,shapes,arrows,shadows,decorations.pathreplacing,shadows.blur}

\pgfkeys{tikz/mymatrixenv/.style={decoration={brace},every left delimiter/.style={xshift=8pt},every right delimiter/.style={xshift=-8pt}}}
\pgfkeys{tikz/mymatrix/.style={matrix of math nodes,nodes in empty cells,left delimiter={(},right delimiter={)},inner sep=1pt,outer sep=0pt,column sep=0pt,row sep=2pt,nodes={minimum width=20pt,minimum height=10pt,anchor=center,inner sep=0pt,outer sep=0pt}}}
\pgfkeys{tikz/mymatrixbrace/.style={decorate,thick}}

\newcommand*\mymatrixbraceleft[4][m]{
    \draw[mymatrixbrace] (#1.west|-#1-#3-1.south west) -- node[left=2pt] {#4} (#1.west|-#1-#2-1.north west);
}
\newcommand*\mymatrixbraceright[4][m]{
    \draw[mymatrixbrace] (#1.east|-#1-#2-1.north east) -- node[right=2pt] {#4} (#1.east|-#1-#3-1.south east);
}
\newcommand*\mymatrixbracetop[4][m]{
    \draw[mymatrixbrace] (#1.north-|#1-1-#2.north west) -- node[above=2pt] {#4} (#1.north-|#1-1-#3.north east);
}
\newcommand*\mymatrixbracebottom[4][m]{
    \draw[mymatrixbrace] (#1.south-|#1-1-#3.north east) -- node[below=2pt] {#4} (#1.south-|#1-1-#2.north west);
}

\usepackage{makecell}

\usepackage{amsmath}

\usepackage{adjustbox}

\usepackage{wrapfig}
\usepackage{xparse}

\tikzfading[name=arrowfading, top color=transparent!0, bottom color=transparent!95]
\tikzset{arrowfill/.style={#1,general shadow={fill=black, shadow yshift=-0.8ex, path fading=arrowfading}}}
\tikzset{arrowstyle/.style n args={3}{draw=#2,arrowfill={#3}, single arrow,minimum height=#1, single arrow,
single arrow head extend=.3cm,}}

\NewDocumentCommand{\tikzfancyarrow}{O{2cm} O{black} O{left color=blue!10, right color=blue!80} m}{
\tikz[baseline=-0.5ex]\node [arrowstyle={#1}{#2}{#3}] {#4};
}

\NewDocumentCommand{\tikzlessfancyarrow}{O{2cm} O{black} O{left color=white, right color=white} m}{
\tikz[baseline=0.5ex]\node [arrowstyle={#1}{#2}{#3},text width=3cm] {#4};
}
\usepackage{parskip}
\usepackage{xargs,ifthen,xstring,xparse,etoolbox,mathtools}
\usepackage{environ}
\usetikzlibrary{quantikz}

\usepackage{tabulary}
\renewcommand{\arraystretch}{1.1}
\usepackage{multirow}

\usepackage[left=1.5cm,right=1.5cm,top=1.5cm,bottom=1.5cm,ignoreheadfoot]{geometry}
\usepackage{array}
\usepackage{hyperref}
\usepackage{amsmath}
\usepackage{amsfonts}
\usepackage{dsfont}
\usepackage{bbm}

\usepackage{caption}
\usepackage{subcaption}

\usepackage{xspace}
\usepackage{tcolorbox}

\usepackage{xcolor,colortbl}
\usepackage{framed}
\newcommand\R[0]{\ensuremath{\mathbb{R}}}

\newcommand{\maxkcut}[1]{MAX $#1$-CUT}
\newcommand{\maxcut}{MAX ($2$-)CUT\xspace}

\usepackage{authblk}

\usepackage{tcolorbox}
\usepackage{soul}
\usepackage{csquotes}

\usepackage[numbers]{natbib}

\definecolor{Gray}{gray}{0.85}
\newcolumntype{a}{>{\columncolor{Gray}}r}

\NewDocumentCommand\citelong{m O{}}{%
    \citeauthor{#1}~(\citeyear{#1})#2\cite{#1}%
}

\title{Efficient encoding of the weighted \maxkcut{k}\\ on a quantum computer using QAOA}

\author[$\dagger$]{Franz G. Fuchs\footnote{On behalf of all authors, the corresponding author states that there is no conflict of interest.}}
\author[$\ddagger$]{Herman Øie Kolden}
\author[$\ddagger$]{Niels Henrik Aase}
\author[$\dagger$]{Giorgio Sartor}
\affil[$\dagger$]{SINTEF AS, Department of Mathematics and Cybernetics, Oslo, Norway}
\affil[$\ddagger$]{NTNU, Department of Physics, Trondheim, Norway}

\date{\today}

\usepackage[printwatermark]{xwatermark}
\newwatermark[allpages,angle=90,scale=.45,xpos=-10cm,ypos=0]{Submitted to SN Computer Science}

\begin{document}

\maketitle
\begin{abstract}
The weighted \maxkcut{k} problem consists of finding a $k$-partition of a given weighted undirected graph $G(V,E)$ such that the sum of the weights of the crossing edges is maximized.
The problem is of particular interest as it has a multitude of practical applications.
We present a formulation of the weighted \maxkcut{k} suitable for running the quantum approximate optimization algorithm (QAOA) on noisy intermediate scale quantum (NISQ)-devices to get approximate solutions.
The new formulation uses a binary encoding that requires only $|V|\log_2k$ qubits.
The contributions of this paper are as follows:
i) A novel decomposition of the phase separation operator based on the binary encoding into basis gates is provided for the \maxkcut{k} problem for $k>2$.
ii) Numerical simulations on a suite of test cases comparing different encodings are performed.
iii) An analysis of the resources (number of qubits, CX gates) of the different encodings is presented.
iv) Formulations and simulations are extended to the case of weighted graphs.
For small $k$ and with further improvements when $k$ is not a power of two, our algorithm is a possible candidate to show quantum advantage on NISQ devices.
\end{abstract}

\section{Introduction and Related Work}
The search for quantum algorithms of practical interest has intensified since the announcement of quantum supremacy in~\cite{Arute2019}.
For the foreseeable future, quantum hardware will limit the depth (length of the computation) and width (number of qubits) of the algorithms that can be run.
Hybrid quantum-classical algorithms based on the variational principle are a promising approach to achieve an advantage over purely classical algorithms.
The variational quantum eigensolver (VQE)~\cite{peruzzo2014variational}/quantum approximate optimization algorithm (QAOA)~\cite{farhi2014quantum} is such a hybrid algorithm for approximately finding the solution of a problem encoded as the ground state of a Hamiltonian.
In this early stage, even small reductions of the depth and/or width of an algorithm can make the difference between success and failure.
In light of this we investigate in this article how QAOA can be used to approximately solve the \maxkcut{k} problem. The problem has interesting applications that make it practically relevant.
These range from placement of television commercials in program breaks, placement of containers on a ship with $k$ bays, partition a set of items (e.g. books sold by an online shop) into $k$ subsets, design of product modules, frequency assignment problems, scheduling problems and pattern matching~\cite{Aardal2007,gaur2008capacitated}.

The problem discussed in this paper falls within the class of Ising models.
An Ising model is a mathematical model of ferromagnetism in statistical mechanics, consisting of discrete variables $s_i$ that represent atomic ``spins'' that can be in one of the two states $\pm1$.
The objective function of an Ising model is given by
\begin{equation}
    \label{eq:Es}
    E(\mathbf{s}) = \sum_{i=1}^N \sum_{j=i+1}^N J_{i,j} s_i s_j + \sum_{i=1}^N h_i s_i,
\end{equation}
where $h_i$ are the biases and $J_{i,j}$ the coupling strengths.
Using the transformation $s_i = 2 x_i -1$ this can be transformed into a quadratic unconstrained binary optimization problem (QUBO) which is given by
\begin{equation}
    \underset{\mathbf{x}\in\{0,1\}^N}{\min} \mathbf{x}^T Q \mathbf{x} = \sum_{i\leq j}^N x_i Q_{i,j} x_j, \quad x_i \in\{0, 1\},
\end{equation}
where the matrix $Q$ is an upper-diagonal $N\times N$ real matrix.
In this way, the Ising model without an external field is equivalently formulated as a \maxcut problem.
For an overview of other Ising-type formulations of NP problems, we refer to~\cite{lucas2014ising}, which includes a discussion of graph coloring, but not of \maxkcut{k}.
A generalization of the Ising model is given by the Potts model, where the spin takes one of $k$ possible values, see~\cite{Wu1982}.
The \maxkcut{k} problem is connected to the search for a ground state in the anti-ferromagnetic k-state Potts model~\cite{welsh1993complexity}.
Using Equation~\eqref{eq:Es} and replacing the terms $s_i$ with Pauli-Z operators, one arrives at an Ising Hamiltonian, which ground states, i.e. solutions of the original problem, can be found (approximatively) by the QAOA.
The QAOA consists of the following main steps:
\begin{center}
\begin{adjustbox}{width=.75\textwidth}
\begin{tikzpicture}[scale=0.7]
        \path (.5,3) node[draw,shade,
      top color=yellow!40,
      bottom color=green!5,
      rounded corners=6pt,
      blur shadow={shadow blur steps=5}](a) {\parbox[c][3\baselineskip][c]{3cm}{\centering Optimization\\ Problem}};
      
        \path (4.5,3.5) node(s1) {(S1)};
        \path (8.5,3) node[draw, ellipse](c) {
        \parbox[c][2\baselineskip][c]{2cm}{\centering Hamiltonian}};

        \path (17.5,5) node[draw,shade,
      top color=blue!40,
      bottom color=blue!5,
      rounded corners=6pt,
      blur shadow={shadow blur steps=5}](d) {\parbox[c][3\baselineskip][c]{5cm}{\centering \ul{\textbf{Quantum Device}:} \begin{itemize}
            \item[(S2)] Prepare trial state 
            \item[(S3)] Measure 
        \end{itemize}}};

        \path (17.5,1) node[draw,shade,
      top color=green!40,
      bottom color=green!5,
      rounded corners=6pt,
      blur shadow={shadow blur steps=5}](e) {\parbox[c][3\baselineskip][c]{5cm}{\centering \ul{\textbf{Classical Device}:} \begin{itemize}
            \item[(S4)] Update control parameters
        \end{itemize}}};
        
        \draw[very thick, -triangle 45] ($(a)+(3,0)$) -- ($(c)+(-2.8,0)$);
        
        \draw[very thick, -triangle 45] ($(c)+(2.5,0)$) -- ($(c)+(4.8-.25,0)$);
        \draw[decoration={brace},decorate, very thick]
  ($(c)+(5.1-.15,-2.25)$) -- node {} ($(c)+(5.1-.15,2.25)$);

        \draw[very thick, dashed, -triangle 45, bend left] ($(e)-(4.3,.2)$) to node [midway,above]{solution} ($(a)+(0,-1.2)$);
        \draw[very thick, -triangle 45]
            (d) edge[bend right] node {} (e)
            (e) edge[bend right] node {} (d);
        \path (20,3) node(o) {optimize $\theta$};
\end{tikzpicture}
\end{adjustbox}
\end{center}
\begin{itemize}
    \item[(S1)] The solution of a problem is formulated as the ground state of a Hamiltonian $H_P$ that encodes a cost function $f$ to be optimized. It acts diagonally on the computational states, i.e., $H_P \ket{z} = f(z)\ket{z}$.
    \item[(S2)] A quantum processor prepares a parametrized quantum state $\ket{\Psi(\theta)}=U_M(\theta_{2p})U_P(\theta_{2p-1}) \cdots U_M(\theta_{2})U_P(\theta_1) \ket{\Phi_0}$, by alternatingly applying phase separation ($U_P$) and mixing ($U_M$) operators on an easy to prepare initial state $\ket{\Phi_0}$.
    \item[(S3)] Through repeated measurement one obtains an estimate of $E(\theta)=\langle H_P \rangle_{\ket{\Psi(\theta)}}\in \R$ as well as a candidate solution $y$ with probability $|\bra{y}\ket{\Psi(\theta)}|^2$.
    \item[(S4)] The cost function $E(\theta)
    \geq E_\text{min}$ serves a classical computer that finds the ground state energy of the cost function, i.e., finds the optimal parameter $\theta$ such that $E(\theta)$ becomes minimal.
    This iterative process provides candidate solutions, that are typically approximate.
\end{itemize}

A general overview of hybrid quantum classical algorithms (VQE/QAOA) is provided in, e.g., \cite{Moll2018}. The article discusses obstacles and how to overcome them in order to achieve quantum advantage on noisy intermediate scale quantum devices.
The QAOA was introduced by \cite{farhi2014quantum} where it was applied to \maxcut.
Solving small problem instances of \maxcut with when QAOA and classical AKMAXSAT solver, the authors in~\cite{guerreschi2019qaoa} extrapolate to large instances and estimate that a quantum speed-up can be obtained with (several) hundreds of qubits.
It has also been shown numerically that the QAOA can achieve solutions of better quality~\cite{crooks2018performance} then the best known classical approximation algorithm.
The authors in \cite{Zhou2020} introduce heuristic strategies inspired by quantum annealing to generate good initial points for the outer optimization loop for the \maxcut problem. They show that this leads to large improvements in the approximation ratio achieved.

Since its inception, there have been several extensions/variants of the QAOA proposed.
A recent approach, dubbed \textit{ADAPT-QAOA}, presented in~\cite{zhu2020adaptive} is to create an iterative version that is problem-tailored and can adapt to specific hardware constraints.
The method is exemplified on a class of \maxcut problems, requiring fewer CNOT gates as the original method.
A non-local version of QAOA is proposed in~\cite{bravyi2019obstacles}.
Dubbed \textit{R-QAOA}, the algorithm recursively removes variables from the Hamiltonian until the remaining instance is small enough to be solved classically.
Numerical evidence is provided that shows this procedure significantly outperforms standard QAOA for frustrated Ising models on random 3-regular graphs for the \maxcut problem.
Another recent approach, dubbed \textit{WS-QAOA} is using the solutions of classical algorithms to improve QAOA, see~\cite{egger2020warm}.
An example is provided with \maxcut, which shows numerically that warm-starting QAOA and R-QAOA provides an advantage at low depth, in the form of a systematic increase in the size of the obtained cut for fully connected graphs with random weights.
Warm starting results in a change of the mixer operator only.

To the best of our knowledge, there are only two papers discussing \maxkcut{k} for $k>2$.
The quantum alternating operator ansatz (also abbreviated as QAOA) presented in \cite{hadfield2019quantum} considers general parametrized families of unitaries.
The paper presents a suite of constrained optimization problems, such as maximum independent set, travelling sales person, and the \textit{un}weighted \maxkcut{k}.
Mixing operators are adapted such that the probability of transitioning from a feasible candidate to another is non-zero and circuit compilations are described.
The paper does not provide numerical simulations and the main focus is on the design of mixing operators.
The one-hot encoding of the \maxkcut{k} is further studied numerically in~\cite{Wang2020}
Two approaches are presented that tackle the enforcement of the hard constraints arising from the encoding scheme.
The first is to keep the $X$-mixer but introduce a penalty term in the phase separating Hamiltonian and the second is to instead design an $XY$-mixer together with consistent $W_k$-initial states to stay within the feasible set of solutions.
Both articles~\cite{hadfield2019quantum,Wang2020} present the \textit{un}weighted \maxkcut{k}, although it is not hard to generalize.

The main contributions of this article are:
\begin{itemize}
    \item A novel decomposition of the unitary phase separation operator $U_P$ based on the binary encoding into basis gates is provided for the \maxkcut{k} problem for the general case $k>2$.
    \item Numerical simulations on a suite of test cases comparing different encodings are performed.
    \item We present an analysis of the resources (number of qubits, CX gates) of the different encodings.
    \item The formulations and simulations are extended to the case of weighted graphs.
\end{itemize}
The \textit{main advantages} as compared to \cite{hadfield2019quantum,Wang2020} are that our approach is efficient in the number of qubits and does not require feasibility constraints to be incorporated into the circuit construction of the mixer operator.
Similar to~\cite{Wang2020} we observe that the resulting energy landscape of the binary encoding might be easier to handle for the outer classical optimization loop due to fewer local minima see Table~\ref{tab:E_Barbell}.
As pointed out in~\cite{Wang2020} the ratio of the size of the feasible subspace, which is spanned by states corresponding to $n$ Hamming-weight one bit strings, to the size of the full Hilbert space is
\begin{equation}\label{eq:feasiblesub}
    \frac{\dim(\mathcal{H}_\text{feas})}{\dim(\mathcal{H})} = \left(\frac{k}{2^k}\right)^n,
\end{equation}
which becomes exponentially small (for $k\geq1$) as the graph size $n$ grows.
In contrast, the binary encoding uses the full Hilbert space as feasible space.

The rest of the article is organized as follows.
We describe the classical problem and classical algorithms in Section~\ref{sec:classicalformulation}.
After describing and comparing one-hot encoding and the proposed binary encoding scheme in Section~\ref{sec:quantum}, we discuss implementation and results are presented in Section~\ref{sec:results}, followed by a conclusion in Section~\ref{sec:conclusion}.

\section{The \texorpdfstring{\maxkcut{k}}{MAX k-CUT} Problem and Classical Algorithms}\label{sec:classicalformulation}

\begin{figure}
\centering
    \begin{subfigure}[b]{.35\textwidth}
    \centering
    \includegraphics[width=0.8\linewidth]{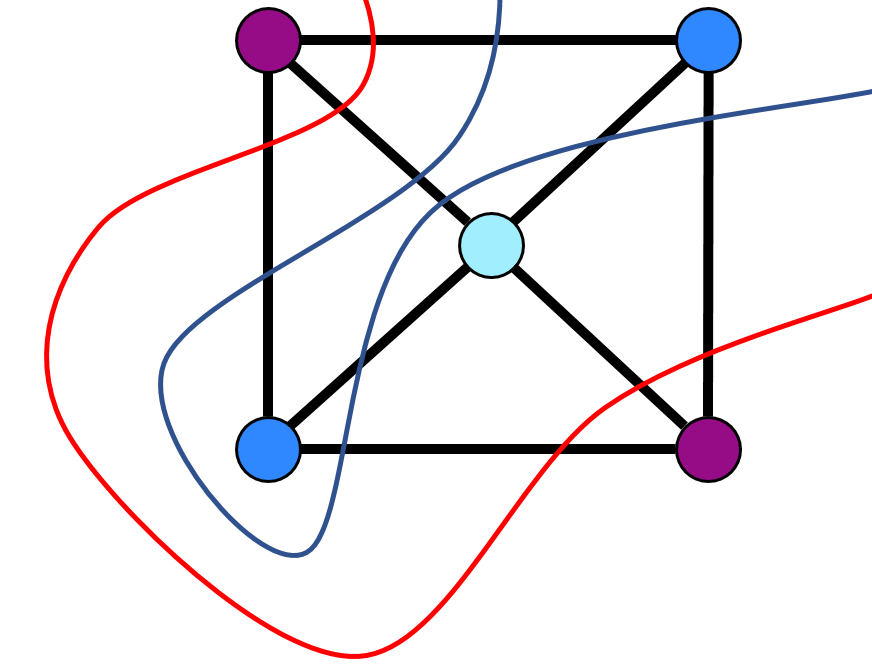}
    \caption{An example of an optimal solution for a \maxkcut{3} problem.}
    \label{fig:example_max4cut}
    \end{subfigure}
    \hspace{1cm}
    \begin{subfigure}[b]{.55\textwidth}
    \centering
    \begin{tabular}{c}
        \begin{tabular}{c|cccc}
            k & 2 & 3 & 4 & 5 \\
            \hline
            $\alpha$ & .878567 & .836008 & .857487 & .876610 \\
        \end{tabular}
        \\
        \\
        \begin{tabular}{c|cccc}
            k & 6 & 7 & 8 & 9 \\
            \hline
            $\alpha$ & .891543 & .903259 & .912664 & .920367 \\
        \end{tabular}
        \\
        \\
    \end{tabular}
        \caption{Approximation guarantees for \maxkcut{k} using classical approximation algorithms, see~\cite{goemans1995improved,Klerk2004}.}
        \label{tab:gw_ratios}
    \end{subfigure}
        \caption{The \maxkcut{k} problem and approximation guarantees for classical algorithms with polynomial runtime.}
        \label{fig:classical_maxkcut}
\end{figure}

The \maxkcut{k} problem is an extension of the well known \maxcut problem (or simply MAX CUT). Given a weighted undirected graph $G=(V,E)$, \maxkcut{k} consists of finding a maximum-weight $k$-cut, that is a partition of the vertices into $k$ subsets such that the sum of the weights of the edges that have end points on different subsets is maximized. Let $w_{ij}$ be the weight assigned to each edge $(i,j)\in E$, and let $\mathcal{P}=P_1,\dots,P_k$ be a partition of the vertices in $V$.
Then the cost function for \maxkcut{k} can be defined as:
\begin{equation}
\label{eq:maxkcut}
    \underset{|\mathcal{P}|=k}{\max} \ \sum_{1\leq r<s\leq k} \sum_{i\in P_r, j\in P_s, (i,j)\in E} w_{ij}.
\end{equation}
Alternatively, one could assign a label $x_i\in \{1, \dots, k\}$ to each vertex $i\in V$, indicating which partition the vertex belongs to. Defining $\mathbf{x} = (x_1,\ldots,x_{|V|}) $ the optimization problem for \maxkcut{k} can be written as:
\begin{equation}
\label{eq:maxkcolor}
    \underset{\mathbf{x}\in\{1,\dots,k\}^n}{\max} C(\mathbf{x}), \qquad  C(\mathbf{x}) = \sum_{(i,j)\in E} w_{ij} [x_i \neq x_j],
\end{equation}
where $C(\mathbf{x})$ is the cost function and $[\cdot]$ is the Iverson bracket, which is $1$ if $x_i \neq x_j$, and 0 otherwise. An example of an optimal solution for \maxkcut{3} is given in Figure~\ref{fig:example_max4cut}.

The \maxkcut{k} problem is NP-complete 
and it has been shown that it does not admit any polynomial-time approximation scheme, for any $k\geq2$, unless P=NP \cite{frieze1997improved}.
By definition, a randomized approximation algorithm for \eqref{eq:maxkcolor} has \textbf{approximation ratio} $\alpha$ if
\begin{equation}\label{eq:approximationratio}
    \mathbb{E}[C(\mathbf{x})] \geq \alpha C(\mathbf{x}^*),
\end{equation}
where $\mathbf{x}^*$ is the optimal solution of \eqref{eq:maxkcolor}.
A trivial algorithm that (uniformly) randomly assigns vertices to partitions has an approximation ratio of $(1-1/k)$, because each edge has a probability of $(1-1/k)$ of having endpoints in different partitions~\cite{frieze1997improved}.
It has also been shown that there can be no polynomial time approximation scheme (PTAS) with approximation ratio $(1 - \frac{1}{34k})$, unless P=NP \cite{kann1996hardness}.
For \maxkcut{2}, the Goemans-Williamson algorithm~\cite{goemans1995improved} exploits the semidefinite programming (SDP) relaxation of the integer programming formulation of \maxkcut{2} to achieve an approximation ratio of 0.878567. The Jerrum-Frieze algorithm~\cite{frieze1997improved} extended this result to \maxkcut{k}, obtaining an approximation ratio of  $(1-\frac{1}{k} + (1+\varepsilon(k))\frac{2ln(k)}{k^2})$, where $\varepsilon(k)$ is a function that approaches 0 as $k\rightarrow \infty$. For small $k$, this approximation ratio was ever so slightly improved in \cite{Klerk2004}. Figure~\ref{tab:gw_ratios} provides an overview of selected approximation ratios achieved in~\cite{goemans1995improved,Klerk2004}.

It is interesting to note that under the Unique Games Conjecture, both the 0.878567 approximation ratio of \cite{goemans1995improved} (for $k=2$) and the $(1-\frac{1}{k} + \frac{2ln(k)}{k^2})$ approximation ratio of \cite{frieze1997improved} (for large $k$) are optimal~\cite{khot2007optimal}. Given that the Unique Games Conjecture is not valid when there are entangled provers~\cite{kempe2007unique,Kempe2010}, it is possible that quantum algorithms may allow for an improvement over classical algorithms.

\section{Quantum Algorithms}\label{sec:quantum}
As a first step, we need to encode the problem described in Section~\ref{sec:classicalformulation} in a way that is suitable for the QAOA.
There are three different possibilities (of which the first two are presented in \cite{hadfield2019quantum}, and the last is proposed in this article):
\begin{itemize}
    \item \emph{Qudit encoding:} Expressing the solutions as strings of $k$-dits (as in Equation~\eqref{eq:maxkcolor}) is a natural extension of the \maxcut problem to $k>2$.  The problem can be formulated using $|V|$ qudits. In order to be practically relevant it requires, however, the realization of a k-level quantum system.
    \item \emph{One-hot encoding:} A second method is to use $k$ bits for each vertex, where the single bit that is $1$ encodes which set/color the vertex belongs to.
    Using this encoding requires $k|V|$ qubits.
    However, the formulation requires the introduction of constraints in order to prevent solutions where a vertex belongs to several sets of a partition or none.
    \item \emph{Binary encoding:} For a given $k$ we encode the information of a vertex belonging to one of the sets by $\ket{i}_L$, which requires $L=\lceil log_2(k)\rceil$ qubits. Here $\lceil \cdot \rceil$ means rounding up to the nearest integer. This formulation can be executed on systems using qubits and requires $L|V|$ qubits.
\end{itemize}
Binary encoding uses exponentially fewer qubits as compared to one-hot encoding.
As an example, for $k=4$ encoding the information of a vertex belonging to one of the four sets using one-hot encoding is done through identifying color 1,2,3,4 with the bit strings $0001, 0010, 0100, 1000$, respectively.
The binary encoding identifies colors 1,2,3,4 with the bit strings $00, 01, 10, 11$, respectively.
Observe that for one-hot encoding there are $2^4-4=12$ possible bit strings in the space that encode infeasible solutions consisting of multiple colors or no color at all, whereas all possible bit strings in the binary encoding are valid encodings, see also Equation~\eqref{eq:feasiblesub}.

In the following we describe the problem Hamiltonian and unitary evolution for the one-hot encoding 
as well as the proposed binary encoding.

\subsection{One-hot Encoding}
Here, we provide a brief description of the one-hot encoding. For details and further discussion we refer to \cite{hadfield2019quantum,Wang2020}.
The one-hot encoding uses $k$ qubits per vertex, which are indexed such that, e.g., $\sigma^{x,y,z}_{i,a}$ applies a Pauli-X,-Y, or -Z gate to qubit number number $ik+a$, for $a\in\{1,k\}$.
The definition of the approximation ratio (Equation~\eqref{eq:approximationratio} needs to be adapted to
\begin{equation}
    \alpha = \frac{\langle P_\text{feas} H_P P_\text{feas} \rangle}{C(\mathbf{x}^*)},
\end{equation}
where $P_\text{feas}$ is the projection operator onto the feasible subspace.
In practice this means that infeasible solutions are assigned zero cost.

\subsubsection{Problem Hamiltonian}
Up to a global phase the problem Hamiltonian is given by
\begin{equation}\label{eq:Hbin}
    H_P = \sum_{(i,j)\in E} w_{i,j}\sum_{a=1}^k \sigma^z_{i,a} \otimes  \sigma^z_{j,a}.
\end{equation}
One way to incorporate the constraint that the feasible subspace consists of only Hamming-weight one bit strings is to introduce a quadratic penalty term that results (up to global phase) in the Hamiltonian
\begin{equation}\label{eq:Hbin_pen}
     H_\text{pen} = \frac{1}{2}\sum_{v=1}^{|V|} \sum_{a=1}^{k}\sum_{b=a+1}^{k}  \sigma^z_{v,a} \otimes  \sigma^z_{v,b}.
\end{equation}
Overall, the phase separating Hamiltonian becomes a weighted sum
$H_P'=H_P + \beta H_\text{pen}$,
where $\beta$ should satisfy $\beta \geq \frac{|V|}{k}$ and $\beta > k|E|$, see~\cite{Wang2020}.

\subsubsection{Unitary Evolution}
The unitary evolution consists of creating an initial state, followed by phase-separating and mixing operators.
The unitary evolution of the phase-separating operator given by the exponentiation of $H_P'$ (see Equations~\eqref{eq:Hbin} and \eqref{eq:Hbin_pen}) consists of terms that can be realized through the following circuit
\begin{equation}
    e^{-i\frac{\theta}{2}\sigma^z_j \sigma^z_k}
    =
    \begin{quantikz}
         \lstick{$q_j$} & \qw & \ctrl{1} & \qw & \ctrl{1} & \qw \\
         \lstick{$q_k$} &\qw & \targ{} & \gate{R_z(\theta)} & \targ{} & \qw  \\
    \end{quantikz},
\end{equation}
where $R_z(\theta) = e^{-i\frac{\theta}{2}\sigma^z}$.
The \textbf{standard} $\mathbf{X}$-\textbf{mixing} operator is given by 
\begin{equation}
    e^{-i H_M}, \text{ where } H_M = \sum_{v=1}^{|V|} \sum_{a=1}^k \sigma^x_{v,a},
\end{equation}
with each individual term realized through $R_x(\theta)=e^{-i\frac{\theta}{2}\sigma^x}$-gates.
The initial state, when using the standard mixing operator is given by $\ket{\Phi_0} = H^{\otimes k|V|}\ket{0}$, where $H$ is the Hadamard gate.
However, this approach does not incorporate the feasibility constraint.

Incorporating the feasibility constraint into the mixer results in the $\mathbf{XY}$-\textbf{mixer} for each vertex $v\in V$ based on the Hamiltonian
\begin{equation}
    H_{XY,v} = \frac{1}{2} \sum_{a,b\in K} \sigma^x_{v,a} \sigma^x_{v,b} + \sigma^y_{v,a} \sigma^y_{v,b},
\end{equation}
where $K$ is a set consisting of certain pairs of colors $(a,b)$.
In this article we use the \textit{parity-partitioned} mixer, which can be represented as two separate Hamiltonians
\begin{equation}
    \begin{split}
        H_\text{odd} &= H^{XY}_{(1,2)} + H^{XY}_{(3,4)} + \dots + H^{XY}_{(k-1,k)}\\
        H_\text{even} &= H^{XY}_{(2,3)} + H^{XY}_{(4,5)} + \dots + H^{XY}_{(k,1)},\\
    \end{split}
\end{equation}
where $H^{XY}_{(j,k)} = \sigma^x_j \sigma^x_{j+1} + \sigma^y_j \sigma^y_{j+1}$.
The resulting unitary operator is easily implemented in terms of two $CX$- and one $R_X$ or $R_Y$ operation.
A feasible initial state for the one-hot encoding consistent with the $XY$ mixer is $\ket{\Phi_0} = \ket{W_k}^{\otimes|V|}$, where the $W_k$-state is given by
\begin{equation}
    \ket{W_k} = \frac{1}{\sqrt{k}}\left(\ket{100\dots000} + \ket{010\dots000} + \dots \ket{000\dots001}\right).
\end{equation}
An efficient algorithm for this with logarithmic (in $k$) time complexity is presented in~\cite{cruz2019efficient}.

\subsection{Binary Encoding}
In the following we describe the problem Hamiltonian for the proposed binary encoding, which is given as the sum of local terms, i.e.,
\begin{equation}
\label{eq:HP}
    H_P = \sum_{(i,j)\in E} w_{i,j} H_{i,j},
\end{equation}
where $w_{i,j}$ is the weight of the edge between vertices $i$ and $j$ as well as the resulting unitary evolution.
\subsubsection{Problem Hamiltonian}
The matrix $H_{i,j}$ is a diagonal matrix modeling the interaction between vertices $i$ and $j$
\begin{equation}
H_{i,j}=  \begin{pmatrix}
    d_{0} & & \\
    & \ddots & \\
    & & d_{2^{2L}-1}
  \end{pmatrix}.
  \label{eq:Hij}
\end{equation}
From this point on we consider the two diagonal matrices $H_P$ and $A=aI+bH_P$ to be equivalent for all $a,b\in\R, b\neq0$.
The reason for this is that when we compare the unitary operators $e^{-i \theta A}$ and $e^{-i \theta B}$, a parameter $a\neq0$ results in applying a ``global phase'' which is irrelevant, and $b\neq0$ can be combined with the parameter $\theta$.
As mentioned in~\cite{hadfield2019quantum} ``an affine transformation of the objective function [...] corresponds simply to a physically irrelevant global phase and a rescaling of the parameter''.
The cost function can be easily evaluated classically, independent of the specific form of $H_P$.

From now on we will adapt the notation that $\ket{m}_{2^n}$ is the $m$-th basis vector of an $n$-qubit system. Note that for a basis vector the decomposition $\ket{m}_{2^n}=\ket{l_0}_{2^{n-1}}\otimes\ket{l_1}_{2^{n-1}}$ both exists and is unique.
The $m$-th entry of the local Hamiltonian $H_{i,j}$ are given by
\begin{equation}
    d_m =
    \begin{cases}
    -1, & \text{ if } l_0\neq l_1 \wedge \neg (l_0\geq k-1 \wedge l_1\geq k-1), \text{ where $l_0, l_1$ are given by } \ket{m}_{2^n}=\ket{l_0}_{2^{n-1}}\otimes\ket{l_1}_{2^{n-1}},\\
    +1, & \text{ otherwise. }
    \end{cases}
\end{equation}
This means that eigenvectors of the local Hamiltonian $H_{i,j}$ corresponding to eigenvalues $d_m=-1$ indicate a cut.
When k is not a power of two the condition $\neg (l_0\geq k-1 \wedge l_1\geq k-1)$ is introduced such that the sets with number $k-1,\dots, 2^L-1$ are not distinguished and become the same set.
Organizing the diagonal entries $d_m$ in a matrix of size $2^L \times 2^L$ we get a particularly simple structure
\begin{equation}
\begin{split}
D &=  \begin{pmatrix}
    d_0 & \cdots & d_{2^L-1} \\
    d_{2^L} & \cdots & d_{2^{L+1}-1} \\
    \dots & \cdots & \dots \\
    d_{2^{2L-1}}& \cdots & d_{2^{2L}-1} \\
  \end{pmatrix}
  =    \begin{adjustbox}{width=.3\textwidth}
    \begin{tikzpicture}[baseline={-.5ex},mymatrixenv]
        \matrix [mymatrix,inner sep=4pt] (m)  
        {
            2I-J & -J \\
            \hline
            -J & J \\
        }; 
        \mymatrixbracetop{1}{2}{$2^L$}
        \mymatrixbracebottom{1}{1}{$k-1$}
        \mymatrixbraceleft{1}{2}{$2^L$}
        \mymatrixbraceright{1}{1}{$k-1$}
        \draw (m-1-2.north west) -- (m-2-2.south west);
    \end{tikzpicture}
    \end{adjustbox}\\
    & =
    \begin{pmatrix}
    (2I-J)_{k-1,k-1} & -J_{k-1,l} \\
    -J_{l,k-1} & J_{l,l} \\
    \end{pmatrix}
    = (2I - J)_{2^L,2^L} + 2\sum_{c,d=k+1, c\neq d}^{2^L} \Gamma_{2^L,2^L}^{c,d}
    ,
\end{split}
    \label{eq:D}
\end{equation}
where $l = 2^L-(k-1)$, $I$ is the identity matrix, $J$ is a matrix of ones, and $\Gamma^{c,d}$ is a matrix that has a one at entry $c,d$ and is zero otherwise.
Sub-indices indicate the size of the matrix.
Observe, that $D=D^T$ and that the sum involving terms $\Gamma$ is zero if $k$ is a power of two. We can construct the matrix $H_{i,j}$ from $D$ through
\begin{equation}
    H_{i,j} = \text{diag}\left(\text{vec}(D^T)\right),
\end{equation}
where $\text{vec}(\dot)$ is a linear transformation which converts a matrix into a column vector by stacking the columns on top of each other, and $\text{diag}(v)$ is a matrix with the entries of the vector $v$ along its diagonal.

Next, we will provide a few examples.

\textbf{\maxcut.}
For $k=2$ we can use $L=\lceil log_2(2)\rceil=1$ qubit per vertex, where  $\lceil \cdot \rceil$ means rounding up to the nearest integer. The matrix $D$ and the local Hamiltonian are given by
\begin{equation}
    D =
            \begin{tikzpicture}[baseline={-.5ex},mymatrixenv]
        \matrix [mymatrix,inner sep=4pt] (m)  
        {
            \phantom{-}1 &-1 \\
            \hline
            -1 &\phantom{-}1 \\
        }; 
        \draw (m-1-2.north west) -- (m-2-2.south west);
    \end{tikzpicture},
    \quad
    H_{i,j} = \begin{pmatrix*}[r]
    D_{1,1} & 0 & 0 & 0\\
    0 & D_{1,2} & 0 & 0\\
    0 & 0 & D_{2,1} & 0\\
    0 & 0 & 0 & D_{2,2}\\
    \end{pmatrix*} = \begin{pmatrix*}[r]
    1 & 0 & 0 & 0\\
    0 & -1 & 0 & 0\\
    0 & 0 & -1 & 0\\
    0 & 0 & 0 & 1\\
    \end{pmatrix*}.
\end{equation}

\textbf{\maxkcut{3}.}
For the case $k=3$ we need  $L=\lceil log_2(3)\rceil=2$ qubits per vertex.
Since two qubits can encode 4 different sets, we need to make two sets indistinguishable.
Choosing sets $2$ and $3$ to represent one set, the entries of the matrix $D$ and the local Hamiltonian are given by
\begin{equation}
    D =
        \begin{tikzpicture}[baseline={-.5ex},mymatrixenv]
        \matrix [mymatrix,inner sep=4pt] (m)  
        {
            \phantom{-}1 &-1 & -1& -1 \\
            -1 &\phantom{-}1 & -1& -1 \\
            \hline
            -1 &-1 & \phantom{-}1& \phantom{-}1 \\
            -1 &-1 & \phantom{-}1& \phantom{-}1 \\
        }; 
        \draw (m-1-3.north west) -- (m-4-3.south west);
    \end{tikzpicture},
    H_{i,j} 
    = \text{diag}\left( \phantom{-}1, -1, -1, -1, \quad -1, \phantom{-}1, -1, -1, \quad -1, -1, \phantom{-}1, \phantom{-}1, \quad -1, -1, \phantom{-}1, \phantom{-}1 \right).
\end{equation}

\textbf{\maxkcut{4}.}
For the case when $k=4$, we need  $L=\lceil log_2(4)\rceil=2$ qubits per vertex.
The entries of the matrix $D$ and the local Hamiltonian are given by
\begin{equation}
    D =
        \begin{tikzpicture}[baseline={-.5ex},mymatrixenv]
        \matrix [mymatrix,inner sep=4pt] (m)  
        {
            \phantom{-}1 &-1 & -1& -1 \\
            -1 &\phantom{-}1 & -1& -1 \\
            -1 &-1 & \phantom{-}1& -1 \\
            \hline
            -1 &-1 & -1& \phantom{-}1 \\
        }; 
        \draw (m-1-4.north west) -- (m-4-4.south west);
    \end{tikzpicture},
    H_{i,j} 
    = \text{diag}\left( \phantom{-}1, -1, -1, -1, \quad -1, \phantom{-}1, -1, -1, \quad -1, -1, \phantom{-}1, -1, \quad -1, -1, -1, \phantom{-}1 \right).
\end{equation}

\subsubsection{Unitary Evolution}

For the binary encoding, there are no constraints on the binary strings to be a valid solution.
Therefore, there is no need to design special mixers, and the mixing Hamiltonian is given by
\begin{equation}
    H_M = \sum_{j=1}^{|V|L} \beta \sigma^x_{j}, \quad L=\lceil log_2(k)\rceil.
\end{equation}
This leads to the unitary operator
\begin{equation}
    U_{M} = e^{-i\beta H_M} = \prod_{j=1}^{|V|L} e^{-i\beta\sigma^x_j}.
\end{equation}
Each term in the above product can be implemented with an $R_x$-gate.

The unitary operator for phase separation is defined by
\begin{equation}
\label{eq:U_P}
    U_{P} = e^{-i\gamma H_P} = e^{-i\gamma \sum_{(i,j)\in E} w_{i,j} H_{i,j}} = \prod_{(i,j)\in E} e^{-i\gamma  w_{i,j} H_{i,j}},
\end{equation}
where the last equality holds because the terms $H_{i,j}$ trivially commute, as they are diagonal matrices.
Furthermore, we can use Equation~\eqref{eq:D} to further decompose the terms of the product
\begin{equation}
\label{eq:Hprod}
    e^{-i\theta H_{i,j}} = e^{-i\theta\text{diag}\left(\text{vec}(2I - J)\right)} \prod_{c,d=k+1, m\neq n}^{2^L} e^{-i 2\theta \text{diag}\left(\text{vec}(\Gamma_{c,d})\right)}.
\end{equation}
Again, equality holds since only diagonal matrices are involved.
The first term in Equation~\eqref{eq:Hprod}, can be realized through the following circuit.
\begin{equation}
    e^{-i \theta \text{diag}\left(\text{vec}(2I-J)\right)} = 
    \begin{quantikz}[row sep={0.5cm,between origins},column sep=1ex]
    \lstick{$q_i^0$} & \ctrl{5}\gategroup[5,steps=13,style={dashed,rounded corners,fill=blue!20, inner xsep=2pt},background]{} & \ \ldots\ \qw & \qw & \ \ldots\ \qw & \qw & \qw & \qw & \qw & \qw & \ \ldots\ \qw &\qw & \ \ldots\ \qw &\ctrl{5} & \qw\\
    \lstick{\vdots\hphantom{a}}&&\ddots&&&&&&&&&&\adots&& \\
    \lstick{$q_i^l$}  & \qw &\ \ldots\ \qw & \ctrl{5} & \ \ldots\ \qw & \qw  & \qw & \qw & \qw & \qw & \ \ldots\ \qw &\ctrl{5} & \ \ldots\ \qw &\qw & \qw \\
    \lstick{\vdots\hphantom{a}}&&&&\ddots&&&&&&\adots&&&& \\
    \lstick{$q_i^{L-1}$}  & \qw &\ \ldots\ \qw & \qw & \ \ldots\ \qw & \ctrl{5} & \qw & \qw & \qw & \ctrl{5} & \ \ldots\ \qw &\qw & \ \ldots\ \qw &\qw & \qw\\[.5cm]
    \lstick{$q_j^0$} & \targ{}\gategroup[5,steps=13,style={dashed,rounded corners,fill=blue!20, inner xsep=2pt},background]{} & \ \ldots\ \qw &\qw & \ \ldots\ \qw & \qw & \gate{X} & \ctrl{4} & \gate{X} & \qw & \ \ldots\ \qw &\qw &\ \ldots\ \qw & \targ{} & \qw\\
    \lstick{\vdots\hphantom{a}}&&\ddots&&&&\vdots&&\vdots&&&&\adots&& \\
    \lstick{$q_j^l$}  & \qw &\ \ldots\ \qw & \targ{} & \ \ldots\ \qw &\qw  & \gate{X} & \ctrl{2} & \gate{X} & \qw & \ \ldots\ \qw &\targ{} & \ \ldots\ \qw &\qw & \qw \\
    \lstick{\vdots\hphantom{a}}&&&&\ddots&&\vdots&&\vdots&&\adots&&&& \\
    \lstick{$q_j^{L-1}$}  & \qw &\ \ldots\ \qw & \qw & \ \ldots\ \qw &\targ{} & \gate{X} & \gate{U_3(0,-\phi,0)} & \gate{X} & \targ{} &\ \ldots\ \qw & \qw & \ \ldots\ \qw &\qw & \qw \\
    \lstick{}&&&&&&&&&&&&&& \\
    \lstick{}&&&&&&&&&&&&&&
    \end{quantikz}
    \label{eq:ep2}
\end{equation}
The qubits are enumerated such that qubits $q_i^0, \cdots, q_i^{L-1}$ correspond to the label that is assigned to vertex enumerated $i$.
The logic behind the circuit shown can be understood from a classical point of view.
Applying $CX$-gates on pairs of qubits acting on basis states between vertex $i$ and $j$ results in the state
$\ket{q_i^0}\cdots\ket{q_i^{L-1}}\ket{q_j^0\oplus q_i^0}\cdots\ket{q_j^L\oplus q_i^L}$, where the $\oplus$ operation is modulo 2. This means the state of the qubits belonging to $j$ has zero entries if and only if all qubits have the same basis state. Negating the state and applying a multi-controlled $'U_3(0,\phi,0)$-gate therefore applies a phase if the original (basis-) states $\ket{q_i^0}\cdots\ket{q_i^{L-1}}$ and $\ket{q_j^0}\cdots\ket{q_j^L}$ differ. After this one can uncompute by applying $X$ and $CX$-gates in reversed order such that the overall change is that of applying a phase.

The remaining terms in Equation~\eqref{eq:Hprod} (which vanish if $k$ is a power of 2) can be implemented, e.g., with the help of two ancillary qubits, $a_0, a_1$, in the following way:
\begin{equation}
e^{-i\theta \text{diag}\left(\text{vec}(2\Gamma^{c,d})\right)} = 
    \begin{quantikz}[row sep={0.5cm,between origins},column sep=1ex]
\lstick{$q_i^0$}       & \gate[5,nwires={2,3,4,5}]{N_1}\gategroup[5,steps=9,style={dashed,rounded corners,fill=green!20, inner xsep=2pt},background]{}& \qw    & \ctrl{10} & \qw      & \qw                            & \qw       & \ctrl{10}   & \qw & \gate[5,nwires={2,3,4,5}]{N_1} &\qw\\
\lstick{\vdots\hphantom{a}} & &  &           &          &                                &           &            &  & & \\
\lstick{$q_i^l$}        &\qw& \qw    & \ctrl{8}  & \qw      & \qw                            & \qw       & \ctrl{8}    & \qw & \qw& \qw\\
\lstick{\vdots\hphantom{a}}& &  &           &          &                                &           &             & &\\
\lstick{$q_i^{L-1}$}    &\qw & \qw       & \ctrl{6}  & \qw      & \qw                            & \qw       & \ctrl{6}    & \qw & \qw & \qw\\[.75cm]
\lstick{$q_j^0$}        &\gate[5,nwires={2,3,4,5}]{N_2}\gategroup[5,steps=9,style={dashed,rounded corners,fill=green!20, inner xsep=2pt},background]{}  & \qw      & \qw       & \ctrl{6} & \qw                            & \ctrl{6}  & \qw         & \qw & \gate[5,nwires={2,3,4,5}]{N_2} & \qw\\
\lstick{\vdots\hphantom{a}}& & &            &          &                                &           &             & &  &  \\
\lstick{$q_j^l$}        &\qw   & \qw     & \qw       & \ctrl{4} & \qw                            & \ctrl{4}  & \qw         & \qw & \qw& \qw\\
\lstick{\vdots\hphantom{a}}& &  &           &          &                                &           &             &  & & \\
\lstick{$q_j^{L-1}$}    &\qw   & \qw     & \qw       & \ctrl{2} & \gate{U_3(0,-\phi,0)}& \ctrl{2}  & \qw         & \qw & \qw & \qw\\[.75cm]
\lstick{$a_0$}          & \qw& \qw    & \targ{}\gategroup[2,steps=5,style={dashed,rounded corners,fill=red!20, inner xsep=2pt},background]{}   & \qw      & \ctrl{-1}                      & \qw       & \targ{}     & \qw& \qw& \qw\\
\lstick{$a_1$}    & \qw& \qw    & \qw       & \targ{}  & \ctrl{-1}                      & \targ{}   & \qw & \qw& \qw& \qw
    \end{quantikz}
    \label{eq:enp2}
    \end{equation}
The gates $N_1, N_2$ in Equation~\eqref{eq:enp2} are of the form $U_0\otimes\cdots\otimes U_{L-1}$, where $U_i\in\{I,X\}$ are chose such that $N_1 \ket{0}_L = \ket{c}_L$, and $N_2 \ket{0}_L = \ket{d}_L$.
The logic behind this circuit is that multi-controlled NOT gates are used to set two ancillary qubits to the state one if $q_i = \ket{c}_L$, and $q_j = \ket{d}_L$. Of both ancillary qubits are one, a multi-controlled $U_3(0,\phi,0)$-gate is applied to change the phase, followed by a uncomputation steps.
The ancillary qubits can be reused for all other pairs $(i,j)\in E$.
An example for \maxkcut{3} is shown in Figure~\ref{fig:circuit_max_34_cut}.

\begin{table}
    \centering
    \begin{tabular}{llararara}
        &$k$&2&3&4&5&6&7&8\\
        \hline
        \multirow{4}{*}{\rotatebox{90}{binary}} 
        &\#qubits& 1$|V|$ & 2$|V|$+2 &  2$|V|$ & 3$|V|$+2 & 3$|V|$+2 & 3$|V|$+2 & 3$|V|$\\
        &\#$CX$ for $\ket{\Phi_0}$ & 0 & 0 & 0 & 0 & 0 & 0 & 0 \\
        &\#$CX$ for $U_M$ & 0 & 0 & 0 & 0 & 0 & 0 & 0 \\
        &\#$CX$ for $U_P$ & 2$E$ & 70$|E|$ &6$|E|$ &$206|E|$ &$142|E|$ &$78|E|$ &$14|E|$\\
        \hline
        \multirow{4}{*}{\rotatebox{90}{one-hot $XY$}} 
        &\#qubits&  1$|V|$&  3$|V|$&  4$|V|$ & 5$|V|$ & 6$|V|$ & 7$|V|$ & 8$|V|$\\
        &\#$CX$ for $\ket{\Phi_0}$ & $2|V|$ & $4|V|$ &$6|V|$ &$8|V|$ &$10|V|$ &$12|V|$ &$14|V|$\\
        &\#$CX$ for $U_M$ & $8|V|$ & $12|V|$ &$16|V|$ &$20|V|$ &$24|V|$ &$28|V|$ &$32|V|$\\
        
        &\#$CX$ for $U_P$ & $4|E|$ & $6|E|$ &$8|E|$ &$10|E|$ &$12|E|$ &$14|E|$ &$16|E|$\\
    \end{tabular}

    \caption{Number of $CX$-operations per layer and number of qubits 
    of the QAOA circuit for a graph $G=(V,E)$ depending on $k$.
    One-hot encoding requires the use of a more complicated $XY$-mixer and the preparation of $W_k$-states.
    }
    \label{tab:depth}
\end{table}

\subsection{Resource Analysis of One-hot and Binary Encoding}
We will give a short analysis of the number of gates required to decompose the basic building blocks of the phase operator $U_P$, the mixing operator $U_M$, and the preparation of the initial state for the two different encoding schemes.
In order to be executable on, e.g., one of IBM's quantum devices, all terms need to be decomposed using gates from the set of basis gates $\{U_3,CX\}$, where
\begin{equation}
U_3(\theta,\phi,\lambda) =
\begin{pmatrix}
\cos(\theta/2) & -e^{i\lambda} \sin(\theta/2) \\
e^{i\phi} \sin(\theta/2) & e^{i(\phi+\lambda)}\cos(\theta/2) \\
\end{pmatrix}, \quad CX =
\begin{pmatrix}
1 & 0 & 0 & 0 \\
0 & 1 & 0 & 0 \\
0 & 0 & 0 & 1 \\
0 & 0 & 1 & 0 \\
\end{pmatrix}.
\end{equation}
Note that $U_3(0,\phi,0) = \text{diag}(1, e^{i\phi})$.
Throughout we assume full connectivity of the qubits, i.e., a $CX$-gate can be executed directly on any pair of qubits, without the need for applying SWAP or Bridge gates~\cite{itoko2020optimization}.
Furthermore, 
(multi-)controlled $U_3(0,\phi,0)$ operations can be implemented in terms of its square root $V = U_3(0,\phi/2,0)$, and $V^\dagger$, using polynomially many $CX$-gates, see, e.g., \cite{Liu2008}.
In order to implement the circuit shown in Equation~\eqref{eq:ep2} one needs $2L$ $CX$-gates, $2L$ $X$-gates and $1$ (multi-)controlled $U_3$-gate.
When $k$ is not a power of two we need to additionally execute the circuits of the form shown in Equation~\eqref{eq:enp2}.
This requires $2L$ $C^{L}X$-gates, $2$ $C^{L}U_3$-gates, and $L$ $X$-gates. 
In general these need to be applied $2(2^L-k)$ times.
In the worst case when $k = 2^n+1$ we need to apply these gates $2(2^n-1)$ times.
Overall, Table~\ref{tab:depth} shows the width (number of qubits) and depth requirements of the complete circuit for \maxkcut{k}. We can see that low depth and width are achieved when $k$ is a power of two.

The analysis shows that one-hot encoding has a strong limitation when it comes to the requirement of number of qubits. In addition, preparation of $W_k$ and $XY$-mixers is more costly than the standard $X$-mixer sufficient for binary encoding.
Furthermore, when the number of qubits are limited to a few hundred or thousand, only one-hot encoding for the cases $k=2,4$ and possibly $k=3$ will be of practical interest when quantum advantage is to be achieved.

\section{Implementation and Results}\label{sec:results}

\begin{figure}[t]
    \centering
    \includegraphics[width=1\textwidth]{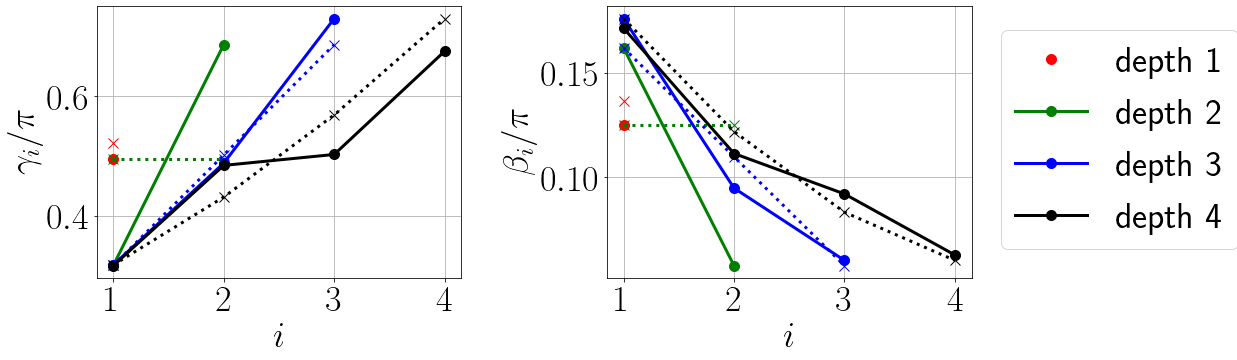}
    \caption{Initial guess (dotted lines with x marker) and locally optimal (solid lines with round marker) parameters $\gamma,\beta$ for the graph shown in Figure~\ref{fig:ERBA10} using the interpolation-based heuristic described in the text. The results indicate a correlation of the parameters between different depths $p$ making this a useful heuristic to avoid global optimization.}
    \label{fig:depth_interp}
\end{figure}

In this section we showcase numerical simulations on different types of graphs.
We start by briefly describing the heuristic we employ for the classical outer optimization loop.
Sampling high-dimensional target functions uniformly quickly becomes intractable for depth $p>1$.
In order to get a good initial guess of the parameters  $(\vec{\gamma}_p, \vec{\beta}_p)$ at level $p$ for the local optimization procedure
we employ the interpolation-based heuristic described in~\cite{Zhou2020}, which is given by the following recursion,
\begin{equation}
    \big[\vec{\gamma}_{(p+1)}^0\big]_i = \frac{i-1}{p}\big[\vec{\gamma}_{(p)}^L\big]_{i-1} + \frac{p -i+1}{p} \big[\vec{\gamma}_{(p)}^L\big]_i, \quad i = 1, 2,..., p + 1.
\label{INTERP}
\end{equation}
In above formula the superscript refers to either the initial parameter (superscript 0), or the local optimum (superscript $L$). The same formula holds for $\vec{\beta}$.
For depth $p=1$ the expectation value is sampled on an $n\times m$ Cartesian grid over the domain $[0,\gamma_\text{max}]\times[0,\beta_\text{max}]$. The initial parameters $(\gamma_1^0, \beta_1^0)$ are then given by identifying a pair of parameters which achieves the lowest expectation value on the grid.
Using the starting point $(\vec{\gamma}_p^0, \vec{\beta}_p^0)$ a local optimization algorithm, e.g. Nelder-Mead or COBYLA, is used to find the local minimum with $(\vec{\gamma}_p^L, \vec{\beta}_p^L)$ .
Figure~\ref{fig:depth_interp} shows that optimal parameters are strongly correlated between different depths $p$, also for non-regular graphs.

\begin{figure}
    \begin{subfigure}[b]{.2\textwidth}
     \centering
        \includegraphics[width=.5\textwidth]{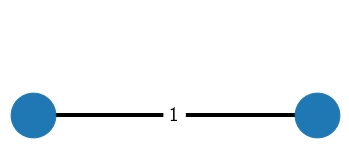}
        \caption{``Barbell'' graph.}
    \end{subfigure}
    \begin{subfigure}[b]{.78\textwidth}
    \centering
    \begingroup
    \setlength{\tabcolsep}{5pt} 
    \renewcommand{\arraystretch}{1} 
    \begin{tabular}{r|rrr|rrr|rrr|rrr}
    &\multicolumn{3}{c|}{binary} &\multicolumn{3}{c|}{one-hot $X$}
    &\multicolumn{3}{c|}{one-hot penalty $X$}&\multicolumn{3}{c}{one-hot $XY$} \\
    $k$ & $\alpha_{1}$ & $\alpha_{2}$ & $\alpha_{3}$
        & $\alpha_{1}$ & $\alpha_{2}$ & $\alpha_{3}$
        & $\alpha_{1}$ & $\alpha_{2}$ & $\alpha_{3}$
        & $\alpha_{1}$ & $\alpha_{2}$ & $\alpha_{3}$\\
    \hline
2 & 1.000 & 1.000 & 1.000 & 0.508 & 0.515 & 0.517  & 0.552 & 0.969 & 0.868  & 1.000 & 1.000 & 1.000 \\
3 & 0.961 & 0.996 & 0.999  & 0.117 & 0.119 & 0.119 & 0.207 & 0.227 & 0.311  & 0.998 & 0.998 & 0.997 \\
4 & 1.000 & 1.000 & 1.000  & 0.052 & 0.052 & 0.055 & 0.179 & 0.175 & 0.185  & 1.000 & 1.000 & 1.000 \\
5 & 0.931 & 0.999 & 0.998 & 0.026 & 0.023 & 0.027  & 0.159 & 0.185 & 0.193  & 1.000 & 1.000 & 1.000 \\
6 & 0.981 & 0.994 & 1.000 & 0.013 & 0.013 & 0.014  & 0.056 & 0.080 & 0.098  & 1.000 & 1.000 & 1.000 \\
7 & 0.996 & 0.999 & 0.999  & 0.007 & 0.007 & 0.007 & 0.116 & 0.185 & 0.076  & 1.000 & 1.000 & 1.000 \\
8 & 1.000 & 1.000 & 1.000  & 0.004 & 0.002 & 0.002 & 0.062 & 0.166 & 0.168  & 1.000 & 1.000 & 1.000 \\
    \end{tabular}
    \endgroup
    \caption{Approximation ratio for 8192 shots. $\alpha_{p}$ is the approximation ratio for depth $p$.}
    \end{subfigure}
    \caption{Results for a graph with two vertices connected by an edge.
    Using one-hot encoding, the $XY$-mixer together with $W_k$ initial states achieves equally good results as the binary encoding. The one-hot encoding with the standard $X$-mixer gets slightly improved by the penalty term. The low values are due to an exponentially small space of feasible solutions.
    The energy landscapes are shown in Table~\ref{tab:E_Barbell}.
    }
    \label{fig:Barbell}
\end{figure}

The first example is a graph with two vertices connected by an edge.
Using an ideal simulator we compare the results for the binary encoding with standard $X$-mixer, the binary encoding with the penalty term and the standard $X$-mixer, as well as the $XY$-mixer without penalty term and the $W_k$ initial state.
The results, shown in Figure~\ref{fig:Barbell}, show that the binary encoding as well as the one-hot encoding with the $XY$-mixer have approximation ratios close to one for all cases.
The pure one-hot encoding becomes increasingly worse for increasing $k$, which is related to the exponentially small feasible subspace, see Equation~\eqref{eq:feasiblesub}. Even adding a penalty term does not improve the situation noteworthy.
The expectation value $E(\theta)=\langle H_P \rangle_{\ket{\Psi(\gamma,\beta)}}$ for different parameters, often referred to as the energy landscape for all cases $k\in{2,\dots,8}$ is given in Table~\ref{tab:E_Barbell}, which seems to indicate that the binary encoding generates optimization problems with fewer (local) minima.

\begin{figure}
    \begin{subfigure}[b]{.22\textwidth}
        \includegraphics[width=1.\textwidth]{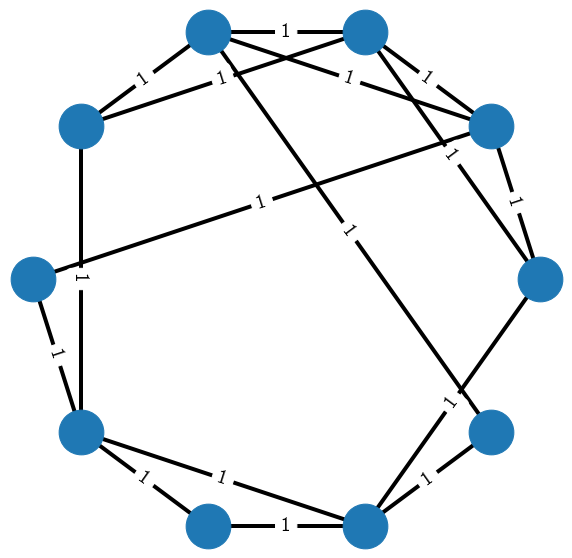}
        \caption{Graph instance.}
    \end{subfigure}
    \begin{subfigure}[b]{.66\textwidth}
     \centering
    \begingroup
    \setlength{\tabcolsep}{5pt} 
    \renewcommand{\arraystretch}{1} 
\begin{tabular}{r|r|r|rrrrr|rrrrr}
 & rand & GW & \multicolumn{5}{c|}{binary} & \multicolumn{5}{c}{one-hot $XY$}\\
    $k$ & $\alpha$& $\alpha$ & \#q & \#$CX$ & $\alpha_1$ & $\alpha_2$ & $\alpha_3$
        & \#q & \#$CX$ & $\alpha_1$ & $\alpha_2$ & $\alpha_3$\\
    \hline
    2 & 0.50 & 0.88 & 10   &   32$p$ & 0.77 & 0.79 & 0.80 &20& 164$p$ & 0.76 & 0.79 & 0.80 \\
    3 & 0.67 & 0.84 & 20+2 & 1120$p$ & 0.73 & 0.75 & 0.77 &30& 256$p$ & 0.75 & 0.76 & 0.77 \\
    4 & 0.75 & 0.86 & 20   &   96$p$ & 0.82 & 0.84 & 0.84 &40& 348$p$ &$-$&$-$&$-$\\
    5 & 0.80 & 0.88 & 30+2 & 3296$p$ & 0.81 & 0.85 & 0.87 &50& 440$p$ &x&x&x\\
    6 & 0.83 & 0.89 & 30+2 & 2272$p$ & 0.87 & 0.89 & 0.90 &60& 532$p$ &x&x&x\\
    7 & 0.86 & 0.90 & 30+2 & 1248$p$ & 0.90 & 0.91 & 0.91 &70& 624$p$ &x&x&x\\
    8 & 0.88 & 0.91 & 30   &  224$p$ & 0.92 & 0.93 & 0.93 &80& 716$p$ &x&x&x\\
\end{tabular}
\endgroup
    \caption{Approximation ratios achieved for graph shown in (a) using 8192 shots.}
    \end{subfigure}
    \begin{subfigure}[b]{.22\textwidth}
        \includegraphics[width=1.\textwidth]{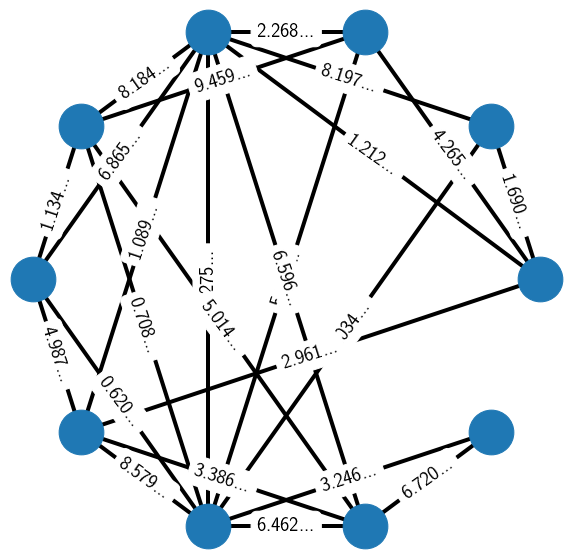}
        \caption{Graph instance.}
    \end{subfigure}
    \begin{subfigure}[b]{.66\textwidth}
    \centering
    \begingroup
    \setlength{\tabcolsep}{5pt} 
    \renewcommand{\arraystretch}{1} 
     \begin{tabular}{r|r|r|rrrrr|rrrrr}
 & rand & GW & \multicolumn{5}{c|}{binary} & \multicolumn{5}{c}{one-hot $XY$}\\
    $k$ & $\alpha$ & $\alpha$ & \#q & \#$CX$ & $\alpha_{1}$ & $\alpha_{2}$ & $\alpha_{3}$
        & \#q & \#$CX$ & $\alpha_{1}$ & $\alpha_{2}$ & $\alpha_{3}$\\
    \hline
    2 & 0.50 & 0.88 & 10   &   48$p$ & 0.73 & 0.75 & 0.76 &20& 196$p$ & 0.73 & 0.75 & 0.75 \\
    3 & 0.67 & 0.84 & 20+2 & 1680$p$ & 0.74 & 0.77 & 0.79 &30& 304$p$ & 0.76 & 0.78 & 0.80 \\
    4 & 0.75 & 0.86 & 20   &  144$p$ & 0.82 & 0.85 & 0.86 &40& 412$p$ &$-$&$-$&$-$\\
    5 & 0.80 & 0.88 & 30+2 & 4944$p$ & 0.82 & 0.86 & 0.89 &50& 520$p$ &x&x&x\\
    6 & 0.83 & 0.89 & 30+2 & 3408$p$ & 0.87 & 0.90 & 0.92 &60& 628$p$ &x&x&x\\
    7 & 0.86 & 0.90 & 30+2 & 1872$p$ & 0.91 & 0.93 & 0.94 &70& 736$p$ &x&x&x\\
    8 & 0.88 & 0.91 & 30   &  336$p$ & 0.93 & 0.95 & 0.95 &80& 844$p$ &x&x&x\\
\end{tabular}
\endgroup
    \caption{Approximation ratios achieved for graph shown in (c) using 8192 shots.}
    \end{subfigure}
    \caption{The results for an unweighted Erdős-Rényi with $|V|=10$ vertices and $|E|=16$ edges and a weighted Barbasi-Albert with $|V|=10$ vertices and $|E|=24$ edges.
    $\alpha_{p}$ is the approximation ratio for depth $p$. \#q is the number of qubits required. The number of $CX$-gates is per layer $p$. Here, $-$ indicates that it is in principle possible to simulate classically (but we only have access to simulator with up to 32 qubits), whereas x indicates that it is infeasible to simulate even on a supercomputer (as 50 qubits is considered the threshold for that).
    Approximation ratios for random and Goemans-Williamson algorithm are shown as reference.
    The energy landscapes are shown in Table~\ref{tab:E_ERBA10}.
    }
    \label{fig:ERBA10}
\end{figure}

The final two examples show numerical examples of larger instances of graphs: An unweighted Erdős-Rényi graph and a weighted Barabási-Albert graph with 10 vertices, presented in Figure~\ref{fig:ERBA10}.
For higher depth we employ the interpolation based heuristic.
In all cases the average approximation ratio achieved is considerably higher than the approximation ratio of randomly drawn a solution or the guarantees of the Goemans-Williamson, which is given as a reference.
Furthermore, the average approximation ratio increases with increasing depth.
One-hot encoding in the case of \maxkcut{4} would already require $40$ qubits, which quickly becomes prohibitive for a simulator.

\section{Availability of Data and Code}
All data, e.g., graphs, and the python/jupyter notebook source code of the \maxkcut{k}-implementation using QAOA for reproducing the results obtained in this article are available at \url{https://github.com/OpenQuantumComputing}.

\section{Conclusion}\label{sec:conclusion}
In this article we provide numerical evidence that NISQ device can be used to (approximately) solve the weighted \maxkcut{k}.
The analysis of the proposed binary encoding shows an exponential improvement of the number of qubits with respect to previously known results.
In addition our results indicate the optimization problem for the one-hot encoding seems to contain many local optima, making it a more demanding problem to solve, see also the discussion in~\cite{Wang2020}.
The circuit depth required is very low when $k$ is a power of two.
When this is not the case, we provide a proof of principle implementation, which requires an exponential number of $CX$ gates with respect to $k$.
Future research directions are therefore to investigate more efficient ways of decomposing the phase separation operators.
Another possibility might be to introduce penalty terms in the mixing operator, similar to the case of one-hot encoding, such that the number of possible sets is limited to $k$, instead of implementing the circuits shown in Equation~\eqref{eq:enp2}.
Applying and testing R-QAOA and WS-QAOA to our formulation provides another future path for investigation.
Finally, the performance of the proposed algorithms could be tested on simulated noise models and real machines.
Another factor is to analyse the balance between number of qubits and circuit depth with respect to extra auxiliary qubits that can be introduced to minimize the number of SWAP/Bridge-gates on hardware without full qubit connectivity.

\begin{appendix}

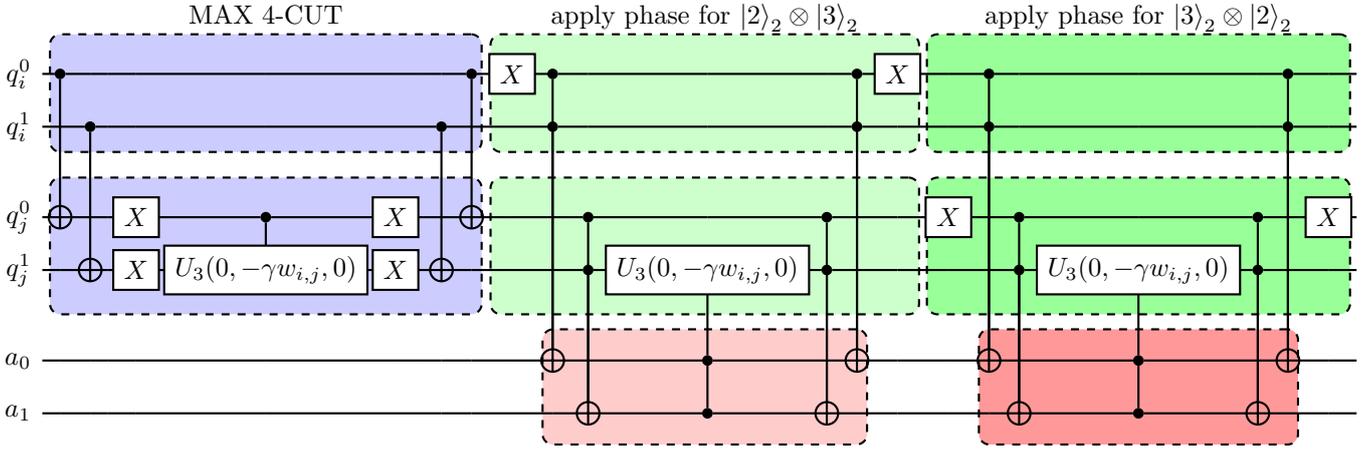
\begin{figure}
    \centering
        \begin{quantikz}[row sep={.7cm,between origins},column sep=2pt]
    \lstick{$q_i^0$} & \ctrl{2}\gategroup[2,steps=9,style={dashed,rounded corners,fill=blue!20, inner xsep=-4pt},background]{\maxkcut{4}} & \qw     & \qw & \qw      & \qw                         & \qw      & \qw & \qw      & \ctrl{2} & 
    \gate{X}\gategroup[2,steps=8,style={dashed,rounded corners,fill=green!20, inner xsep=-4pt},background]{apply phase for $\ket{2}_2\otimes\ket{3}_2$}  & \ctrl{4} & \qw & \qw & \qw & \qw  & \ctrl{4} & \gate{X} &
    \qw\gategroup[2,steps=7,style={dashed,rounded corners,fill=green!40, inner xsep=-4pt},background]{apply phase for $\ket{3}_2\otimes\ket{2}_2$} & \ctrl{4} & \qw & \qw & \qw   & \ctrl{4} & \qw & \qw 
    \\
    \lstick{$q_i^1$} & \qw     & \ctrl{2} & \qw & \qw      & \qw                         & \qw      & \qw & \ctrl{2} & \qw      &
    \qw & \ctrl{3} & \qw & \qw & \qw & \qw  & \ctrl{3} & \qw &
    \qw & \ctrl{3} & \qw & \qw & \qw &\ctrl{3} & \qw  &  \qw 
    \\[.5cm]
    \lstick{$q_j^0$} & \targ{}\gategroup[2,steps=9,style={dashed,rounded corners,fill=blue!20, inner xsep=-4pt},background]{} & \qw      & \qw & \gate{X} & \ctrl{1}                    & \gate{X} & \qw & \qw      & \targ{}  & \qw\gategroup[2,steps=8,style={dashed,rounded corners,fill=green!20, inner xsep=-4pt},background]{} &
    \qw & \qw  & \ctrl{3} &\qw & \ctrl{3} & \qw  & \qw &
    \gate{X}\gategroup[2,steps=7,style={dashed,rounded corners,fill=green!40, inner xsep=-4pt},background]{}  & \qw & \ctrl{3} & \qw & \ctrl{3} & \qw  & \gate{X} & \qw
    \\
    \lstick{$q_j^1$} & \qw     & \targ{}  & \qw & \gate{X} & \gate{U_3(0,-\gamma w_{i,j},0)} & \gate{X} & \qw & \targ{}  & \qw      & \qw &
    \qw &  \qw & \ctrl{2} & \gate{U_3(0,-\gamma w_{i,j},0)} & \ctrl{2} & \qw  & \qw &
    \qw &  \qw & \ctrl{2} & \gate{U_3(0,-\gamma w_{i,j},0)} & \ctrl{2} & \qw  & \qw & \qw
    \\[.5cm]
    \lstick{$a_0$} & \qw & \qw & \qw & \qw & \qw & \qw & \qw & \qw & \qw & \qw &
     \targ{}\gategroup[2,steps=6,style={dashed,rounded corners,fill=red!20, inner xsep=-4pt},background]{} & \qw & \qw & \ctrl{-1} & \qw & \targ{}  & \qw &
    \qw & \targ{}\gategroup[2,steps=5,style={dashed,rounded corners,fill=red!40, inner xsep=-4pt},background]{}  & \qw & \ctrl{-1} & \qw & \targ{}  & \qw & \qw
    \\
    \lstick{$a_1$} & \qw & \qw & \qw & \qw & \qw & \qw & \qw & \qw & \qw & \qw &
    \qw & \qw & \targ{} & \ctrl{-1} & \targ{} & \qw  & \qw &
    \qw & \qw & \targ{} & \ctrl{-1} & \targ{} & \qw  & \qw & \qw
    \end{quantikz}
\caption{Implementation of the phase operator for the edge $(i,j)$ for \maxkcut{3}, which consists of the circuit for \maxkcut{4} plus two extra circuits.}
    \label{fig:circuit_max_34_cut}
\end{figure}
\begin{table}[]
    \centering
    \begingroup
    \setlength{\tabcolsep}{0pt} 
    \renewcommand{\arraystretch}{0} 
    \begin{tabular}{ccccc}
         $k$ & binary & one-hot $X$ & one-hot penalty $X$ & one-hot $XY$ \\
         2&\raisebox{-.5\height}{\includegraphics[trim=120 20 120 20,clip,width=.25\linewidth]{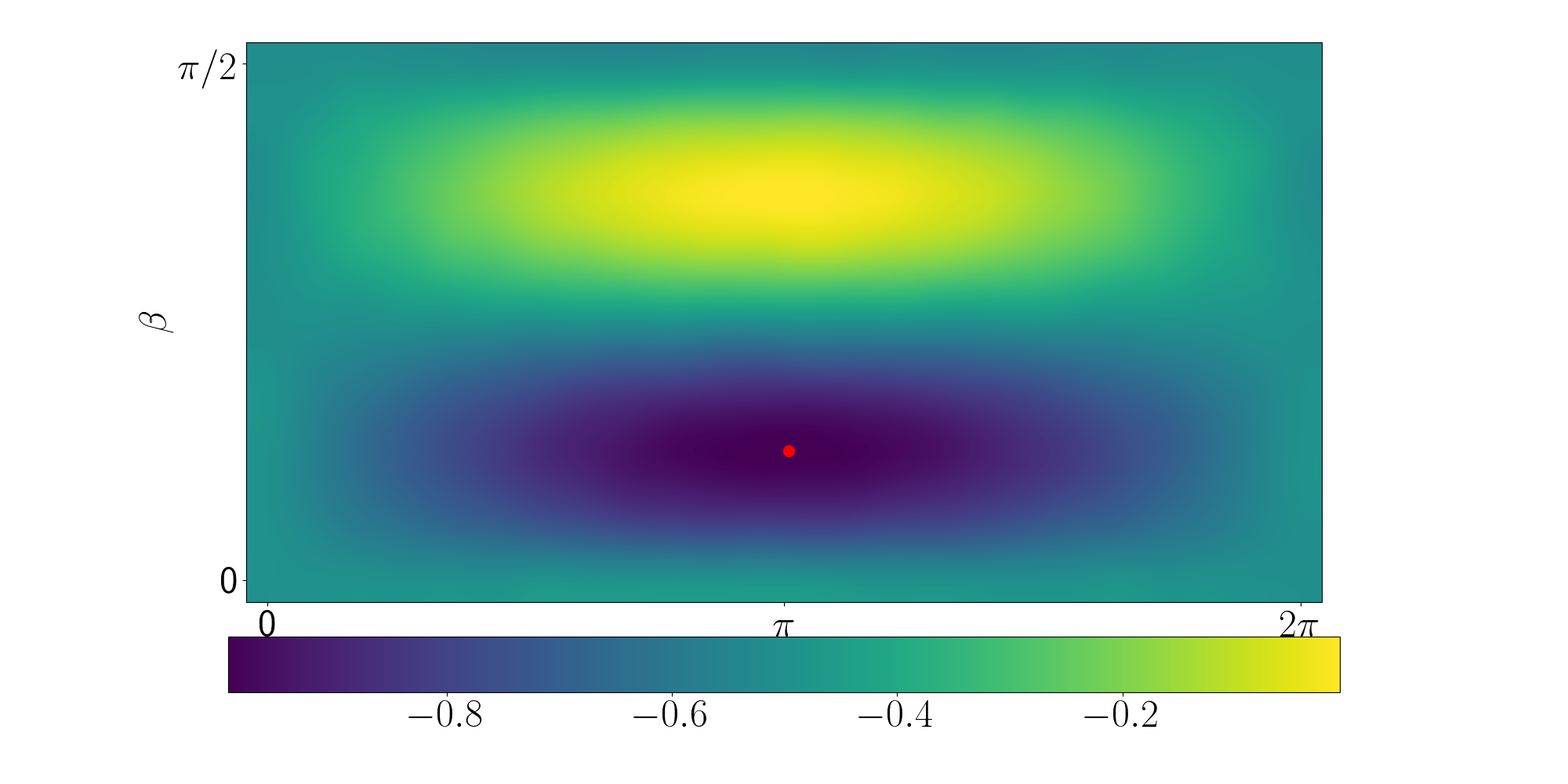}}
         &\raisebox{-.5\height}{\includegraphics[trim=120 20 120 20,clip,width=.25\linewidth]{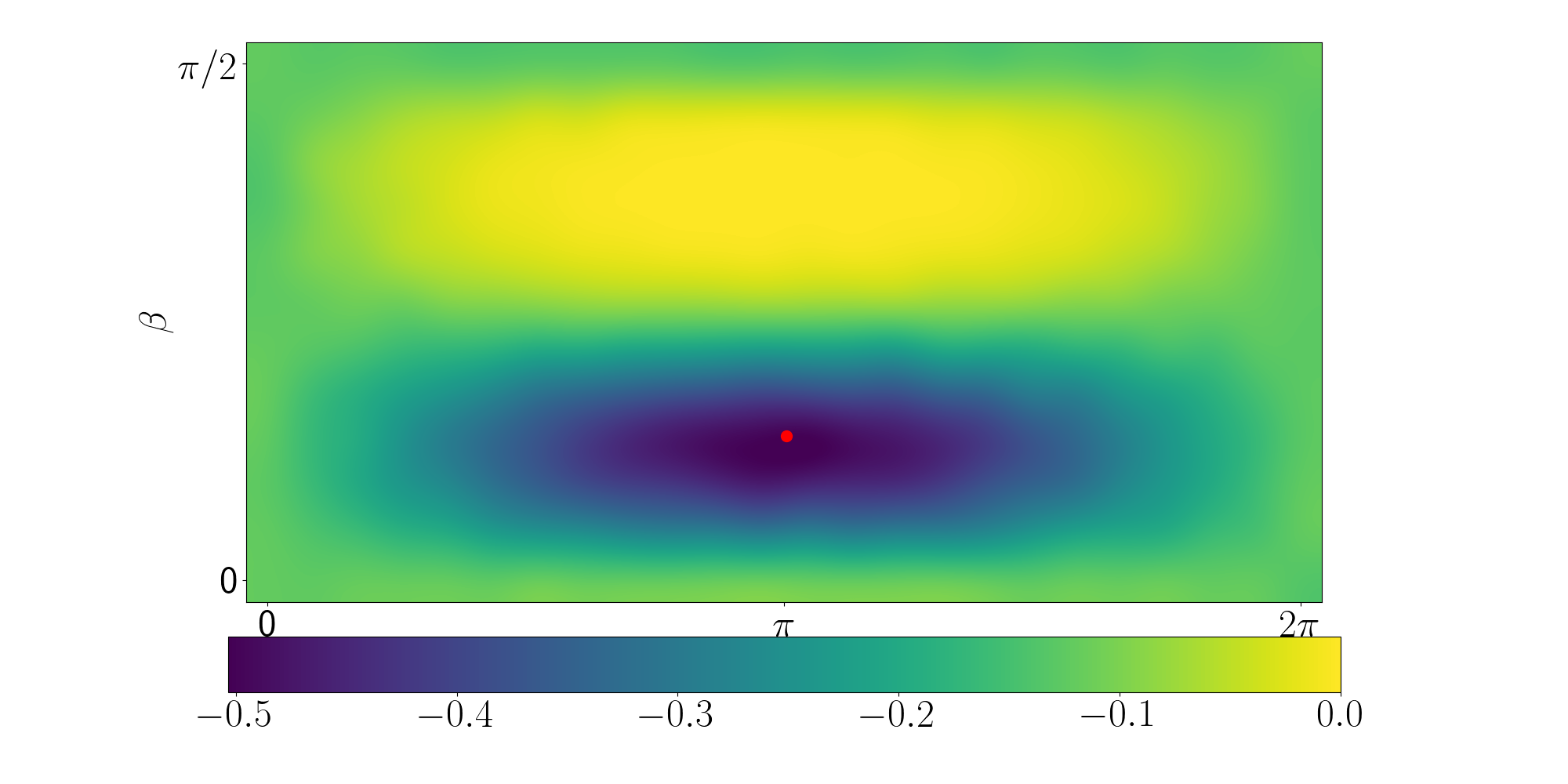}}
         &\raisebox{-.5\height}{\includegraphics[trim=120 20 120 20,clip,width=.25\linewidth]{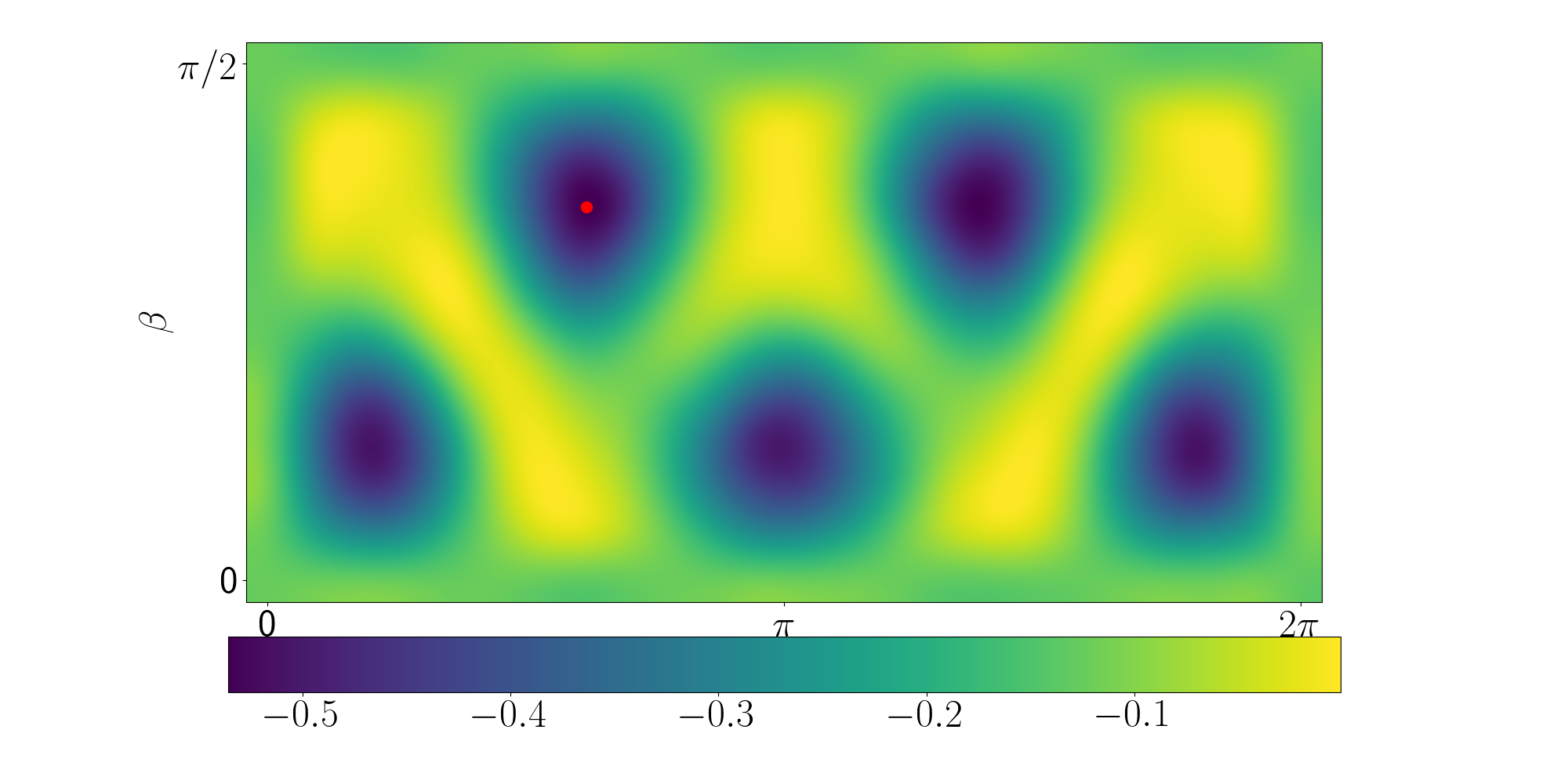}}
         &\raisebox{-.5\height}{\includegraphics[trim=120 20 120 20,clip,width=.25\linewidth]{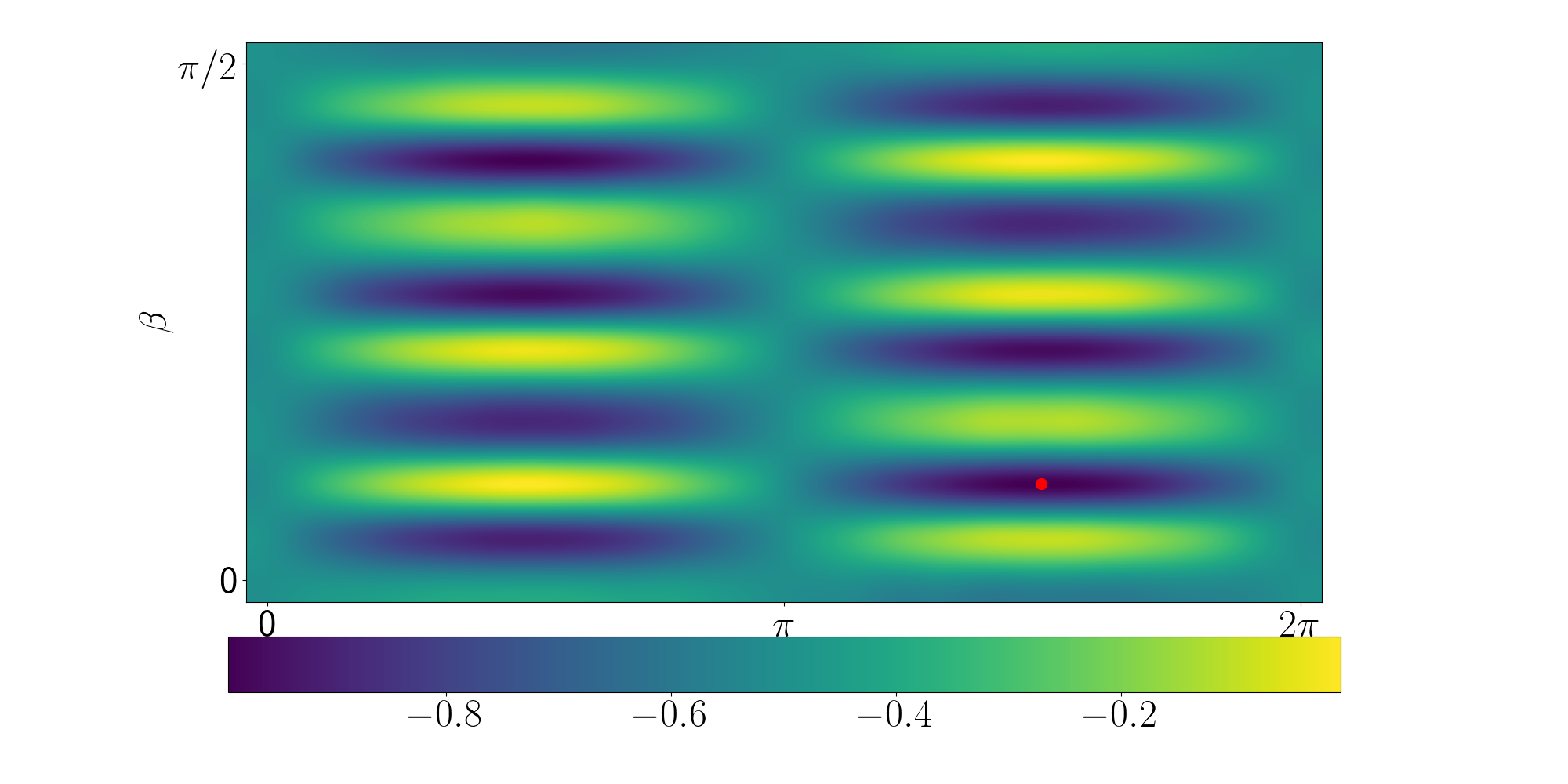}}\\
         3&\raisebox{-.5\height}{\includegraphics[trim=120 20 120 20,clip,width=.25\linewidth]{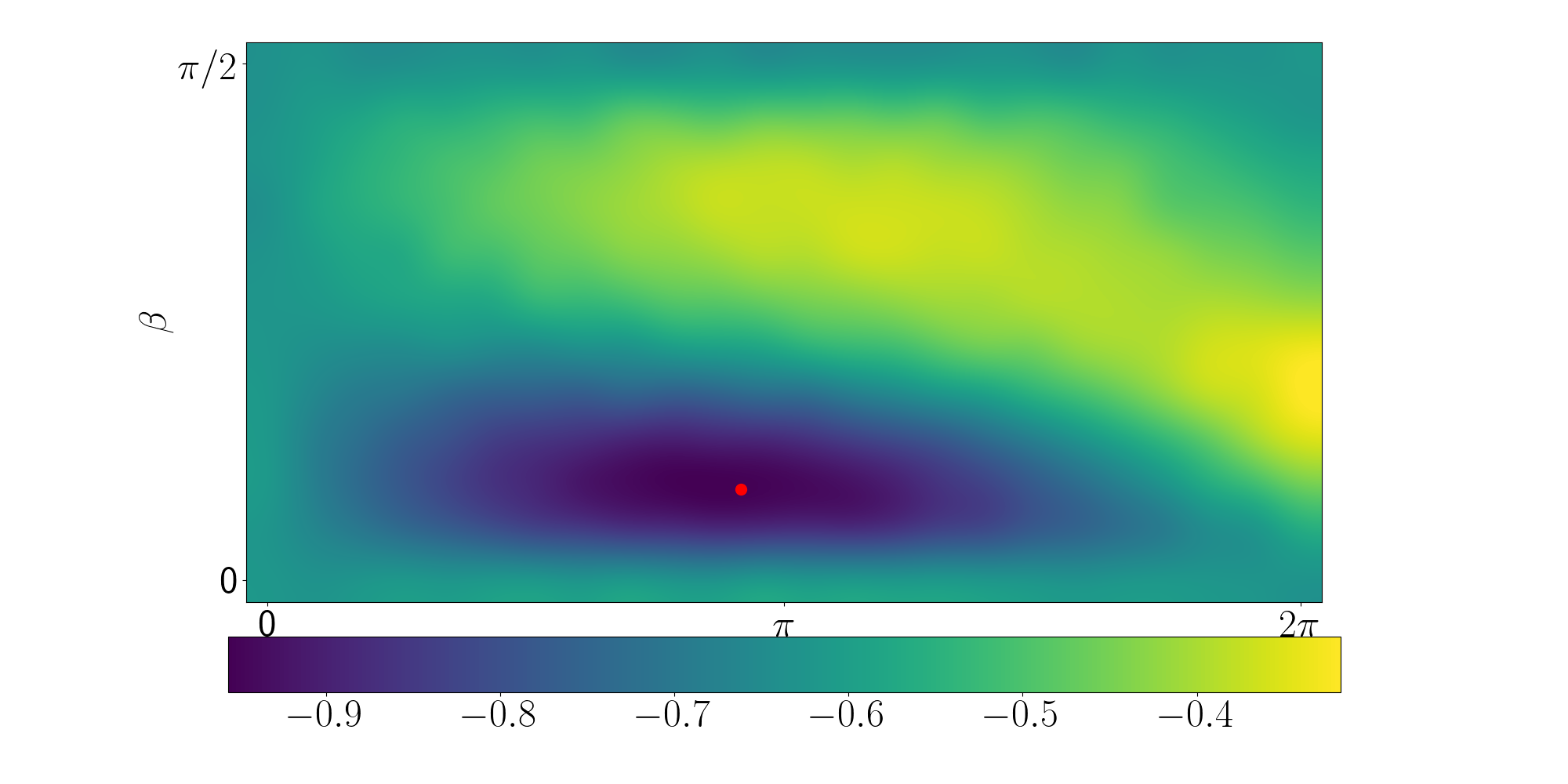}}
         &\raisebox{-.5\height}{\includegraphics[trim=120 20 120 20,clip,width=.25\linewidth]{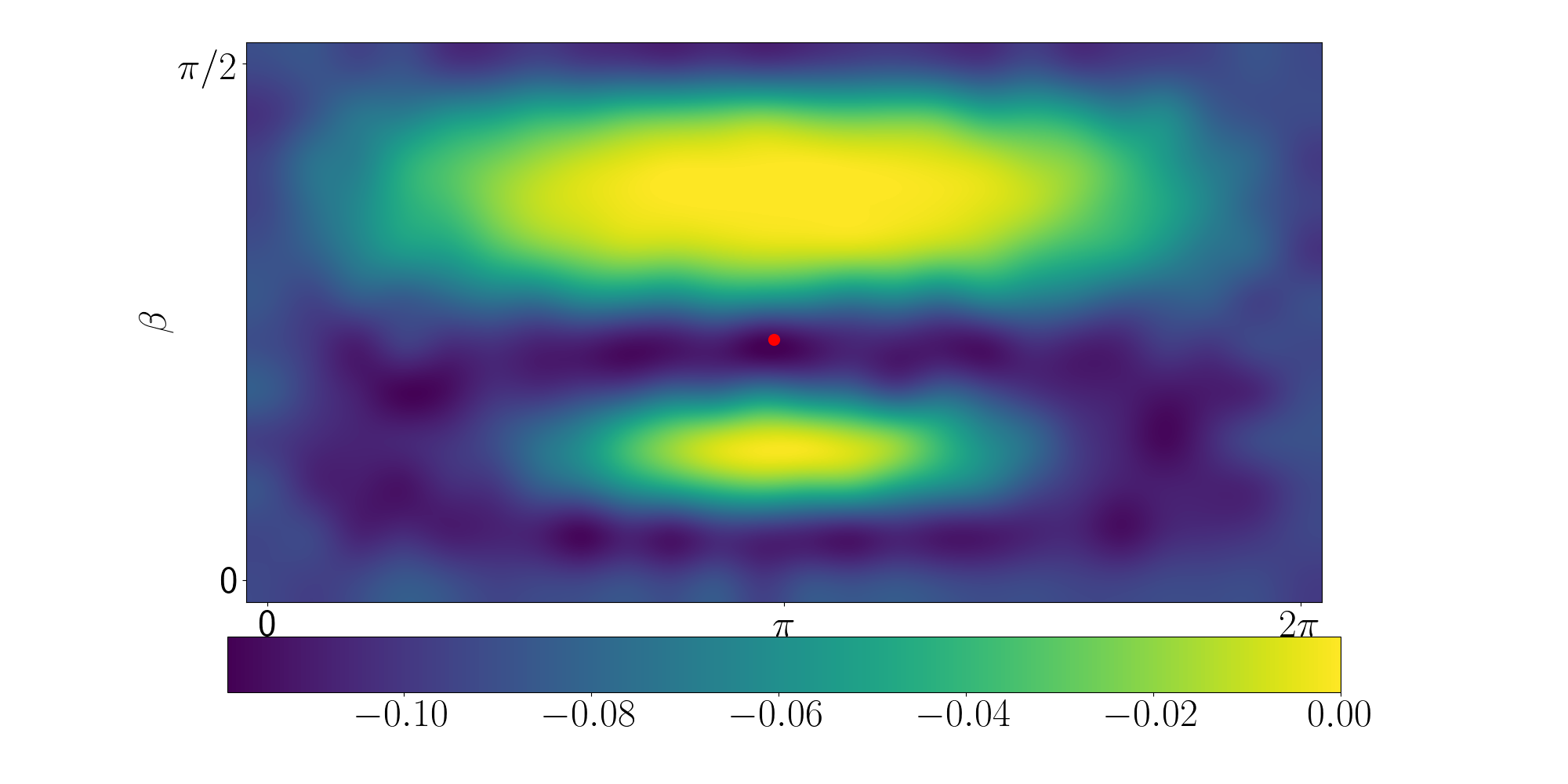}}
         &\raisebox{-.5\height}{\includegraphics[trim=120 20 120 20,clip,width=.25\linewidth]{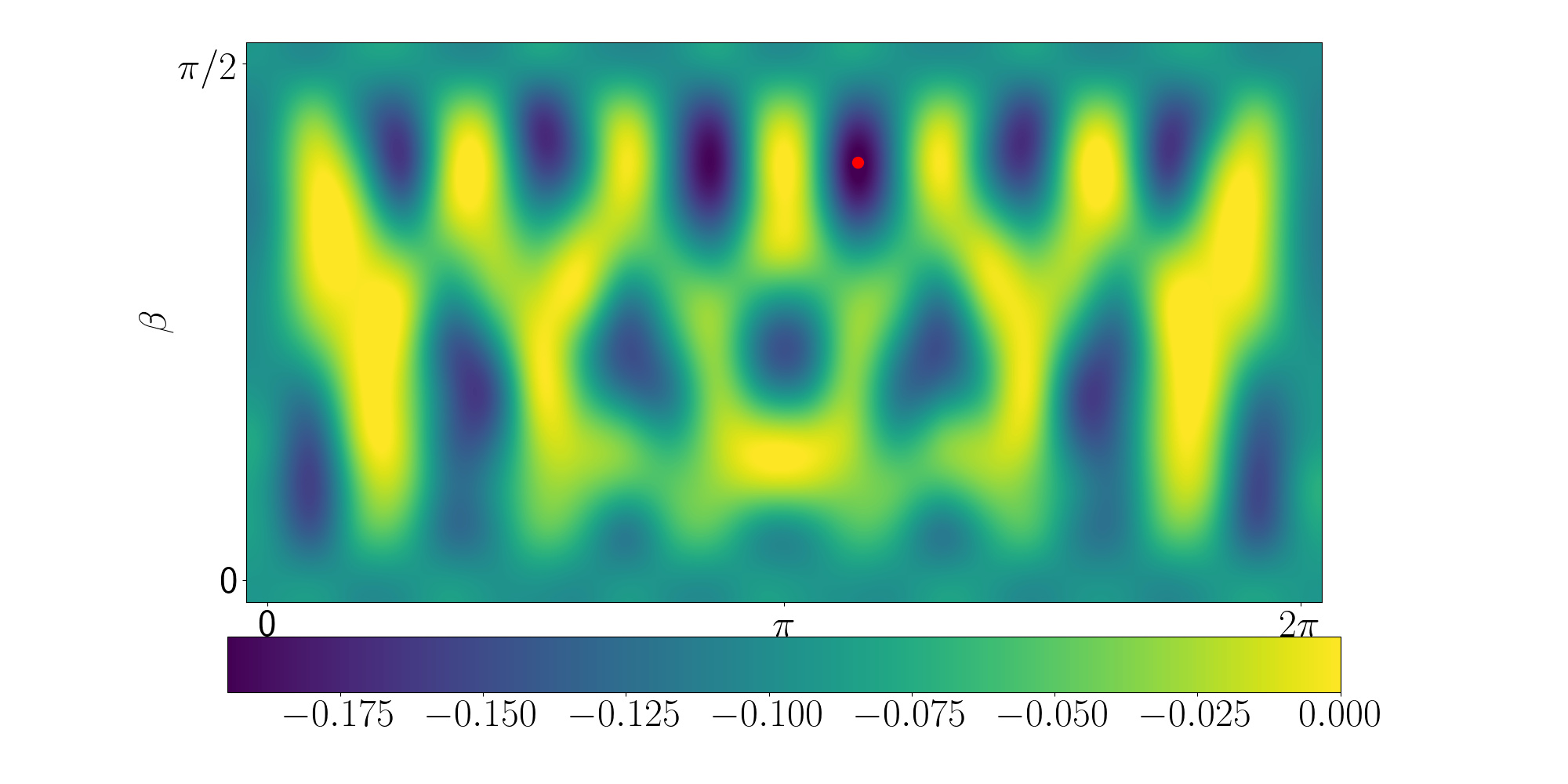}}
         &\raisebox{-.5\height}{\includegraphics[trim=120 20 120 20,clip,width=.25\linewidth]{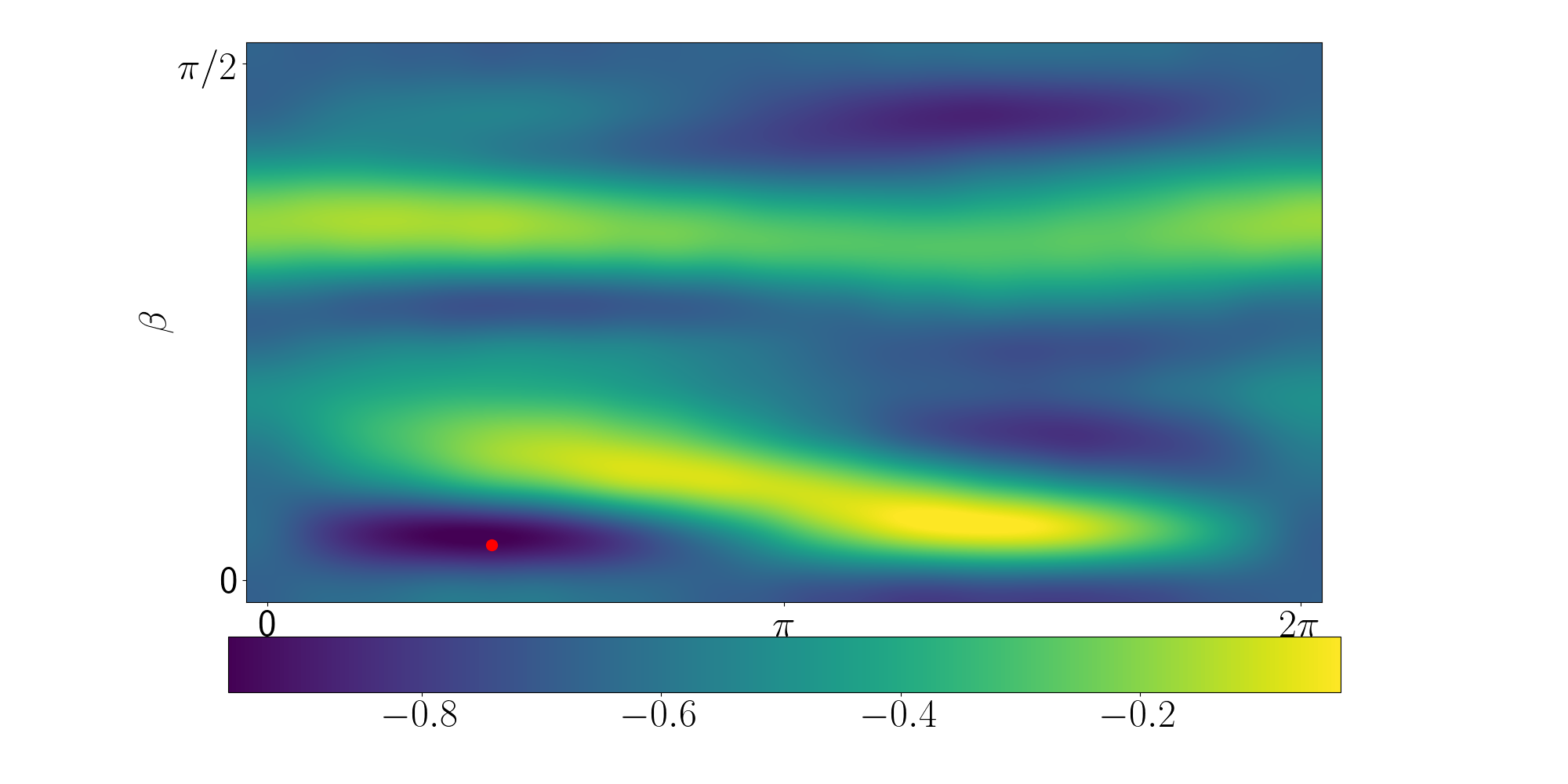}}\\
         4&\raisebox{-.5\height}{\includegraphics[trim=120 20 120 20,clip,width=.25\linewidth]{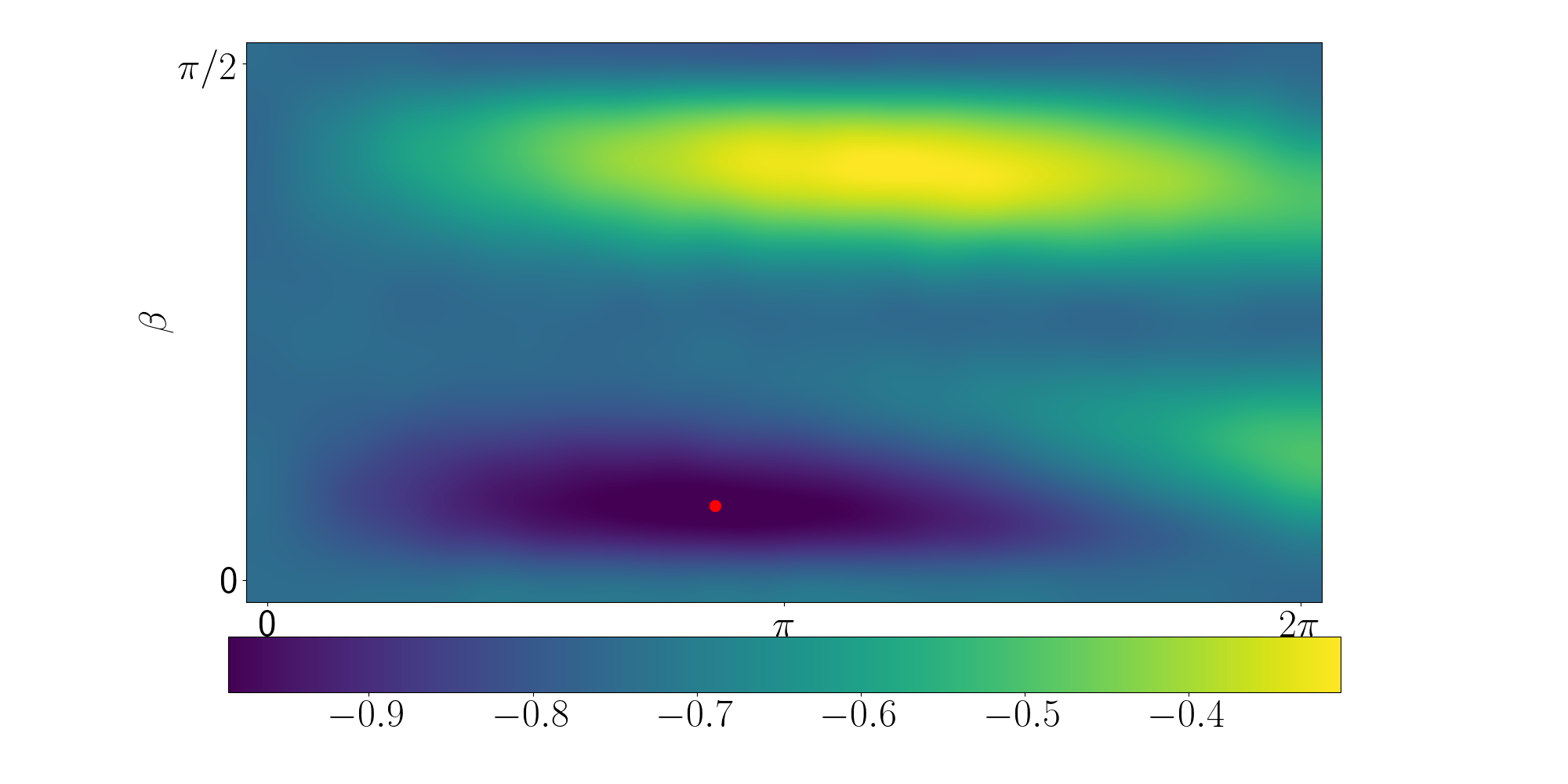}}
         &\raisebox{-.5\height}{\includegraphics[trim=120 20 120 20,clip,width=.25\linewidth]{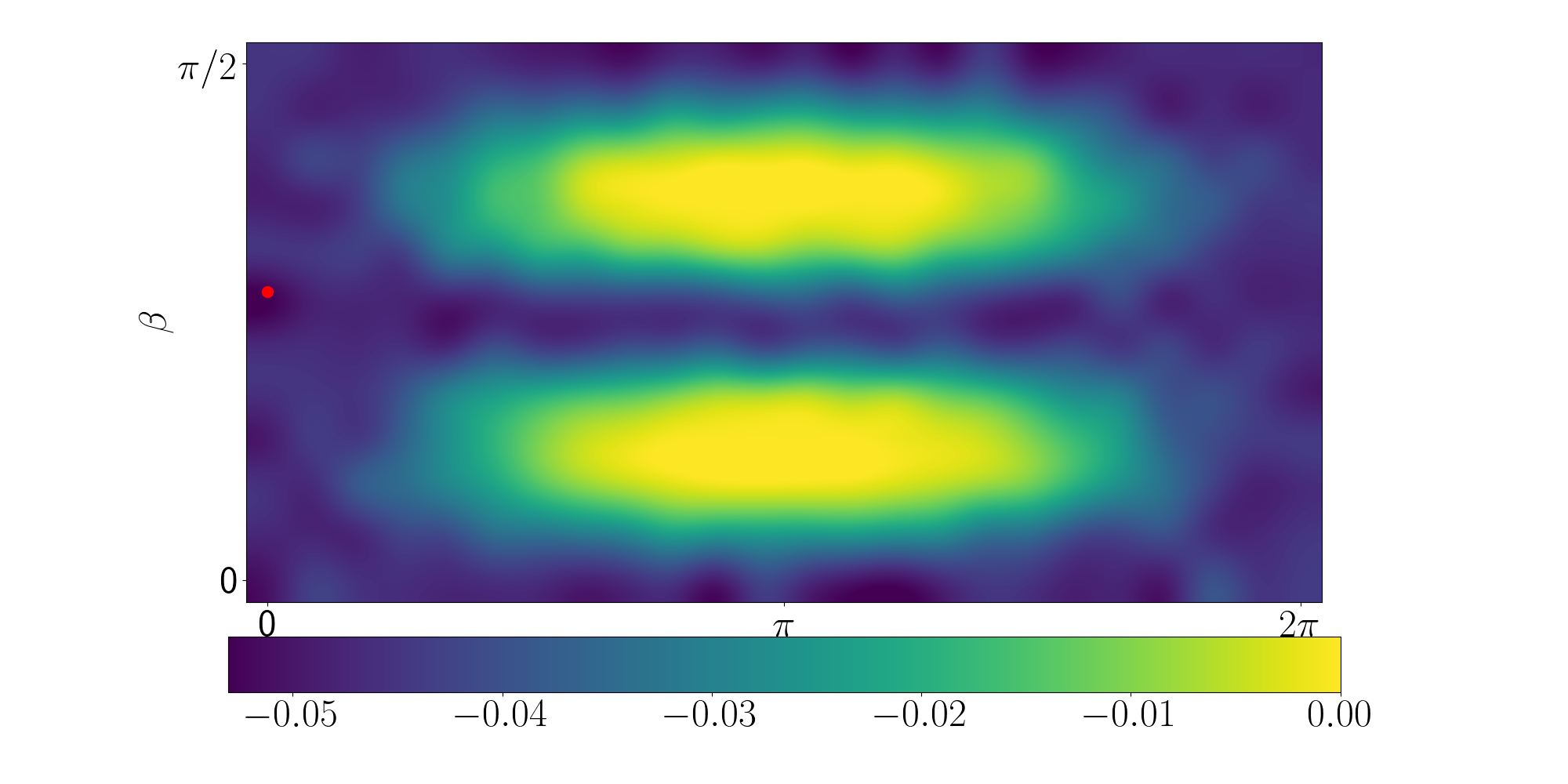}}
         &\raisebox{-.5\height}{\includegraphics[trim=120 20 120 20,clip,width=.25\linewidth]{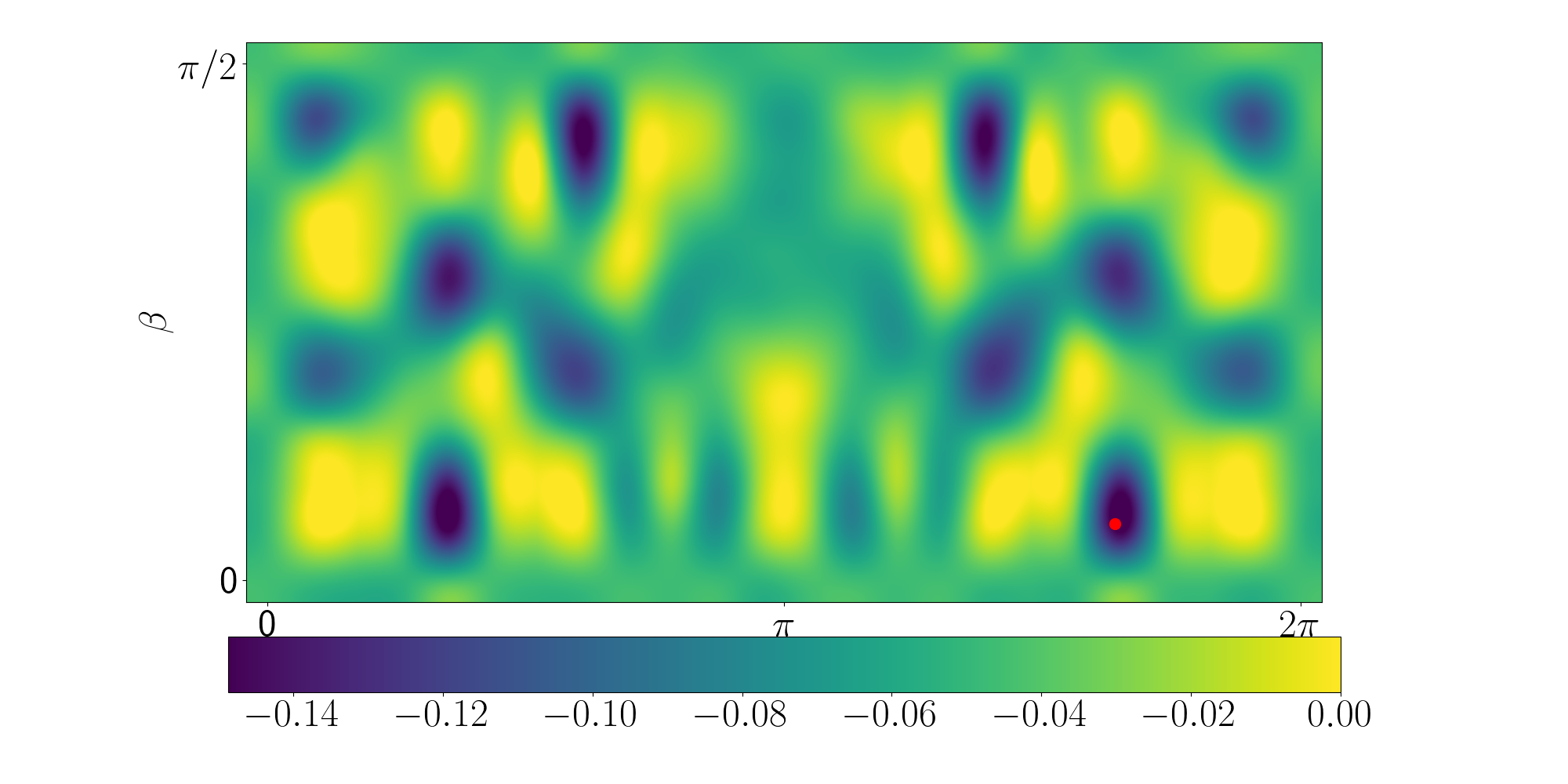}}
         &\raisebox{-.5\height}{\includegraphics[trim=120 20 120 20,clip,width=.25\linewidth]{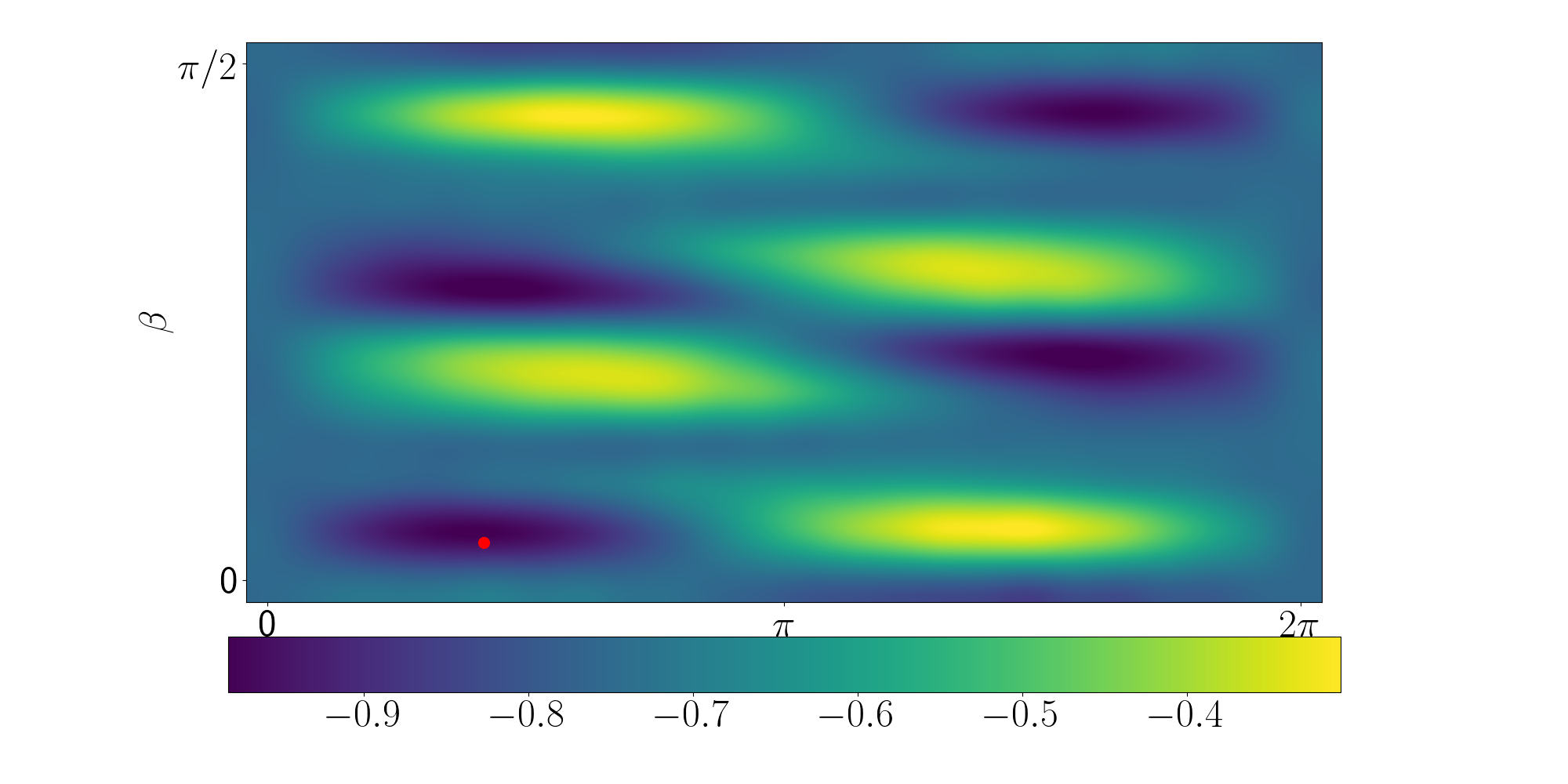}}\\
         5&\raisebox{-.5\height}{\includegraphics[trim=120 20 120 20,clip,width=.25\linewidth]{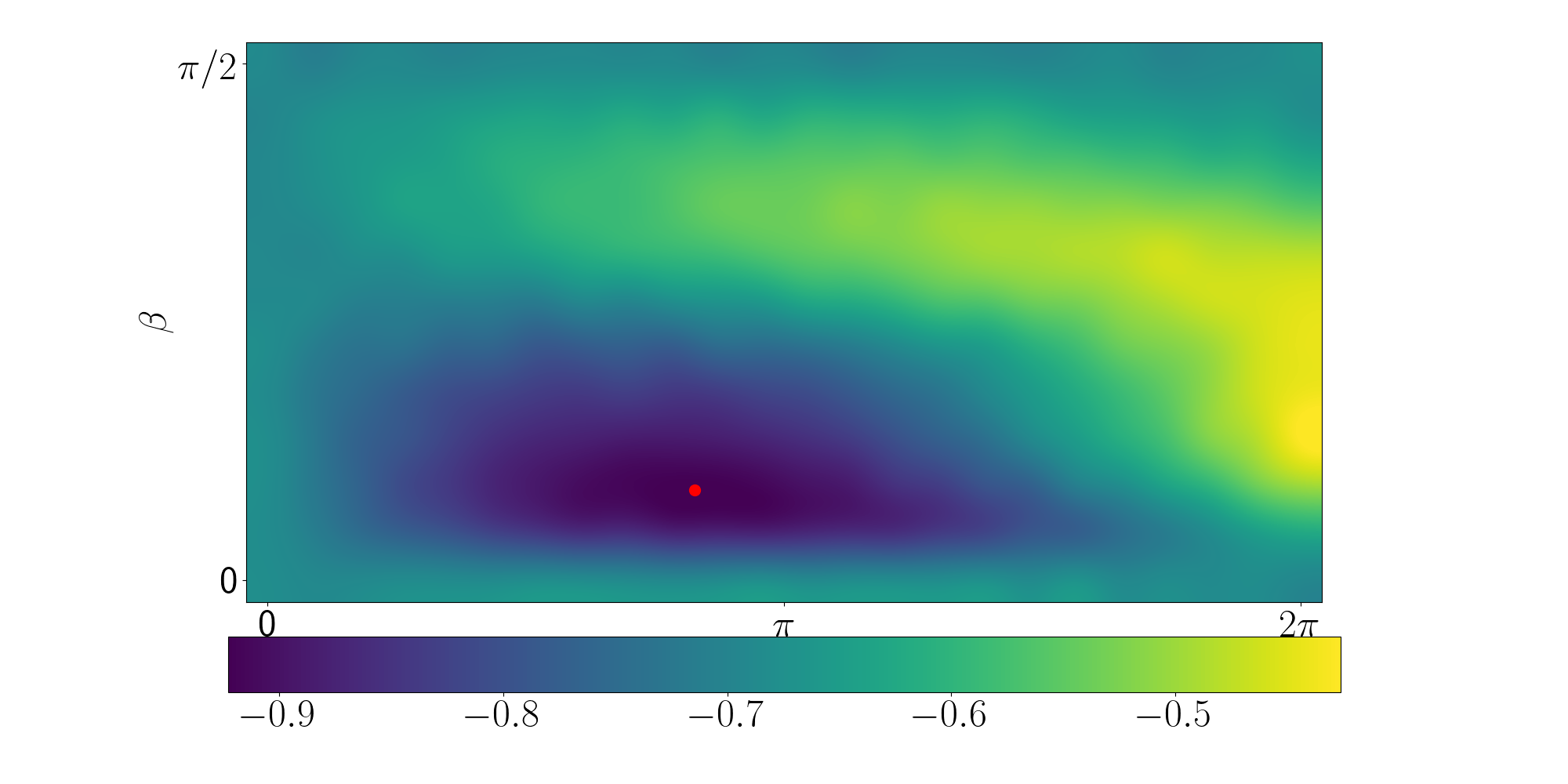}}
         &\raisebox{-.5\height}{\includegraphics[trim=120 20 120 20,clip,width=.25\linewidth]{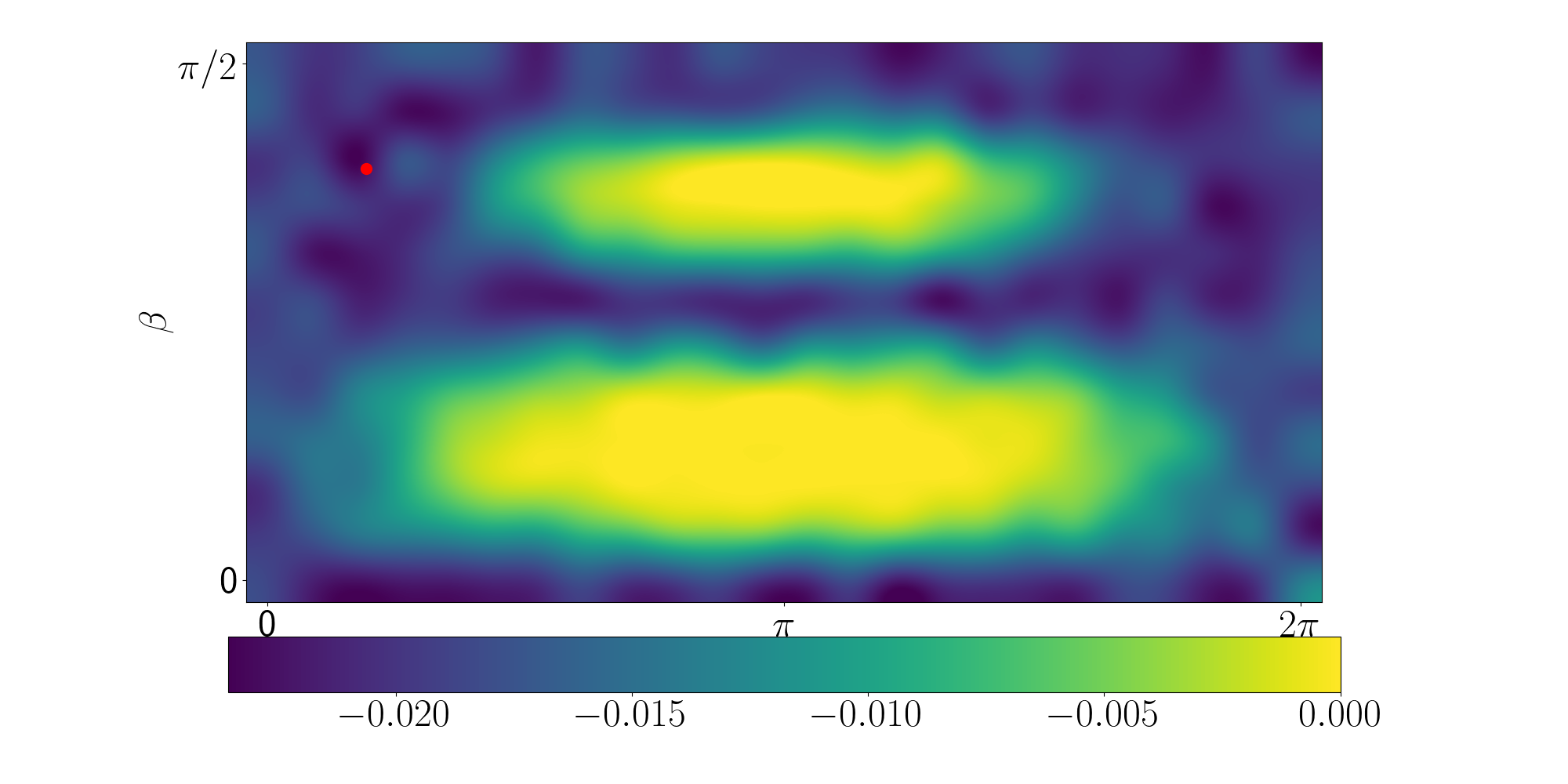}}
         &\raisebox{-.5\height}{\includegraphics[trim=120 20 120 20,clip,width=.25\linewidth]{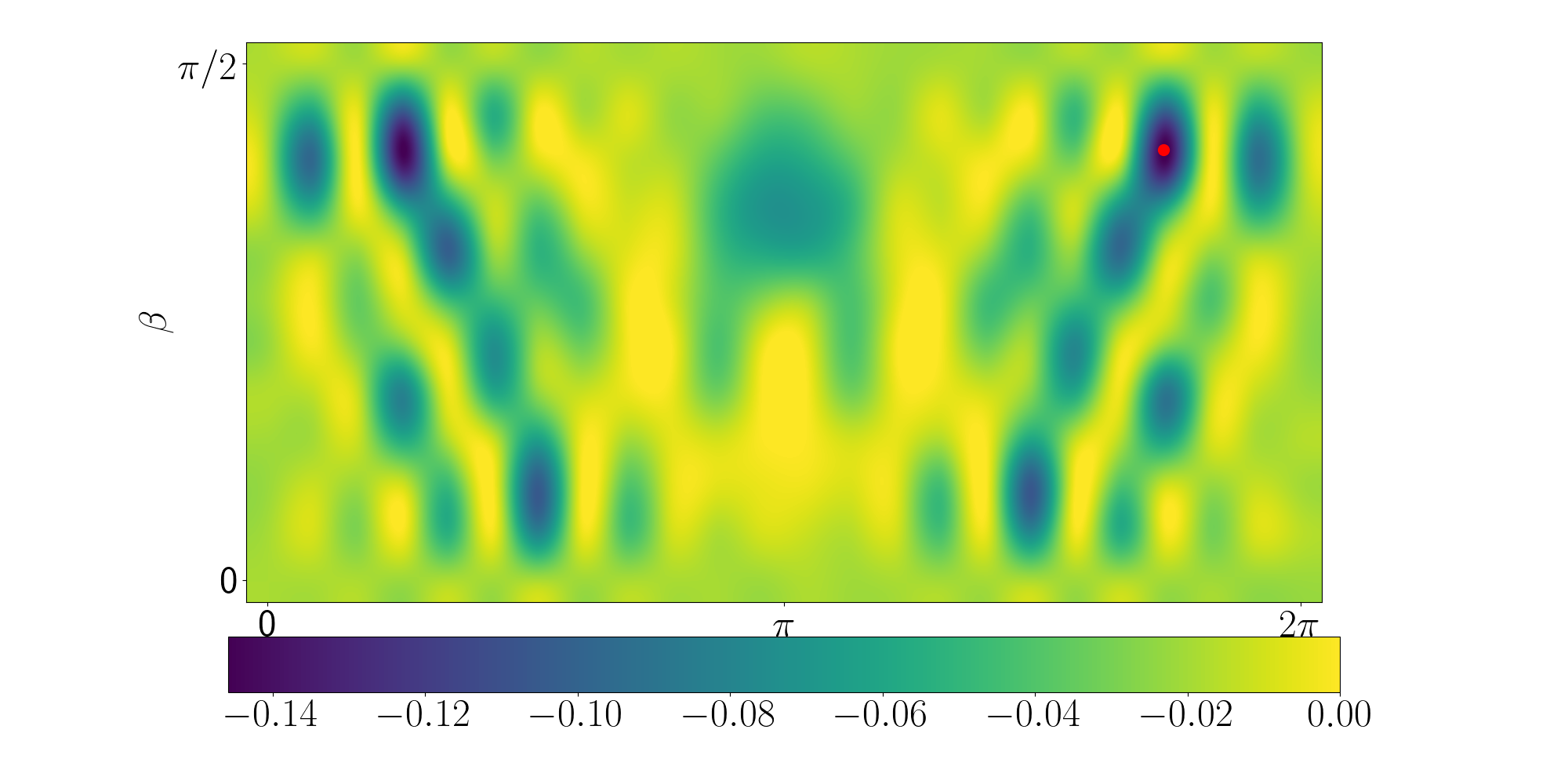}}
         &\raisebox{-.5\height}{\includegraphics[trim=120 20 120 20,clip,width=.25\linewidth]{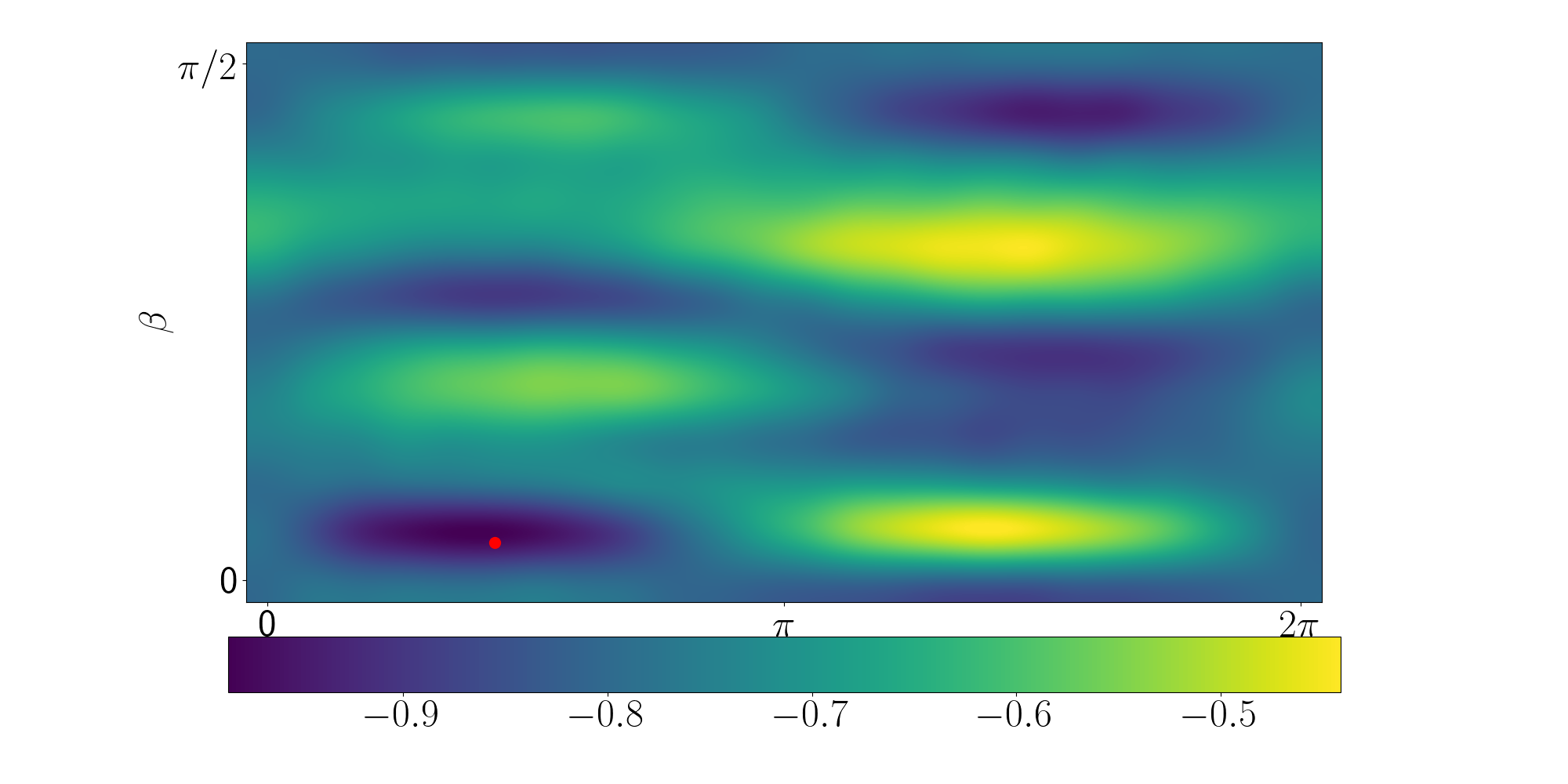}}\\
         6&\raisebox{-.5\height}{\includegraphics[trim=120 20 120 20,clip,width=.25\linewidth]{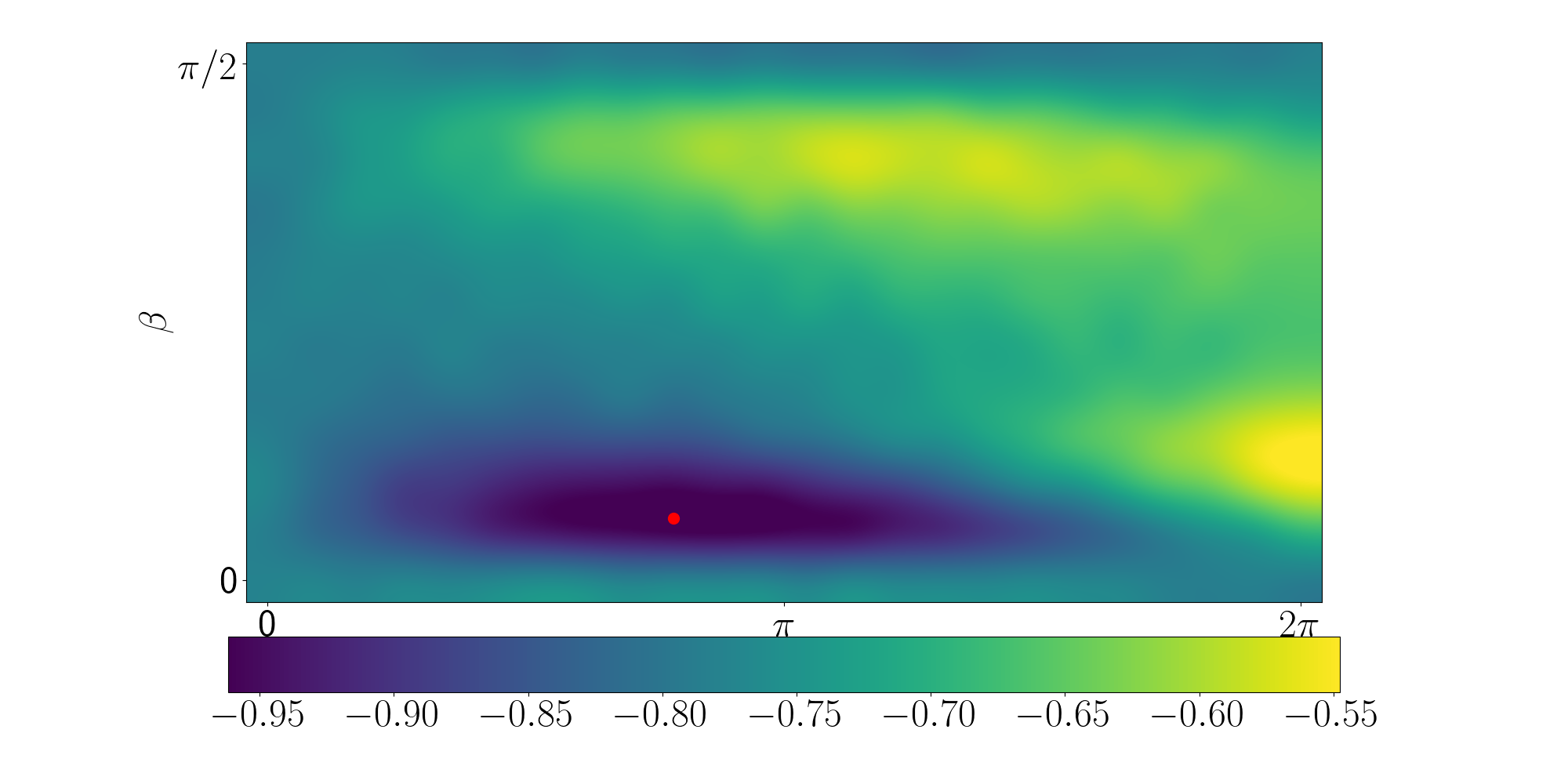}}
         &\raisebox{-.5\height}{\includegraphics[trim=120 20 120 20,clip,width=.25\linewidth]{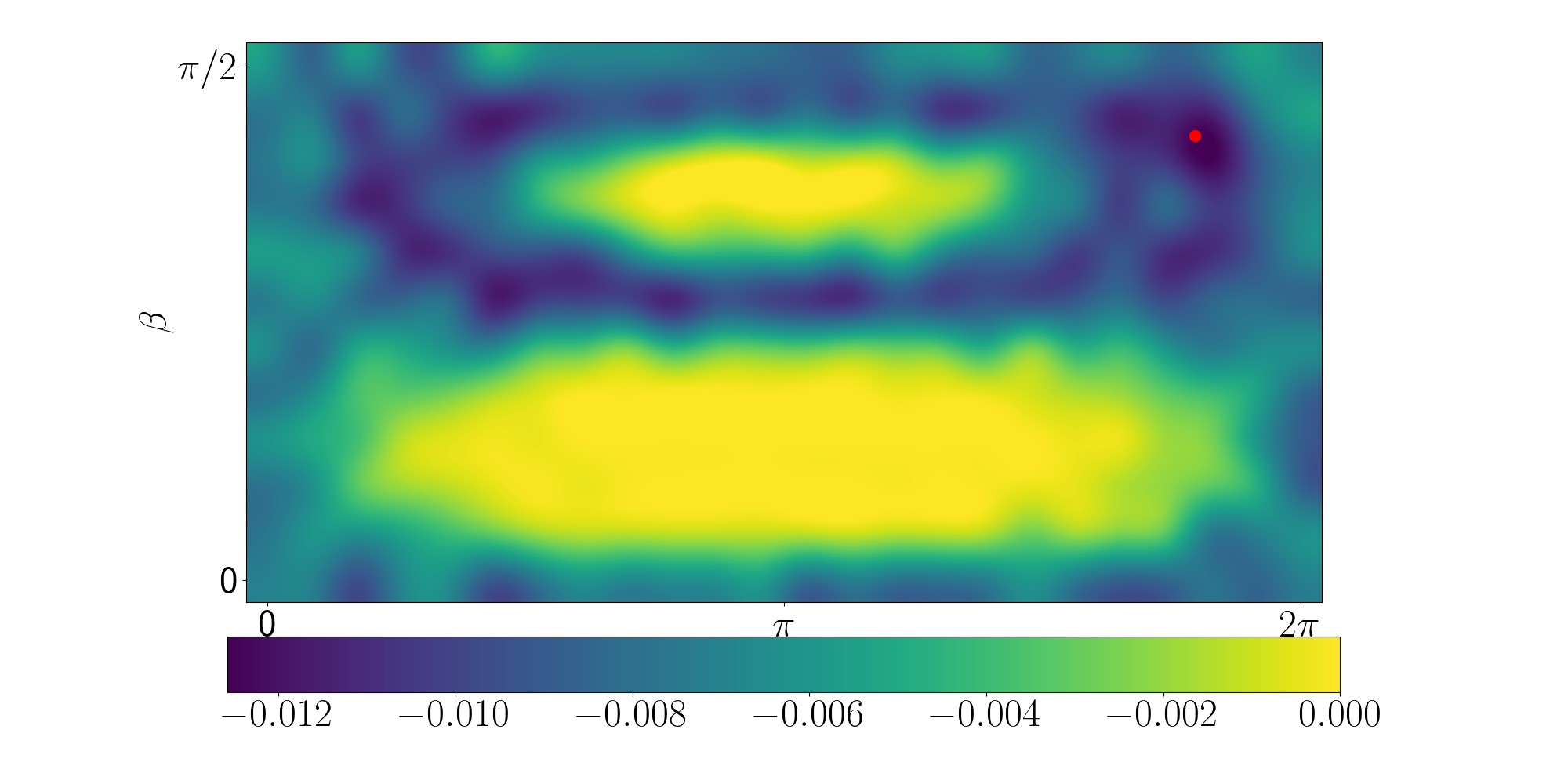}}
         &\raisebox{-.5\height}{\includegraphics[trim=120 20 120 20,clip,width=.25\linewidth]{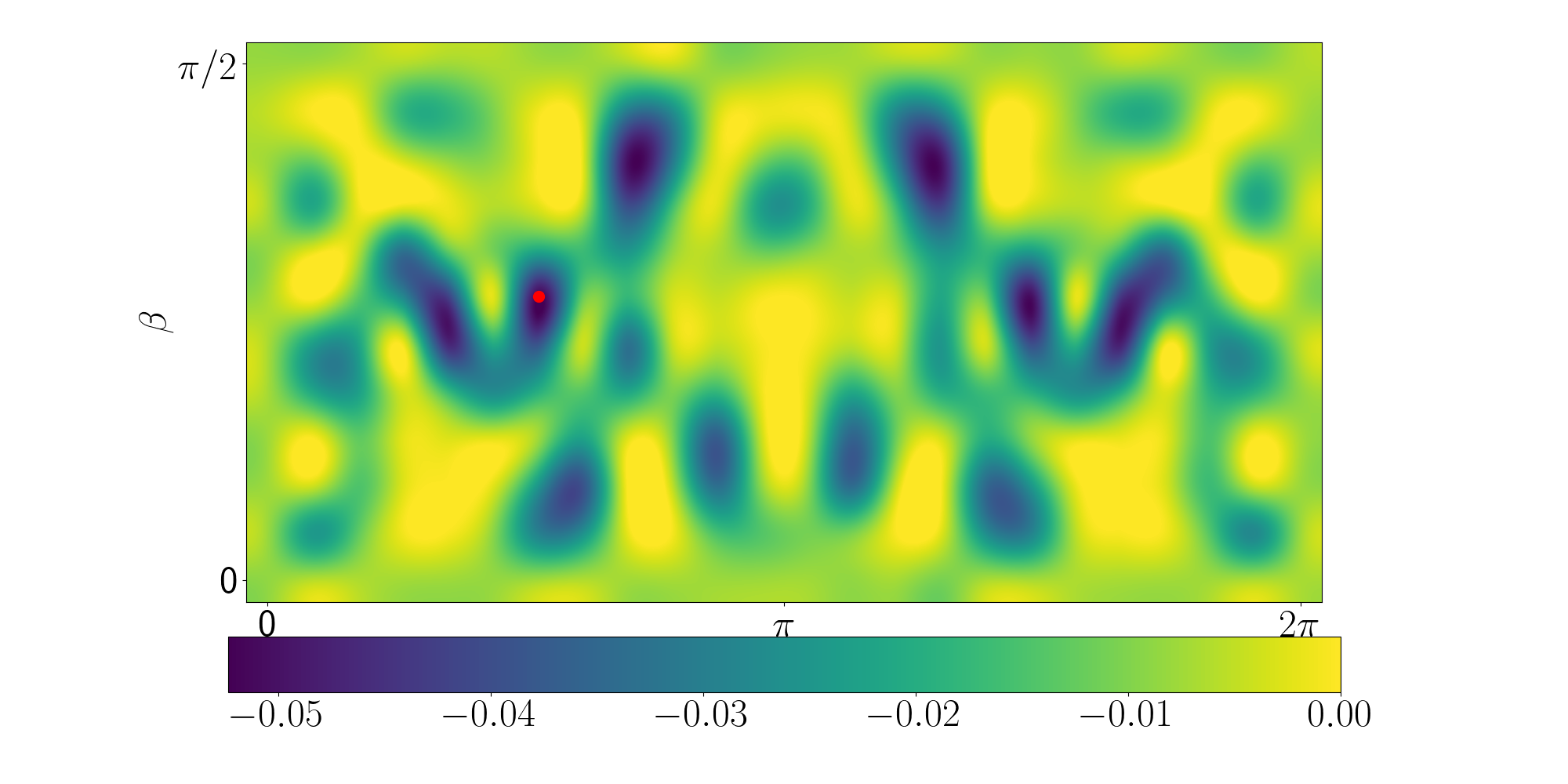}}
         &\raisebox{-.5\height}{\includegraphics[trim=120 20 120 20,clip,width=.25\linewidth]{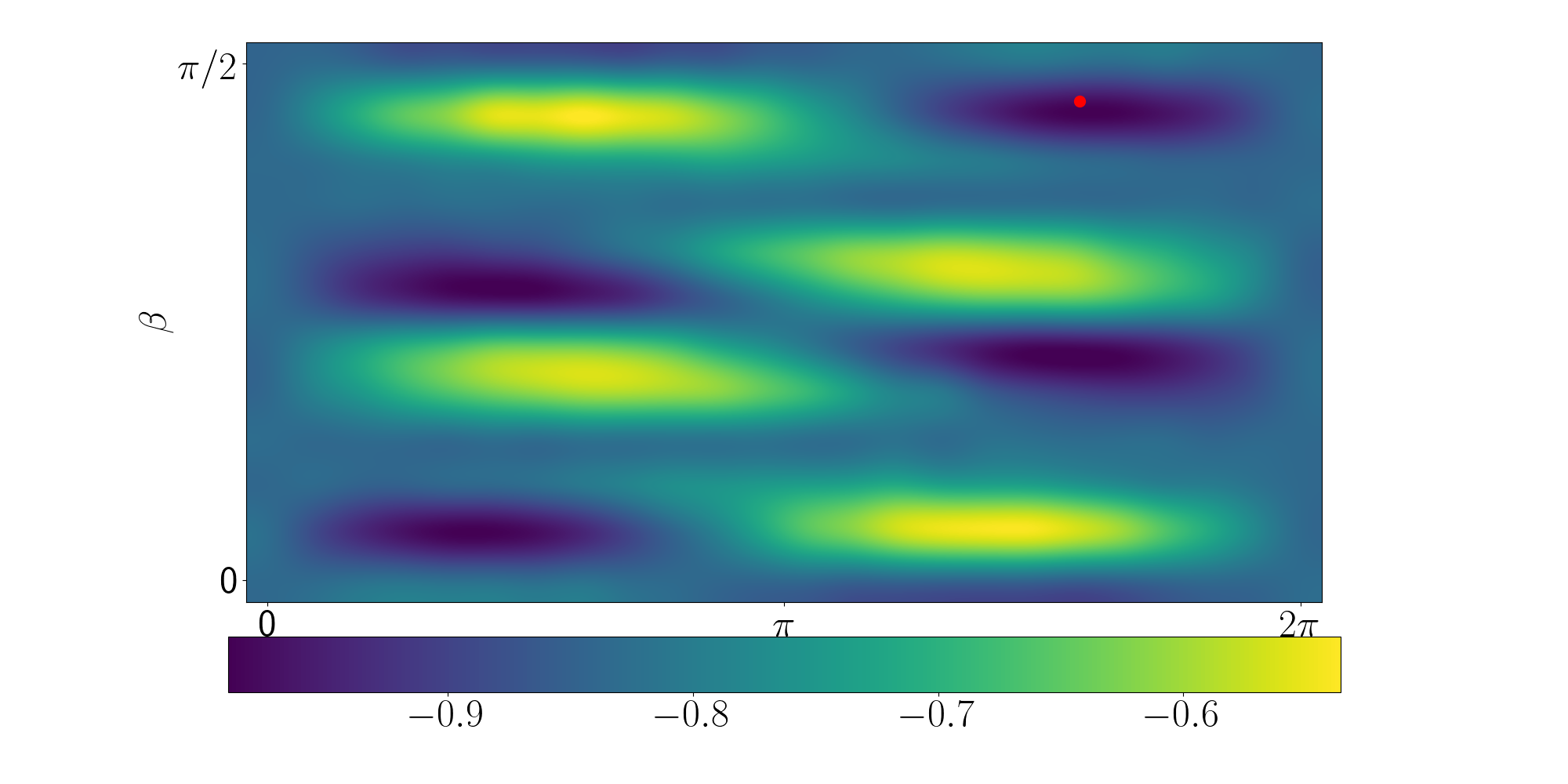}}\\
         7&\raisebox{-.5\height}{\includegraphics[trim=120 20 120 20,clip,width=.25\linewidth]{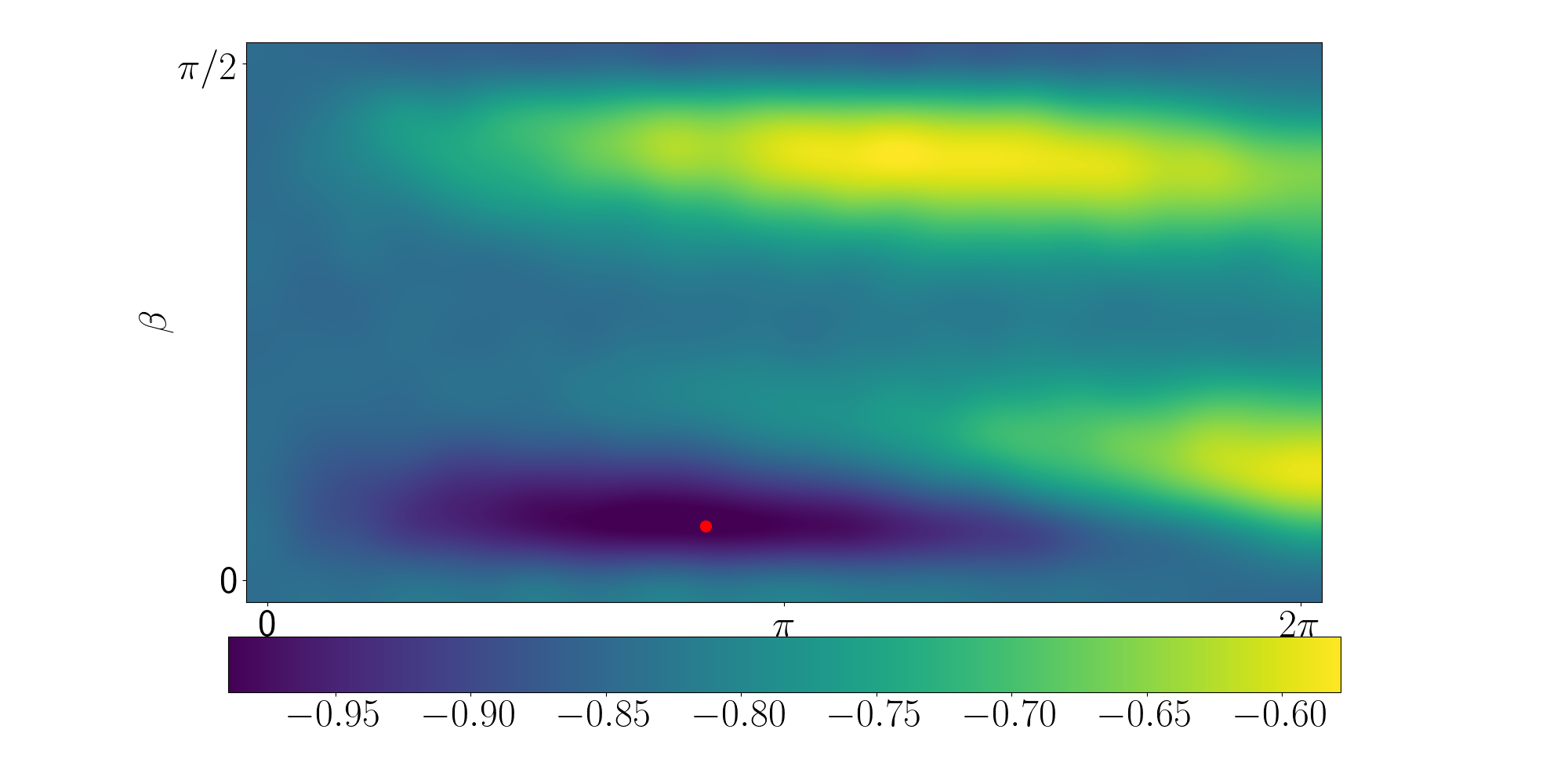}}
         &\raisebox{-.5\height}{\includegraphics[trim=120 20 120 20,clip,width=.25\linewidth]{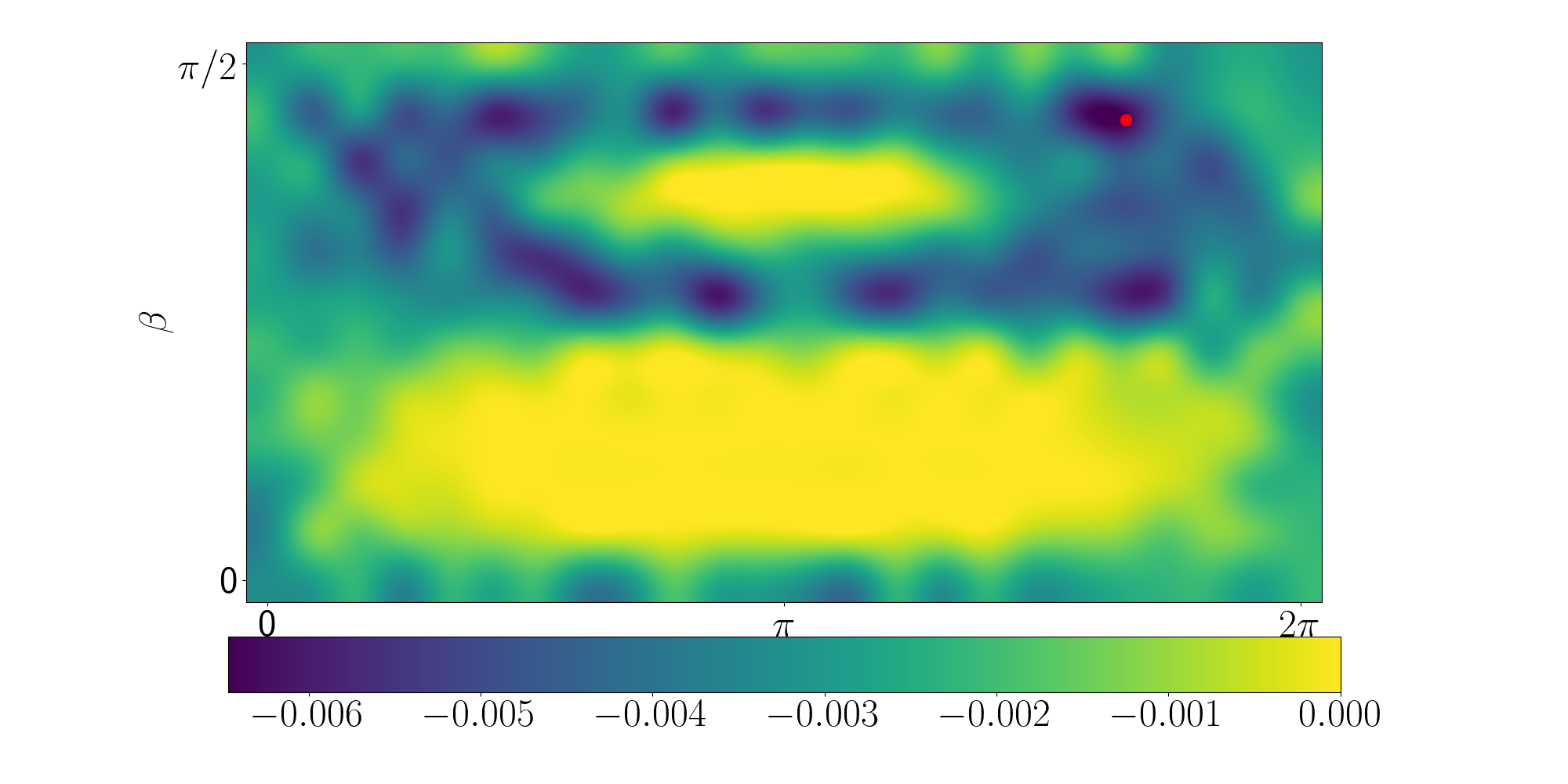}}
         &\raisebox{-.5\height}{\includegraphics[trim=120 20 120 20,clip,width=.25\linewidth]{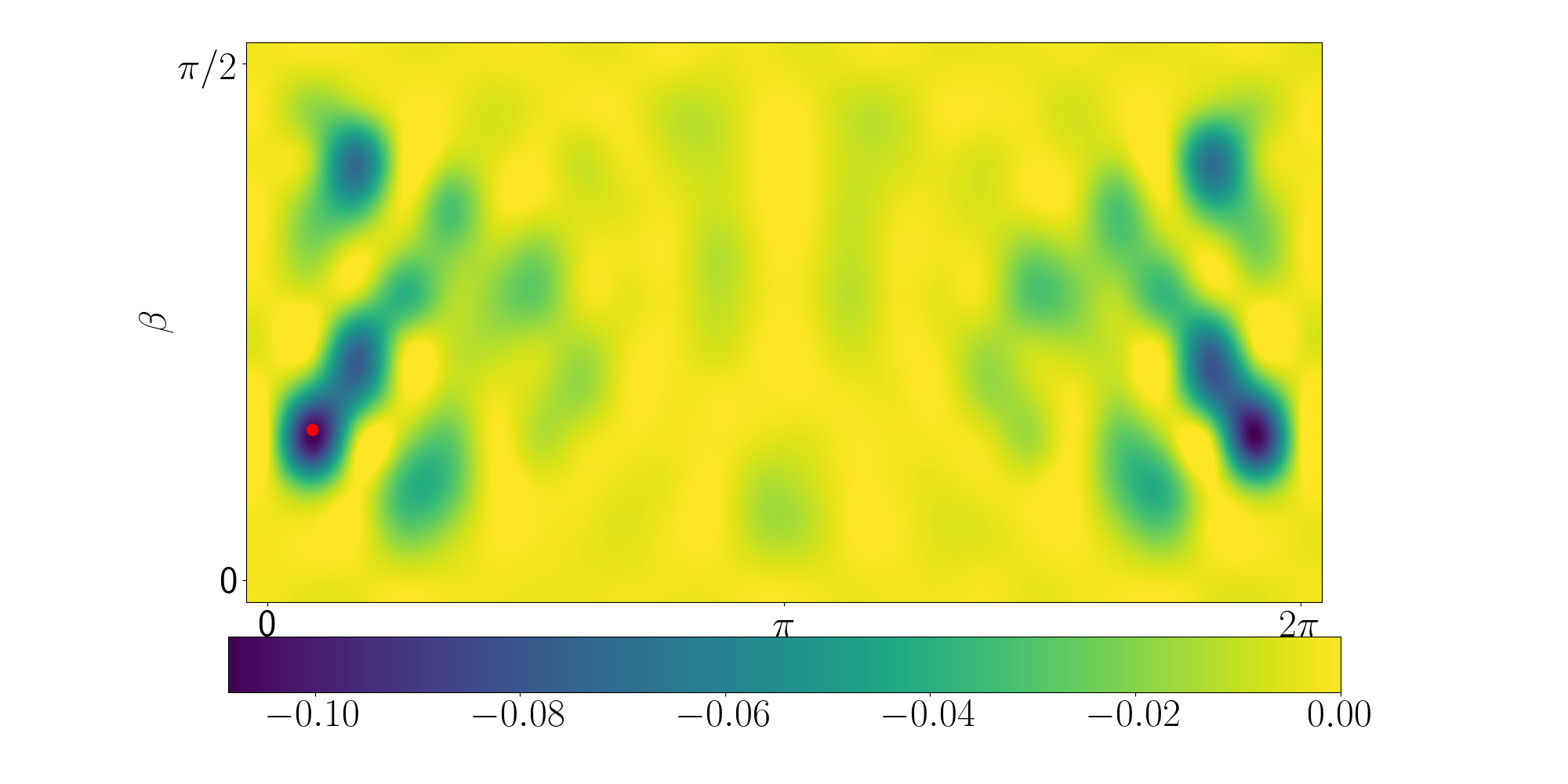}}
         &\raisebox{-.5\height}{\includegraphics[trim=120 20 120 20,clip,width=.25\linewidth]{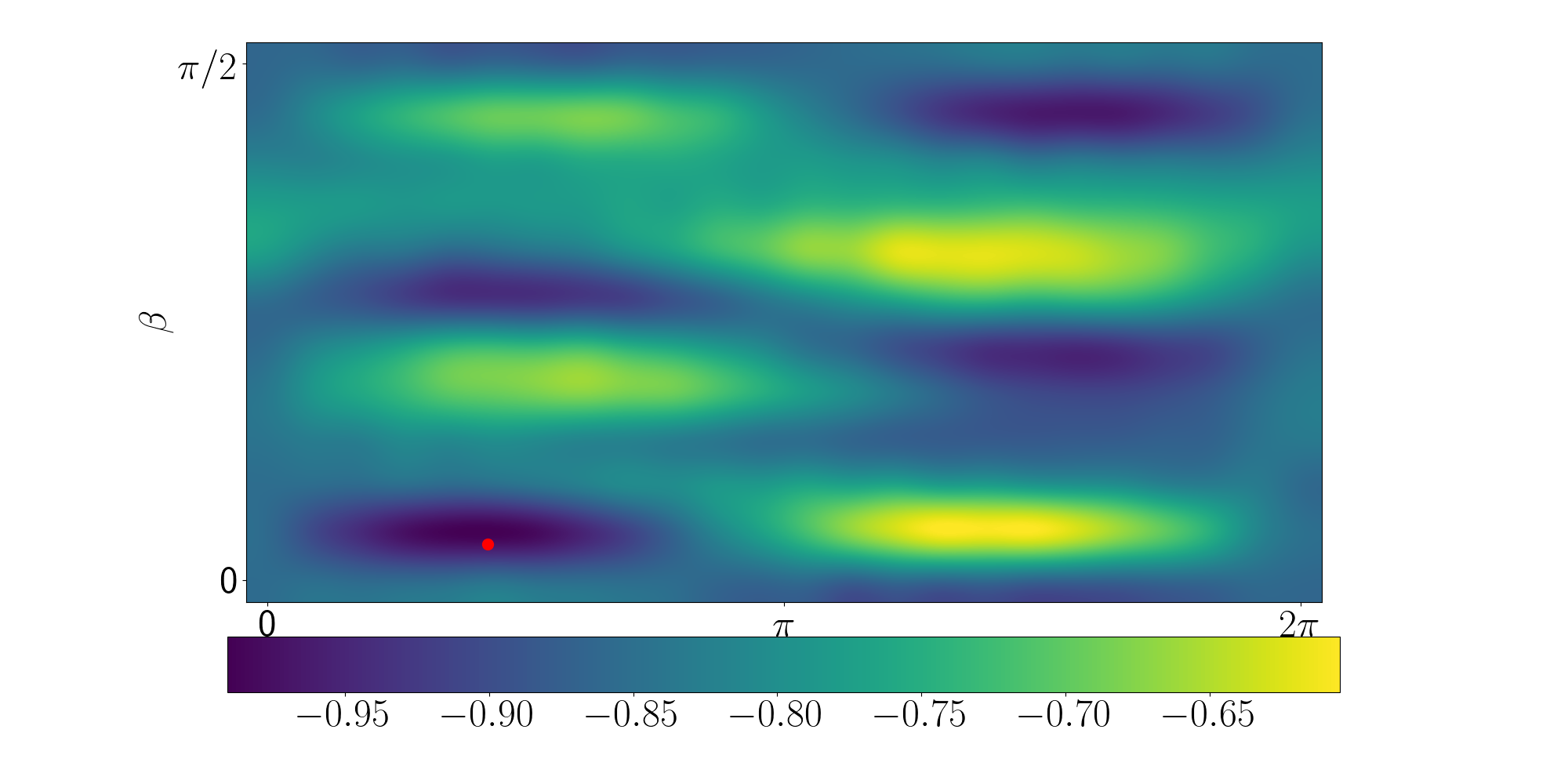}}\\
         8&\raisebox{-.5\height}{\includegraphics[trim=120 20 120 20,clip,width=.25\linewidth]{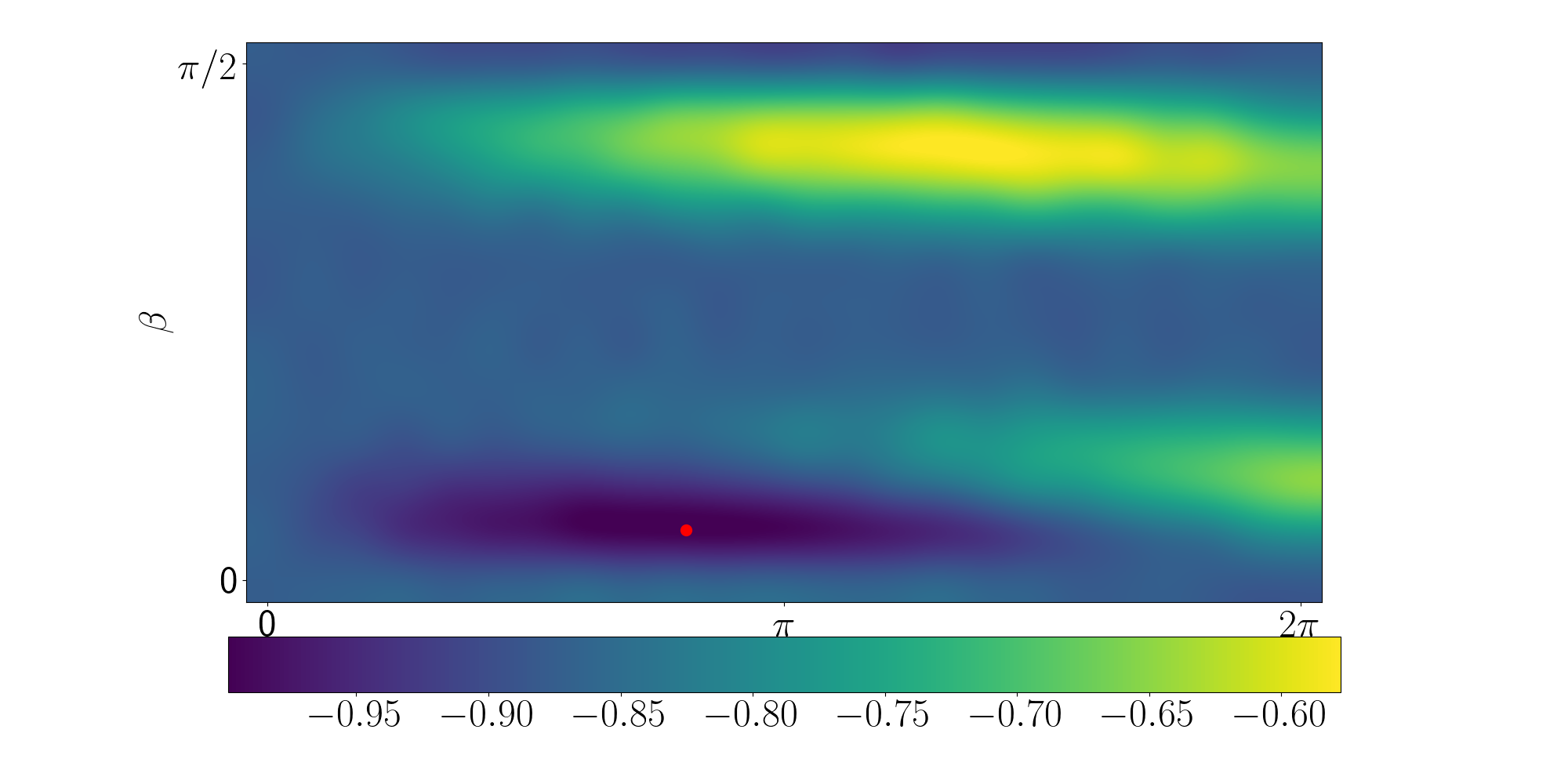}}
         &\raisebox{-.5\height}{\includegraphics[trim=120 20 120 20,clip,width=.25\linewidth]{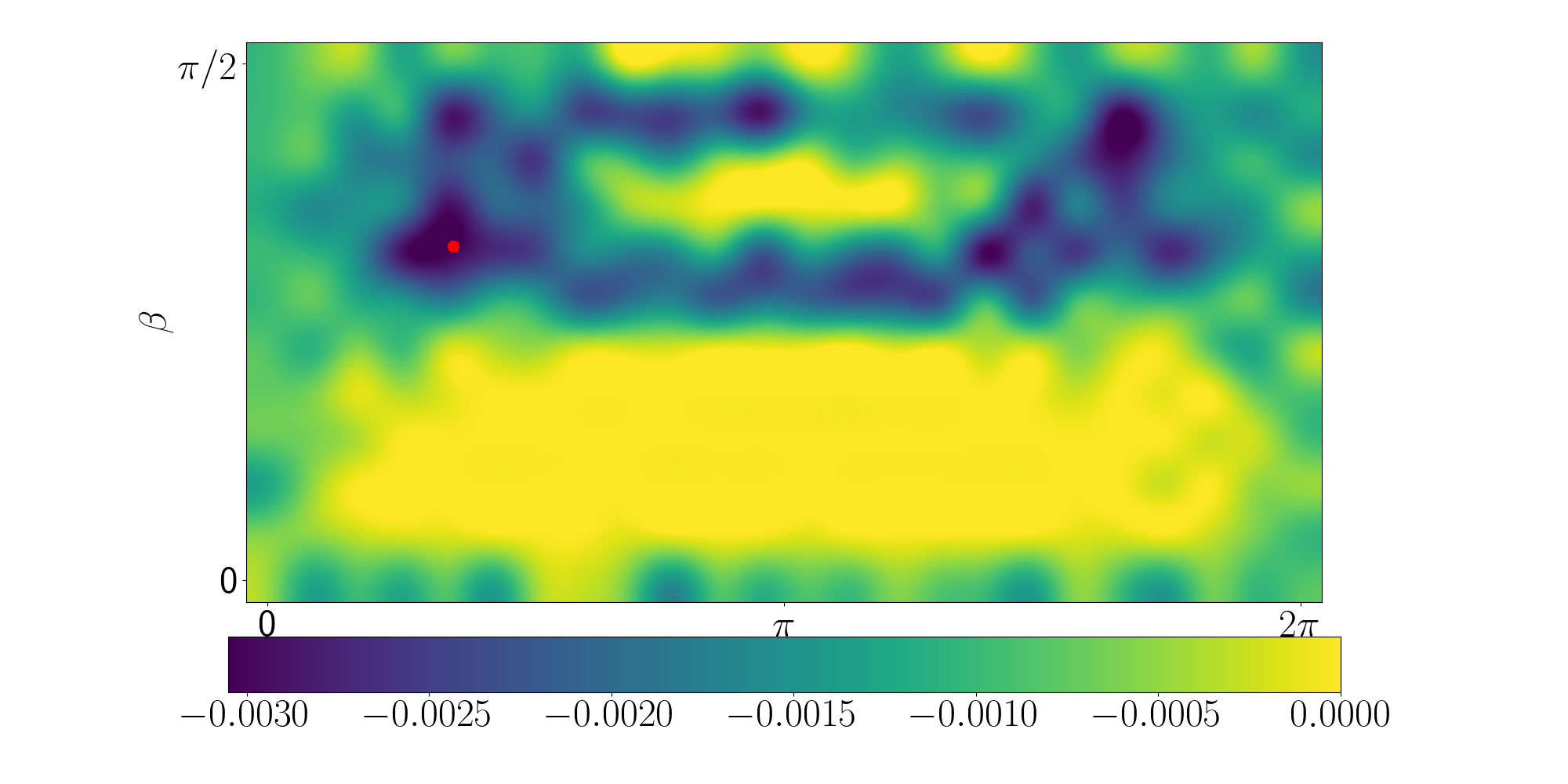}}
         &\raisebox{-.5\height}{\includegraphics[trim=120 20 120 20,clip,width=.25\linewidth]{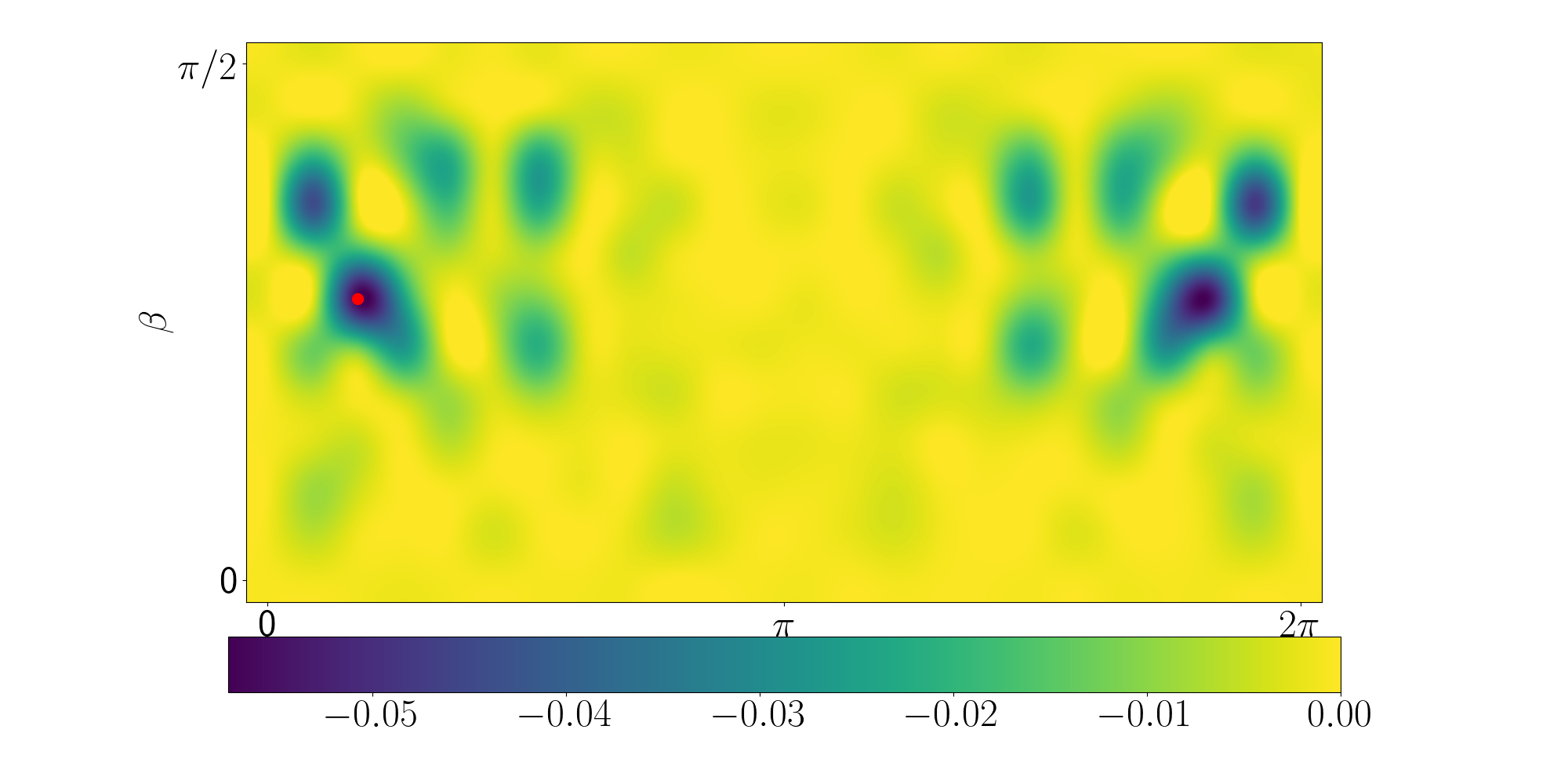}}
         &\raisebox{-.5\height}{\includegraphics[trim=120 20 120 20,clip,width=.25\linewidth]{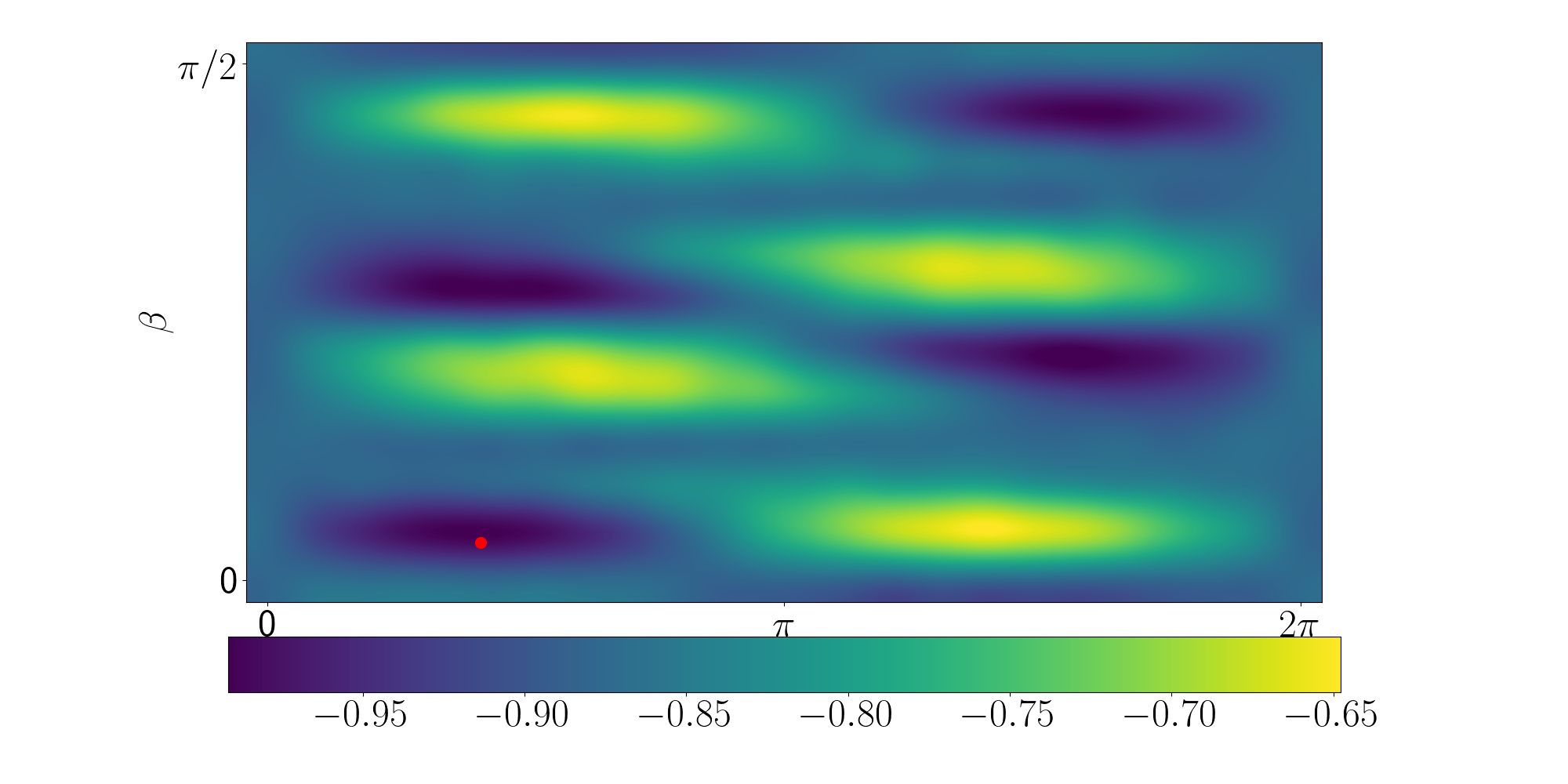}}\\
    \end{tabular}
    \endgroup
    \caption{Energy landscapes for the ``Barbel''-graph shown in Figure~\ref{fig:Barbell}. Generally, the binary encoding seems to generates easier optimization problems.}
    \label{tab:E_Barbell}
\end{table}

\begin{table}[]
    \centering
    \begingroup
    \setlength{\tabcolsep}{0pt} 
    \renewcommand{\arraystretch}{0} 
    \begin{tabular}{ccccc}
    &\multicolumn{2}{c}{Erdős-Rényi un-weighted}&\multicolumn{2}{c}{Barabási-Albert weighted}\\
         $k$ & binary & one-hot $XY$ & binary & one-hot $XY$ \\
         2&\raisebox{-.5\height}{\includegraphics[width=.25\linewidth]{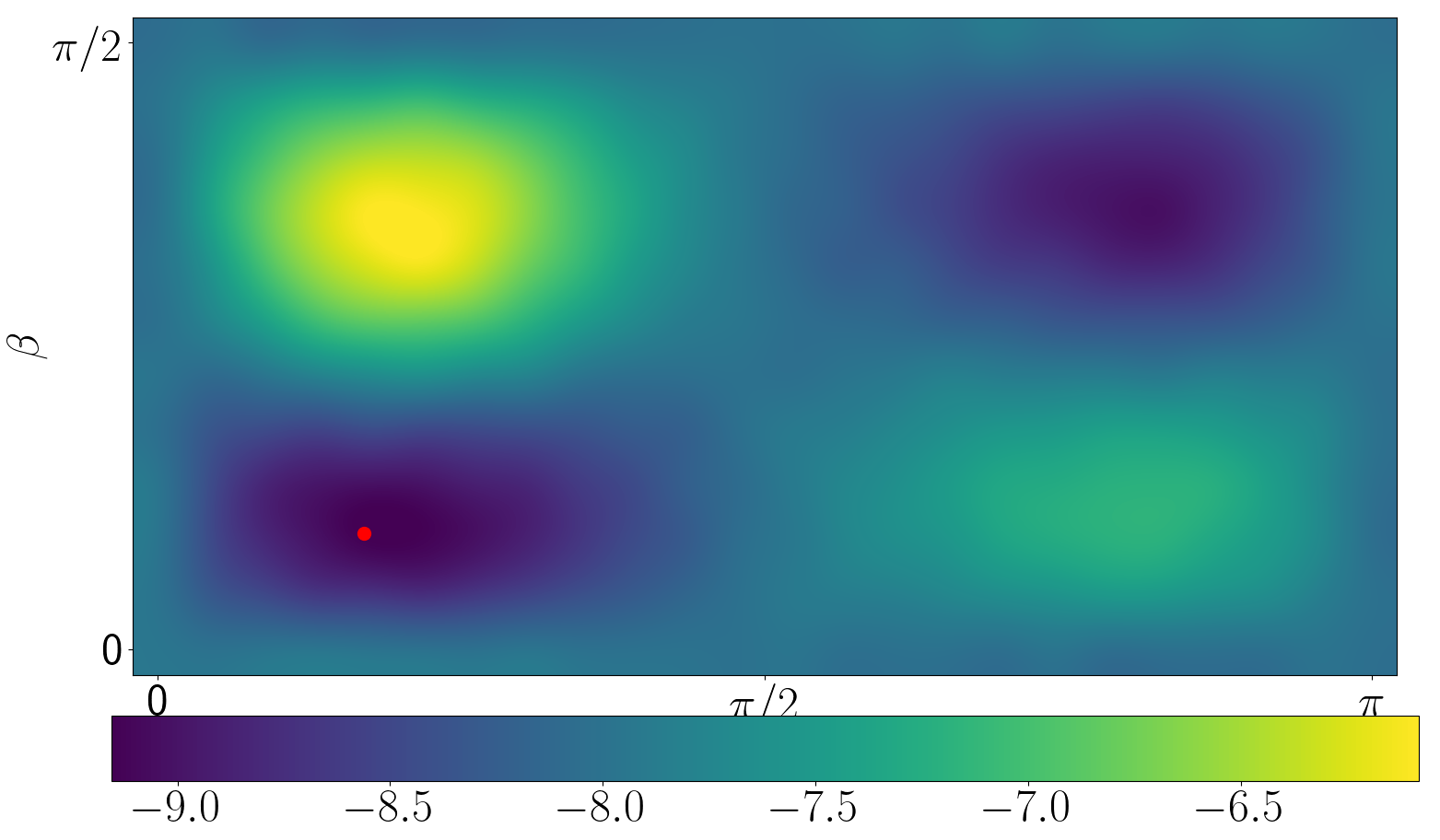}}
         &\raisebox{-.5\height}{\includegraphics[trim=120 20 120 20,clip,width=.25\linewidth]{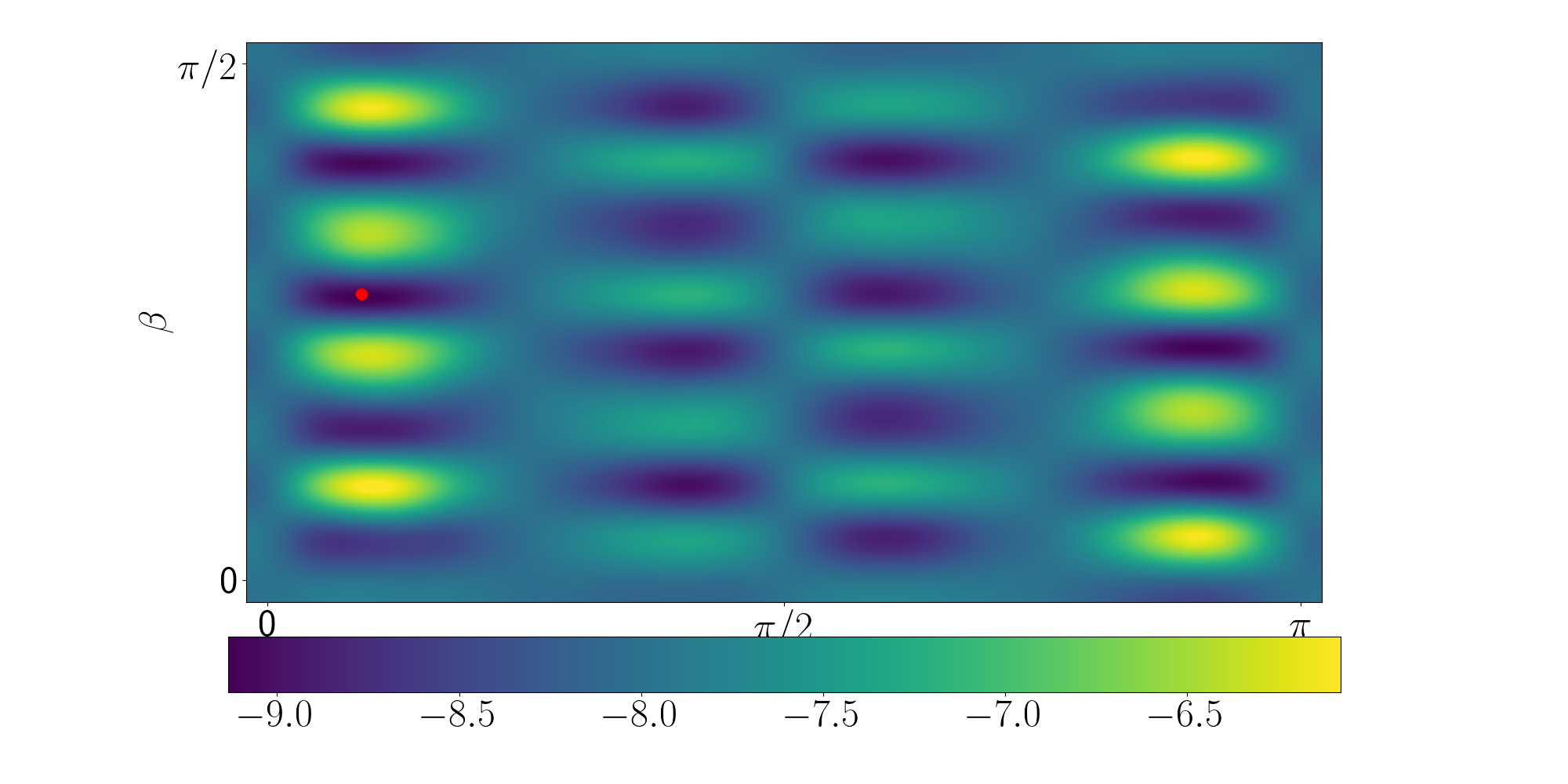}}
         &\raisebox{-.5\height}{\includegraphics[width=.25\linewidth]{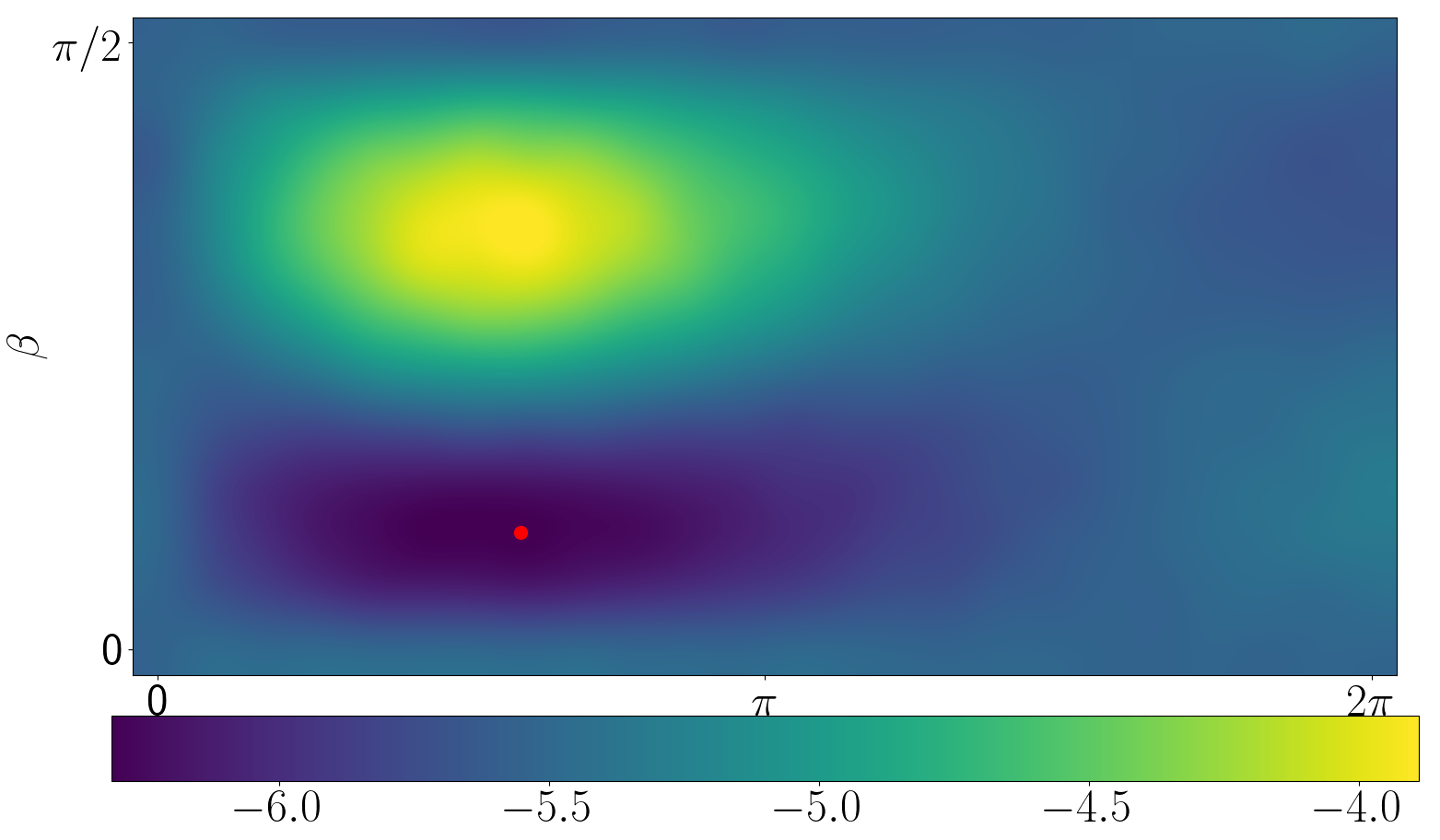}}
         &\raisebox{-.5\height}{\includegraphics[trim=120 20 120 20,clip,width=.25\linewidth]{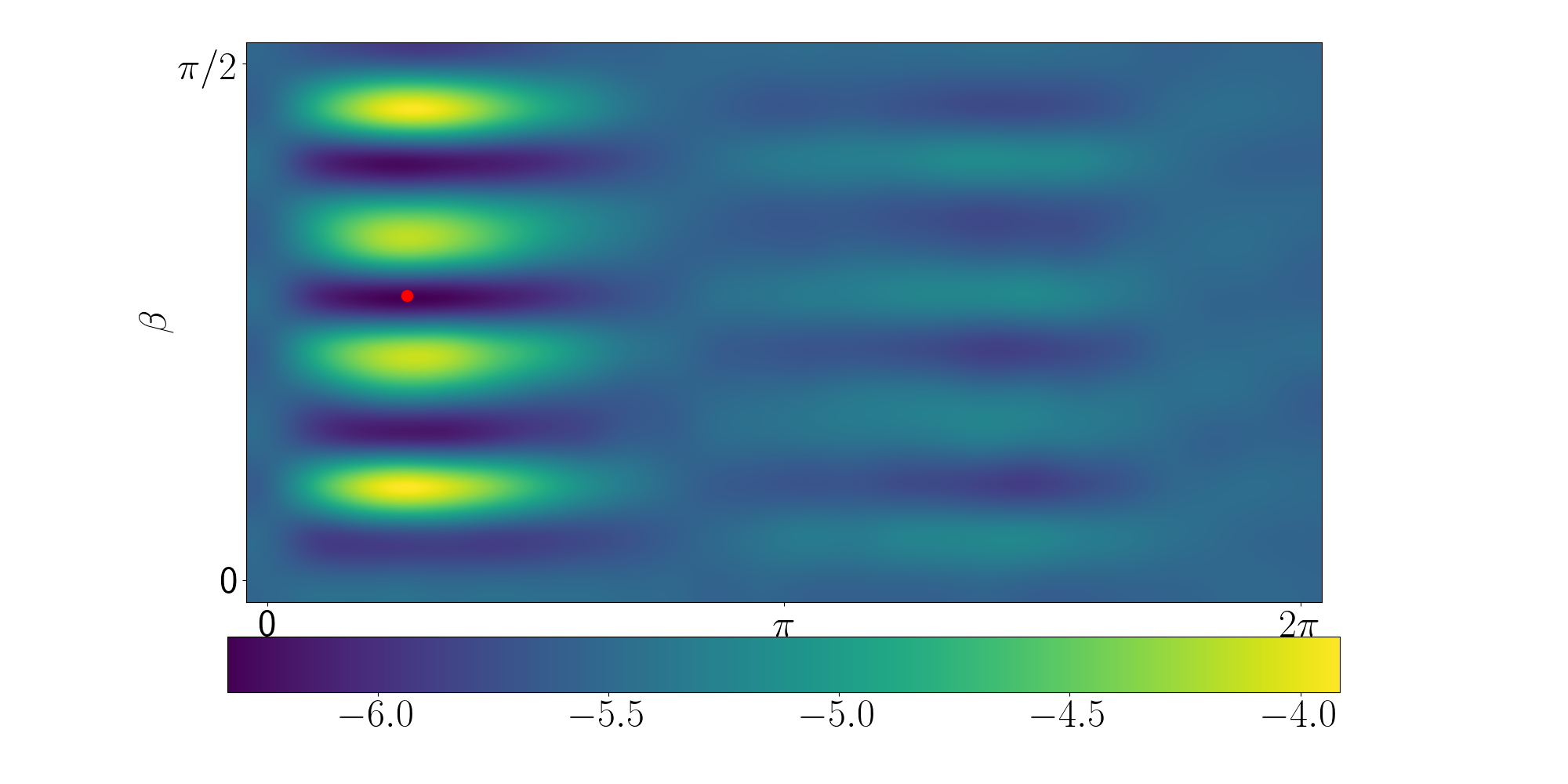}}\\
         3&\raisebox{-.5\height}{\includegraphics[width=.25\linewidth]{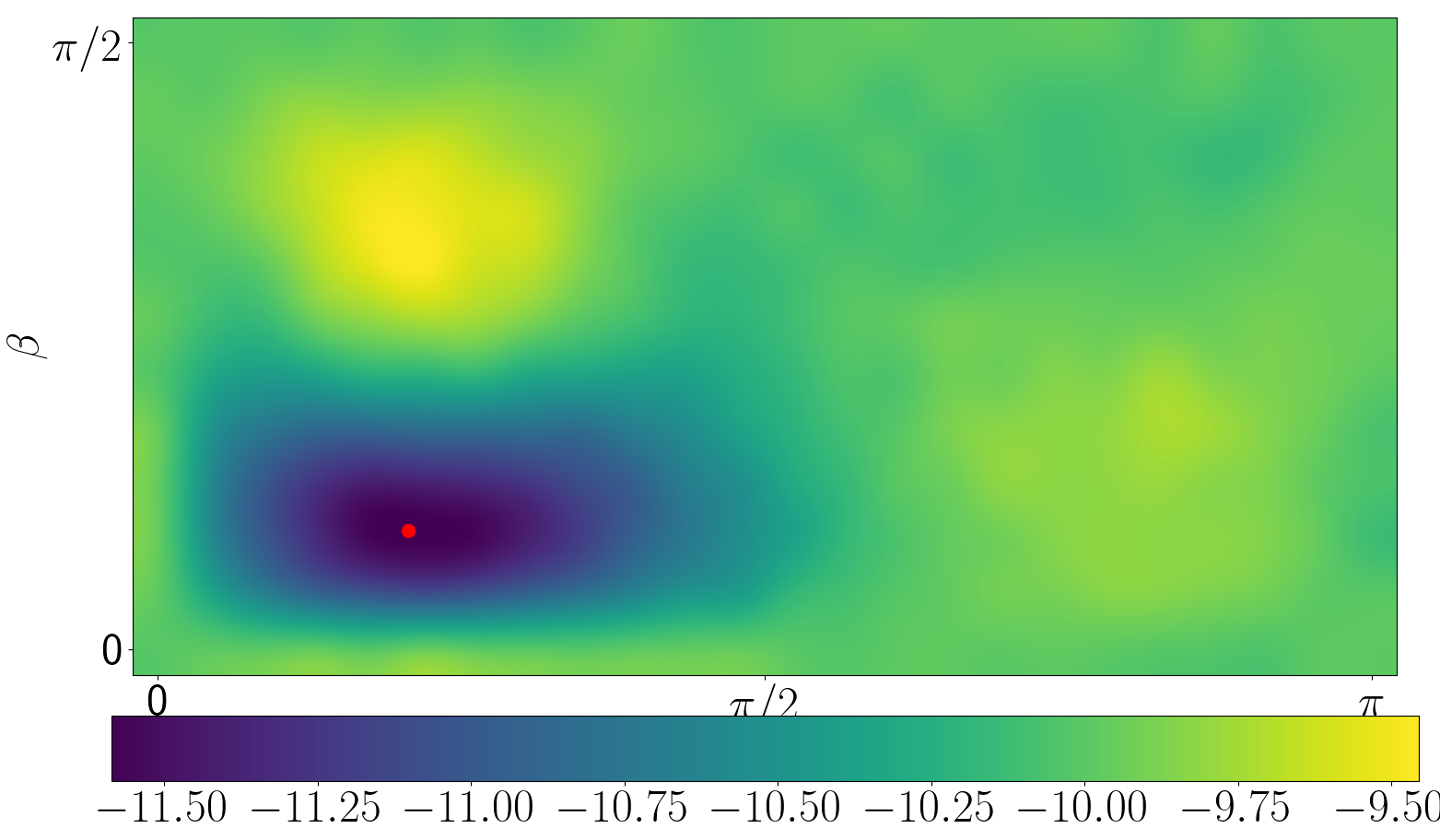}}
         &\raisebox{-.5\height}{\includegraphics[trim=120 20 120 20,clip,width=.25\linewidth]{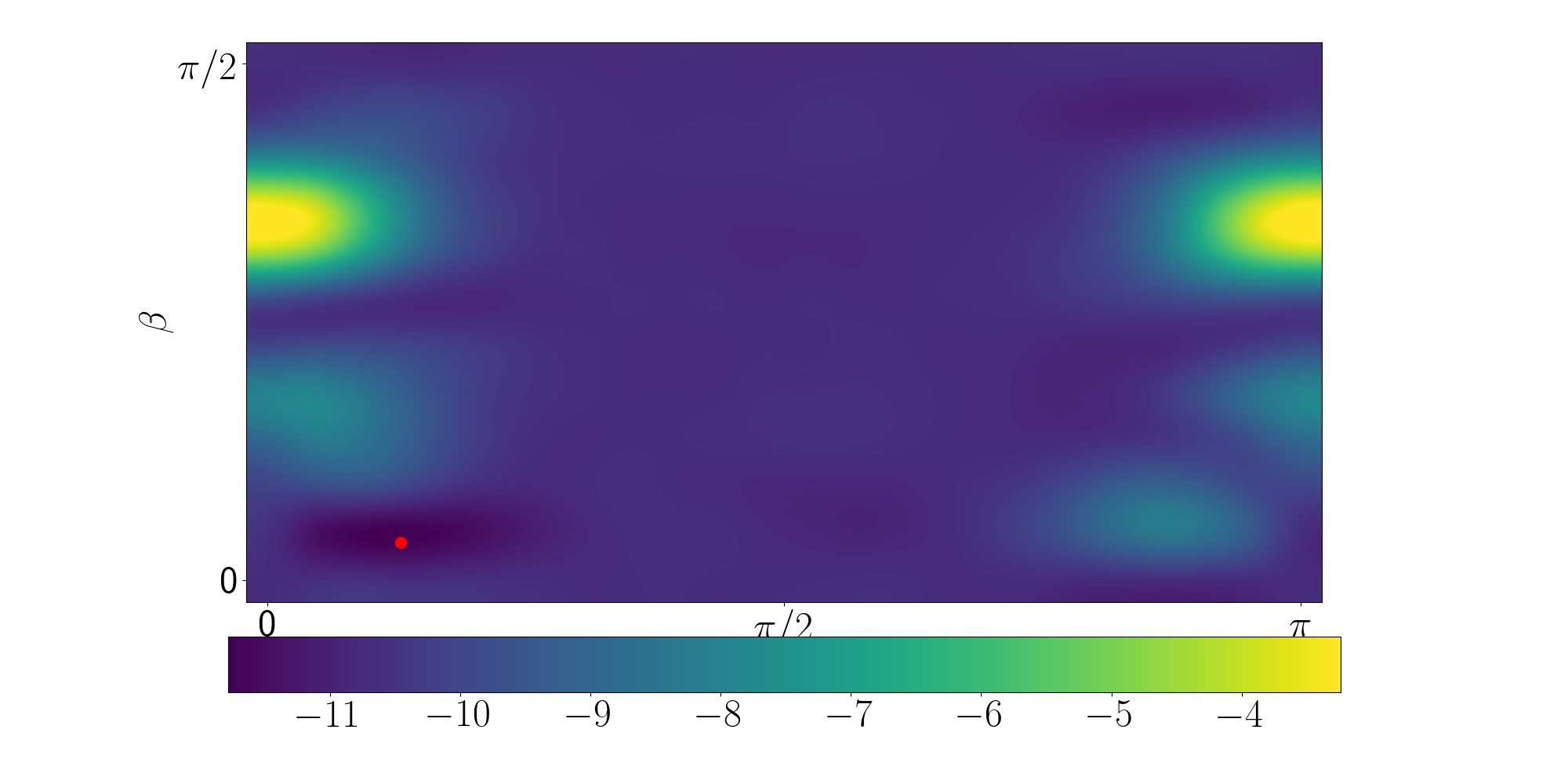}}
         &\raisebox{-.5\height}{\includegraphics[width=.25\linewidth]{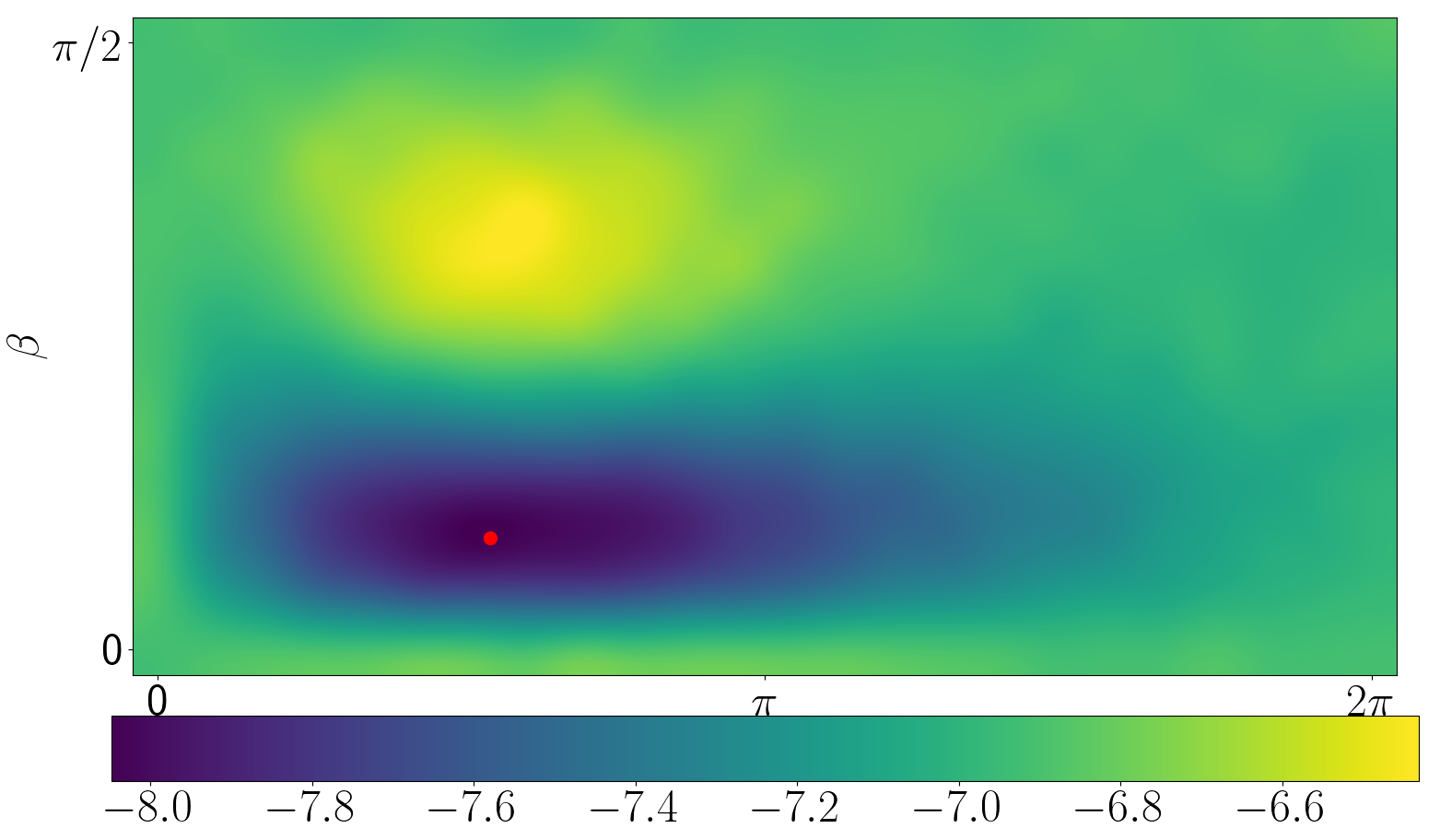}}
         &\raisebox{-.5\height}{\includegraphics[trim=120 20 120 20,clip,width=.25\linewidth]{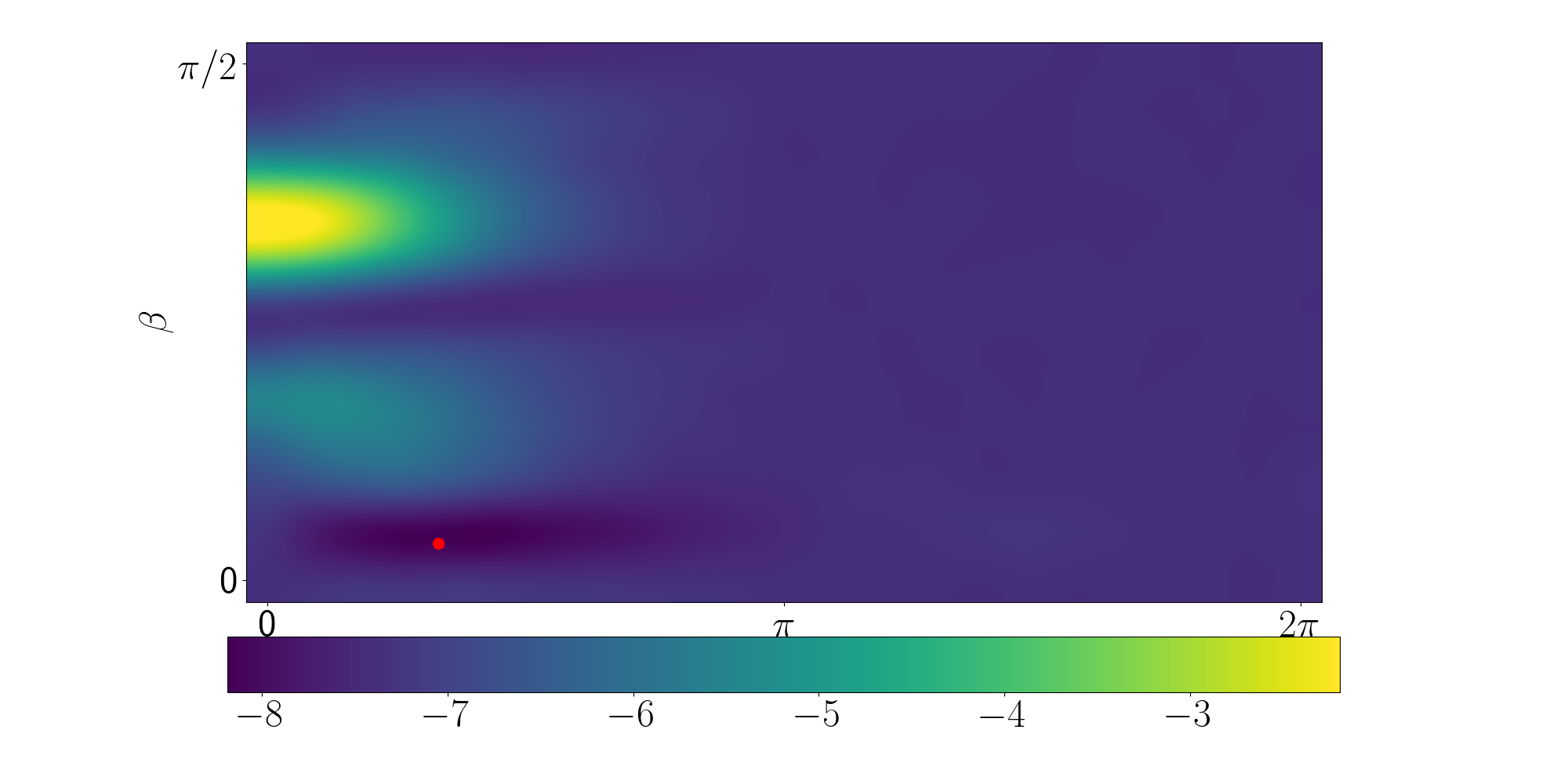}}\\
         4&\raisebox{-.5\height}{\includegraphics[width=.25\linewidth]{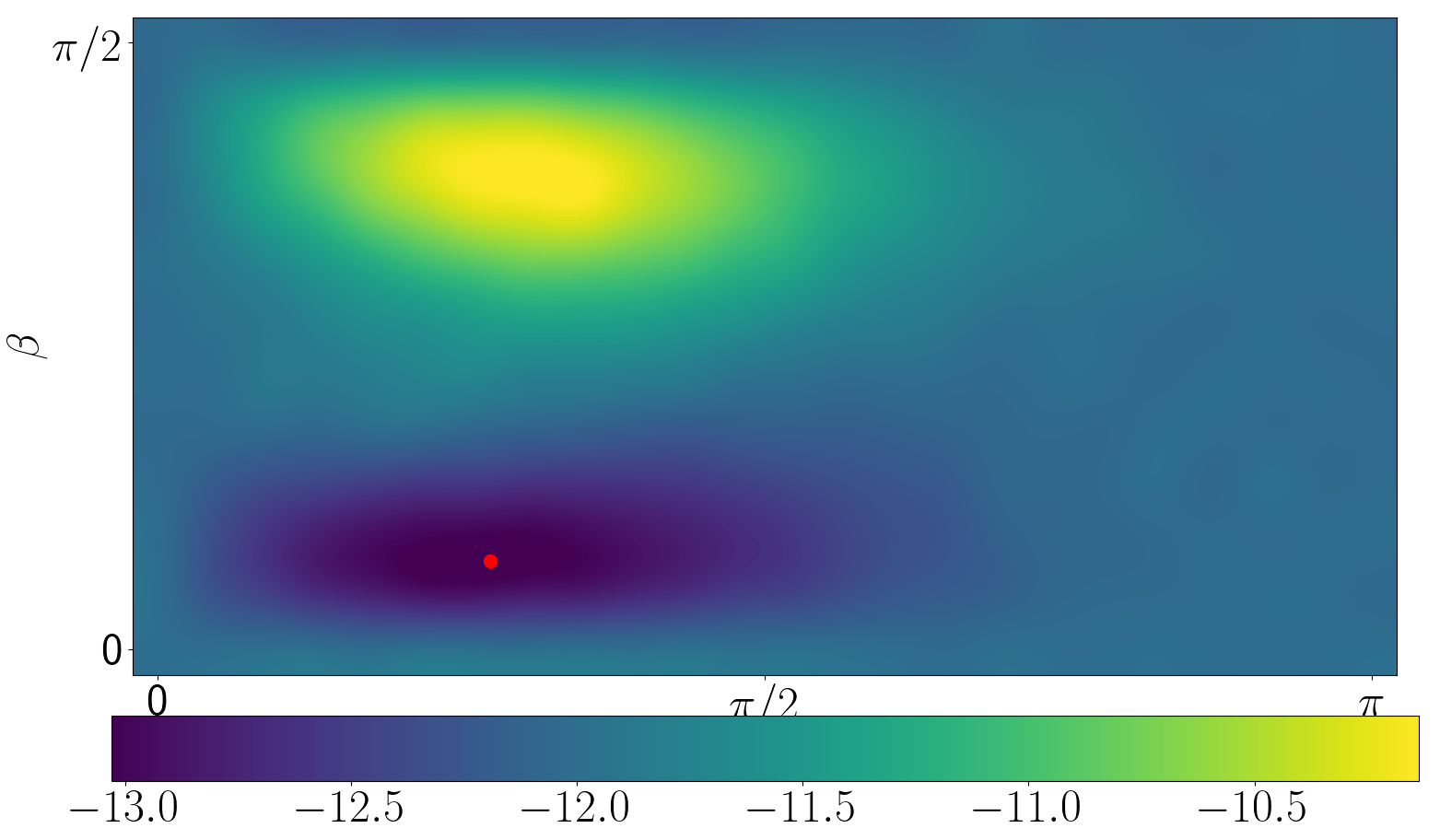}}
         &&
         \raisebox{-.5\height}{\includegraphics[width=.25\linewidth]{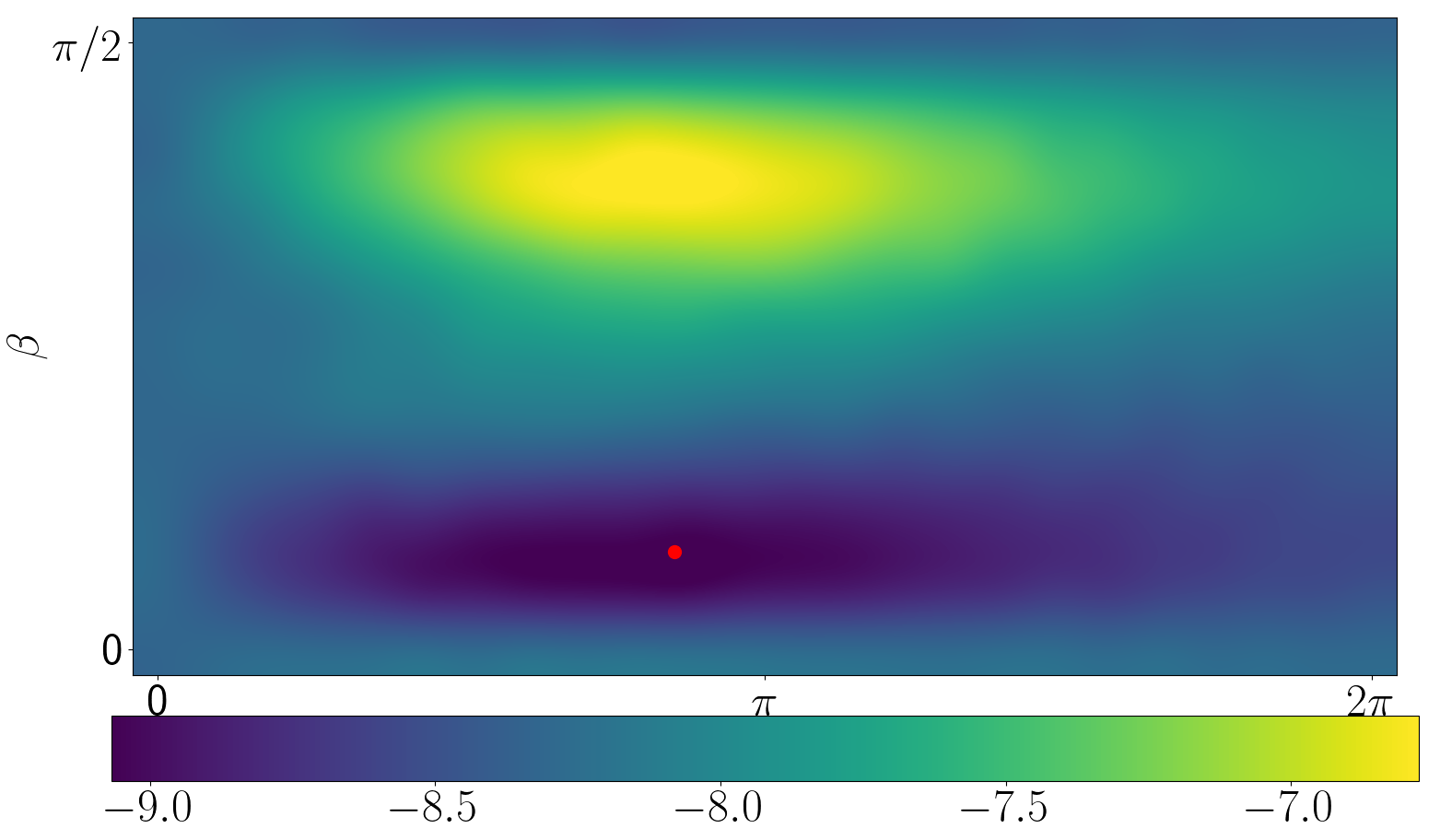}}
         &\\
         5&\raisebox{-.5\height}{\includegraphics[trim=120 20 120 20,clip,width=.25\linewidth]{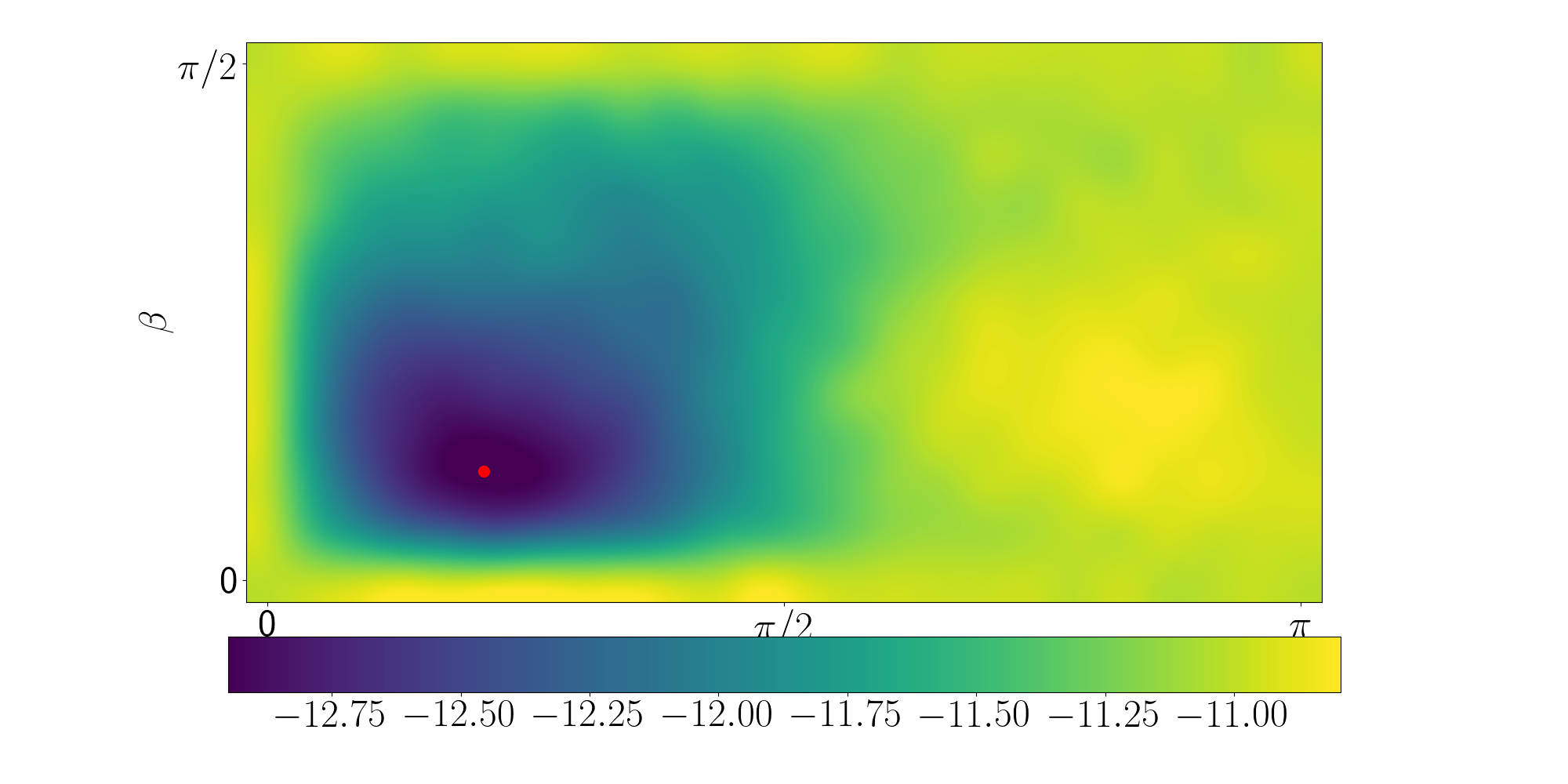}}
         &&
         \raisebox{-.5\height}{\includegraphics[trim=120 20 120 20,clip,width=.25\linewidth]{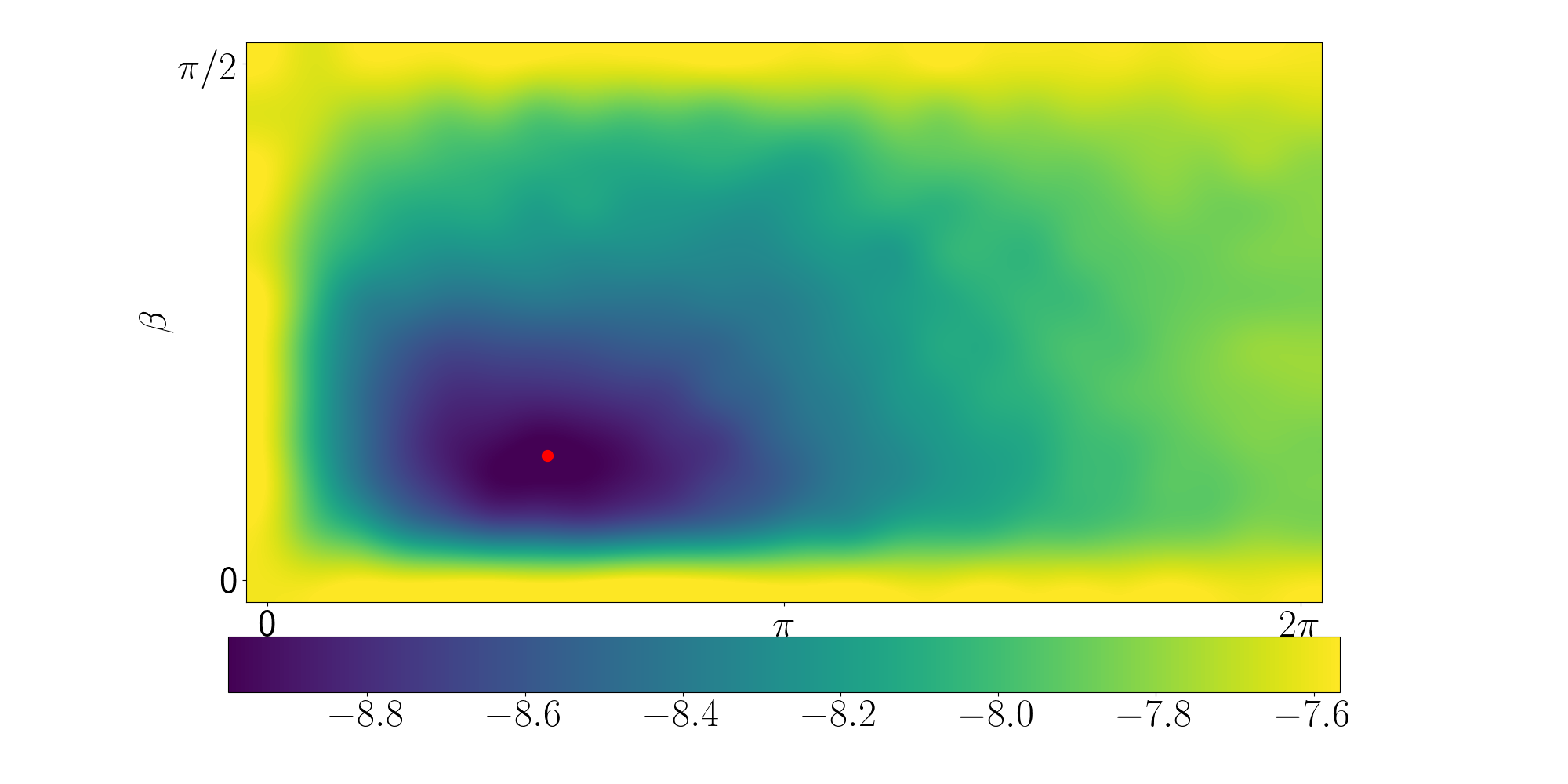}}
         &\\
         6&\raisebox{-.5\height}{\includegraphics[trim=120 20 120 20,clip,width=.25\linewidth]{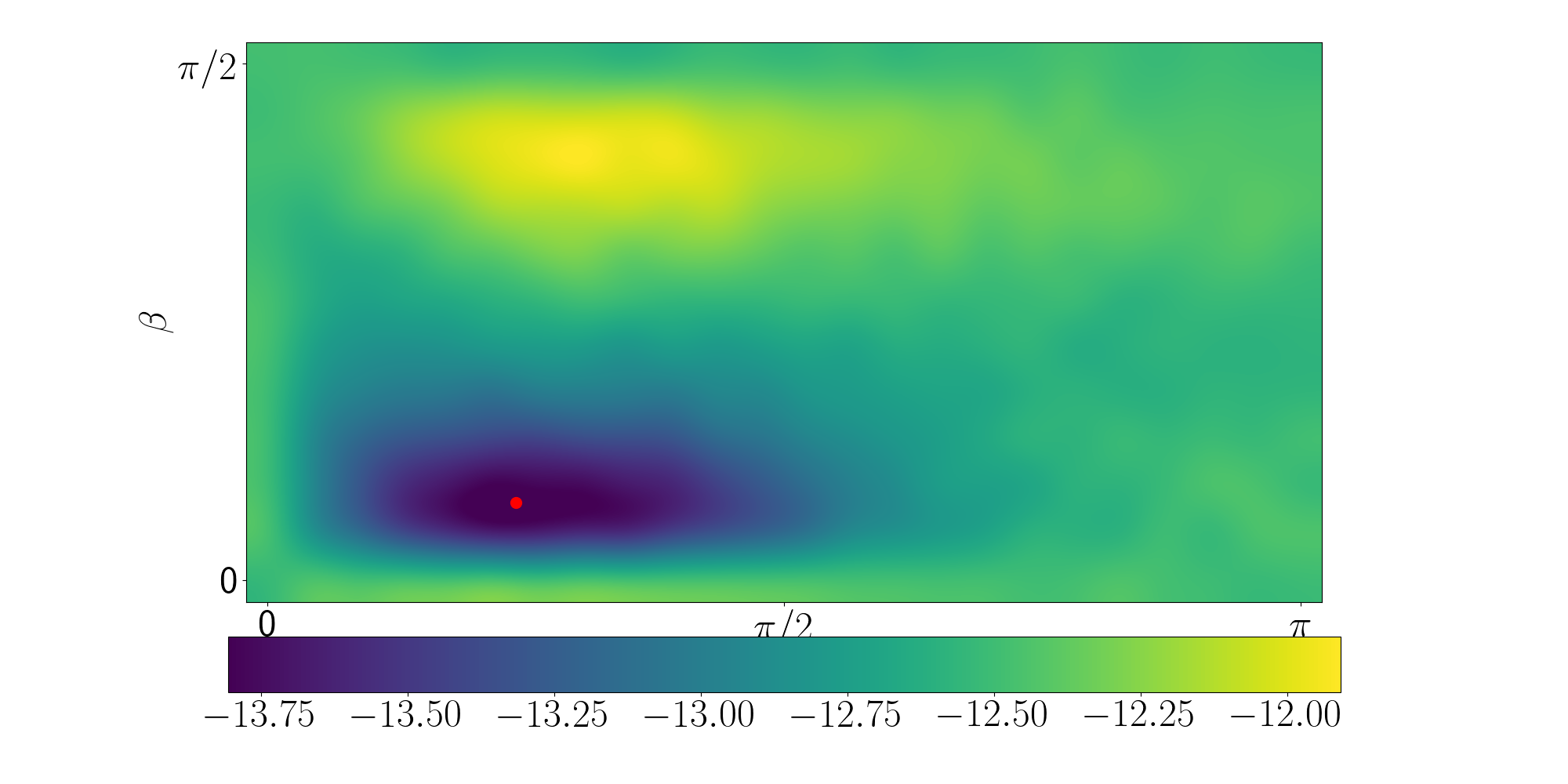}}
         &&
         \raisebox{-.5\height}{\includegraphics[trim=120 20 120 20,clip,width=.25\linewidth]{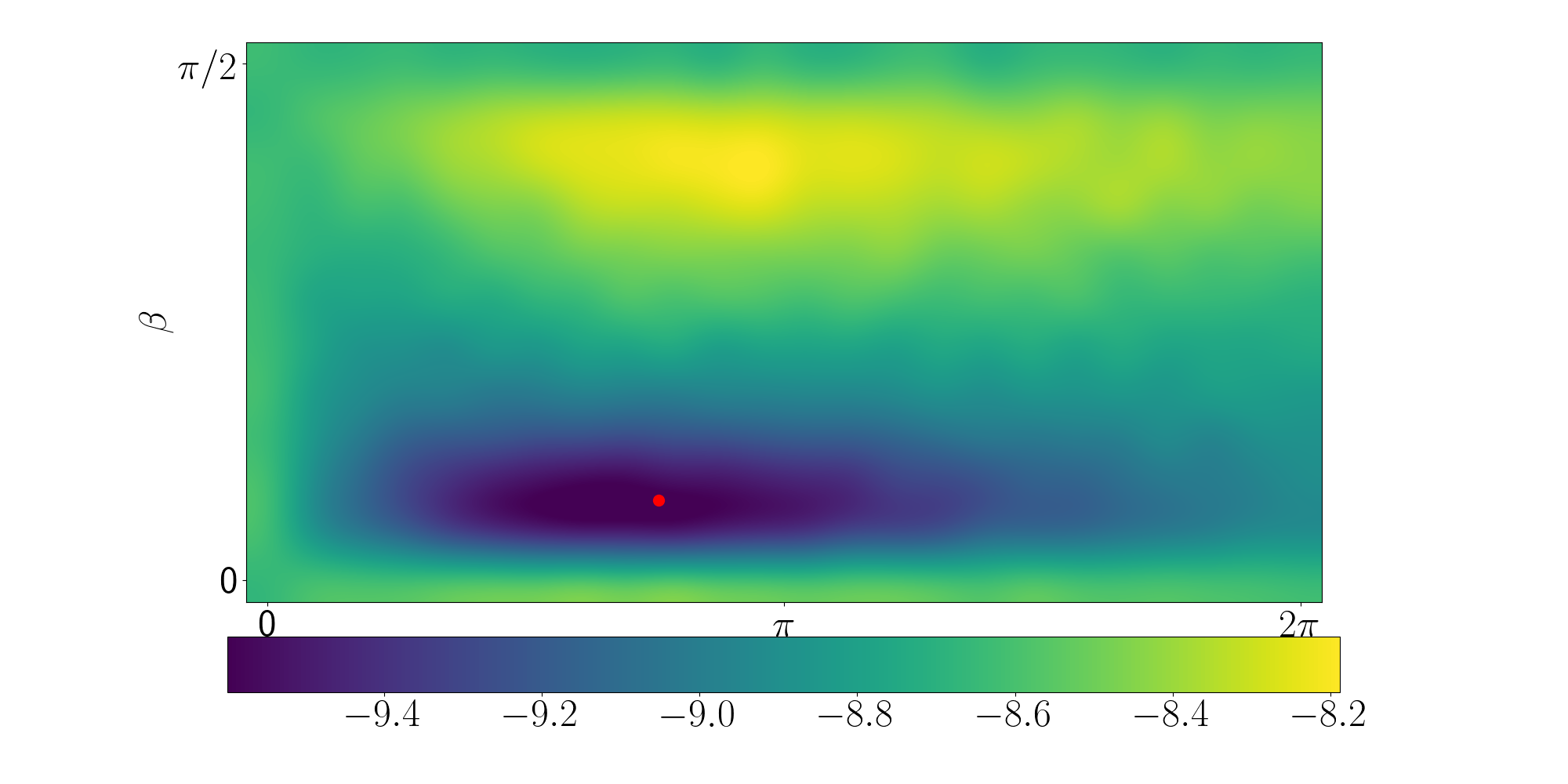}}
         &\\
         7&\raisebox{-.5\height}{\includegraphics[trim=120 20 120 20,clip,width=.25\linewidth]{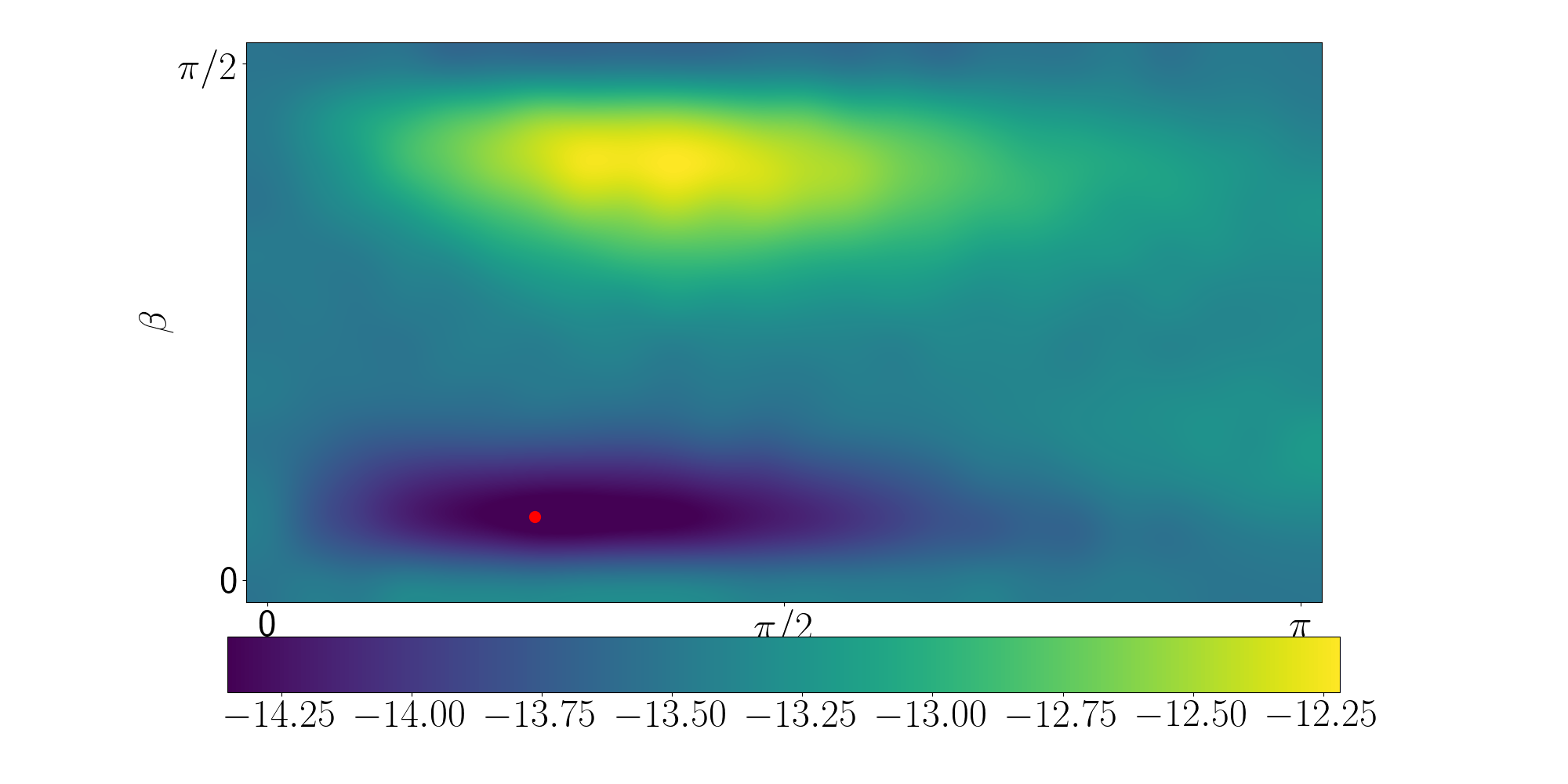}}
         &&
         \raisebox{-.5\height}{\includegraphics[trim=120 20 120 20,clip,width=.25\linewidth]{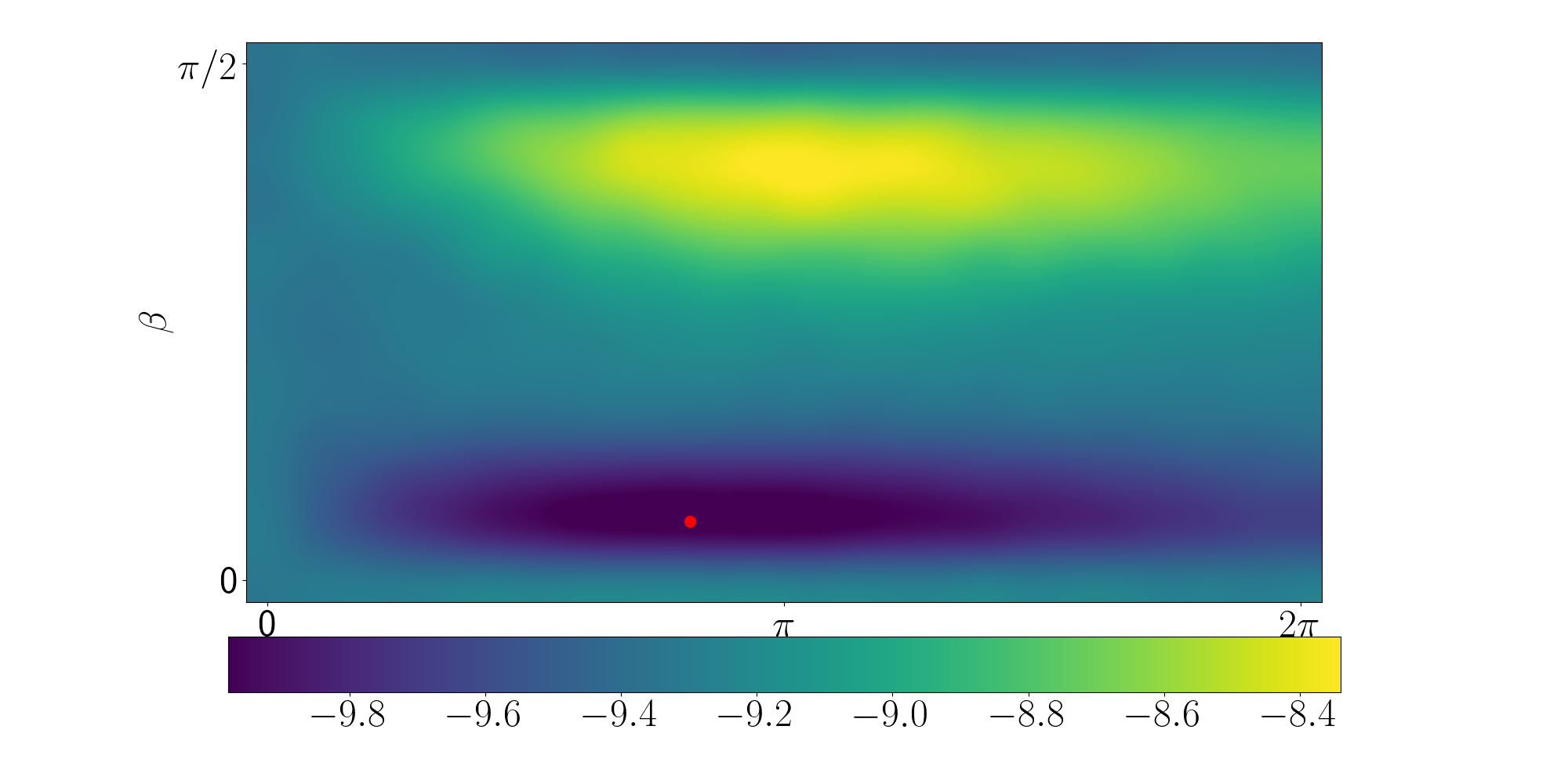}}
         &\\
         8&\raisebox{-.5\height}{\includegraphics[width=.25\linewidth]{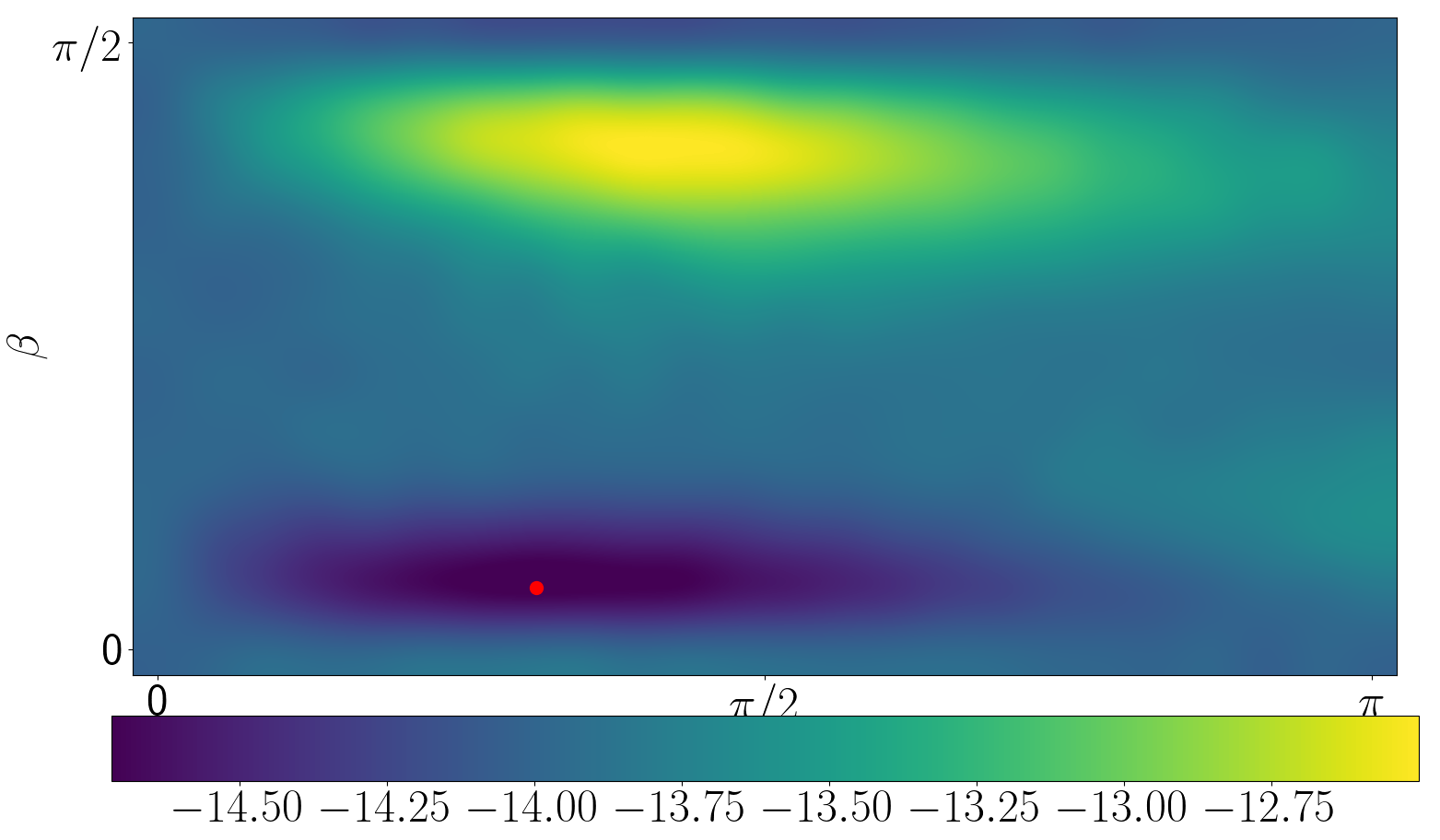}}
         &&
         \raisebox{-.5\height}{\includegraphics[width=.25\linewidth]{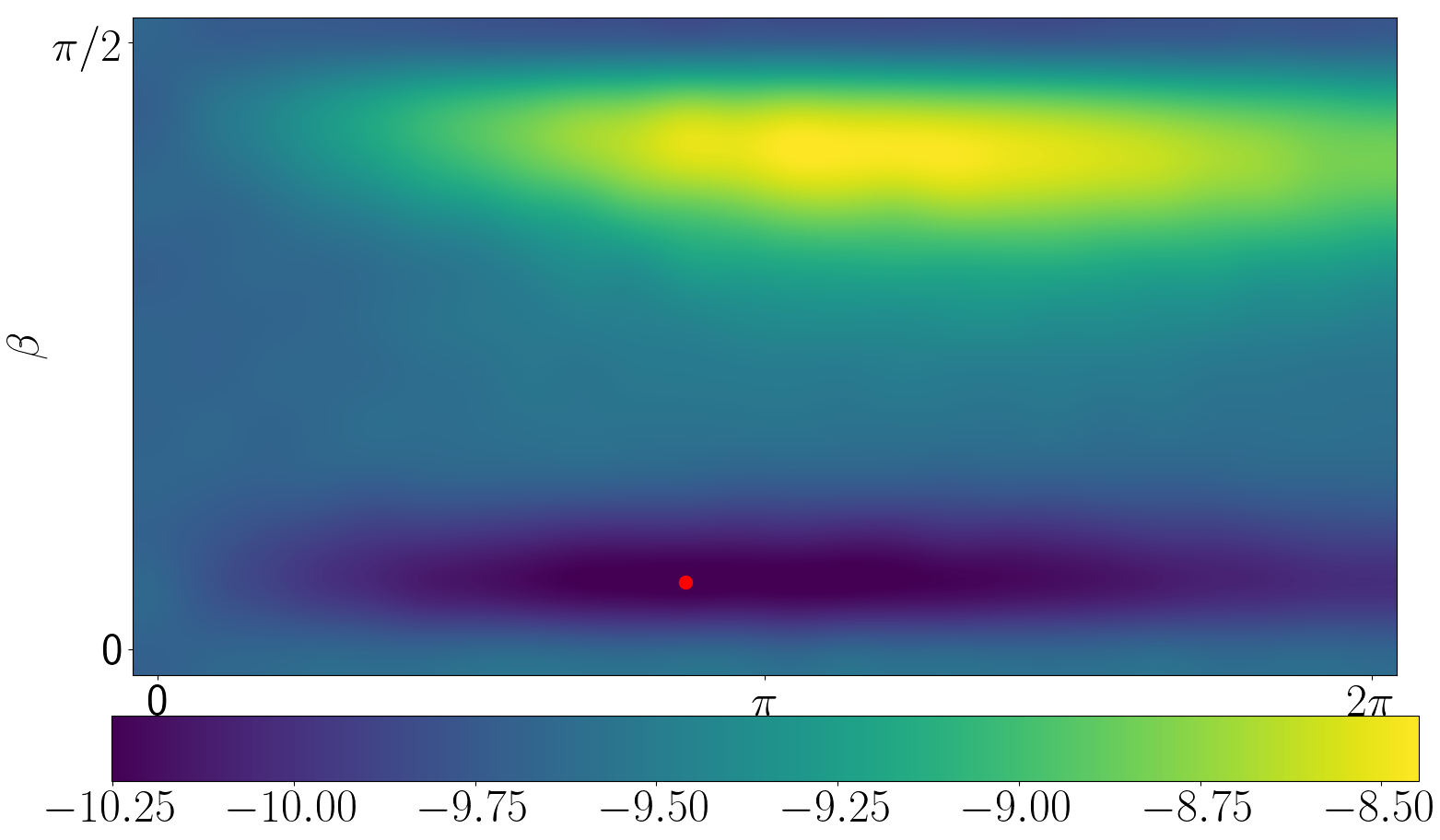}}
         &\\
    \end{tabular}
    \endgroup
    \caption{Energy landscapes for the Erdős-Rényi and Barabási-Albert graphs shown in Figure~\ref{fig:ERBA10}. The missing pictures required 40 qubits and more to generate and are therefore missing.
    Generally, the binary encoding seems to generates easier optimization problems.}
    \label{tab:E_ERBA10}
\end{table}
\end{appendix}

\bibliographystyle{plainnat}
\bibliography{references}

\end{document}